\documentclass[11pt]{JHEP3}
\usepackage{subfigure,axodraw, amsmath,graphicx}

\renewcommand{\vec}[1]{{\mathbf{#1}}}

\newcommand\gA{\mathcal{A}}
\newcommand\gB{\mathcal{B}}

\newcommand{\be}{\begin{equation}}
\newcommand{\ee}{\end{equation}}
\newcommand{\bq}{\begin{eqnarray}}
\newcommand{\eq}{\end{eqnarray}}

\title{S-matrix equivalence restored}

\author{ Chih-Hao Fu,${}^\dag$ Jonathan Fudger,${}^\ddag$  Paul R.W. Mansfield,${}^\dag$ Tim R. Morris${}^\ddag$ and  Zhiguang Xiao${}^\ddag$\\
${}^\dag$Department of Mathematical Sciences, University of Durham\\
South Road, Durham, DH1 3LE, U.K.\\ 
${}^\ddag$School of Physics and Astronomy,  University of Southampton\\
Highfield, Southampton, SO17 1BJ, U.K.\\ 
E-mails:
 \email{chih-hao.fu@durham.ac.uk}, \email{j.p.fudger@phys.soton.ac.uk},
 \email{P.R.W.Mansfield@durham.ac.uk}, \email{T.R.Morris@soton.ac.uk}, 
\email{z.g.xiao@phys.soton.ac.uk}}

\maketitle

\preprint{SHEP 09-06,DCPT-09/31}

\abstract{The canonical transformation that maps light-cone Yang-Mills
theory to a Lagrangian description of the MHV rules is non-local,
consequently the two sets of fields do not necessarily generate the
same S-matrix. By deriving a new recursion relation for the canonical
transformation expansion coefficients, we find a direct map between
these coefficients and tree level light-cone diagrams. We use this to
show that, at least up to one-loop with dimensionally regularised MHV
vertices, the only difference is the omission of the one-loop
amplitudes in which all gluons have positive helicity.}

\begin{document}
\section{Introduction}

The MHV rules of Cachazo, Svr\v{c}ek and Witten \cite{CSW} are
equivalent to a new set of Feynman rules for QCD tree-level scattering
amplitudes that are particularly efficient. Initially conjectured on
the basis of an analogy with strings moving on twistor space
\cite{Witten:2003nn} they were proven by recursion relations
\cite{Britto:2005fq}. They emerge from  gauge fixing a twistor space
action for Yang-Mills \cite{Boels:2006ir}-\cite{Boels:2008du} and can
also be derived using a canonical transformation applied to the
light-cone gauge Yang-Mills Lagrangian
\cite{Gorsky:2005sf,Mansfield:2005yd}.  To generalise these rules to
loop level requires the introduction of a regulator, for example some
variant of dimensional regularisation.  Although much of the
mathematical structure underlying this approach to Yang-Mills theory,
such as conformal invariance and twistor space, is broken by the
passage to arbitrary dimension there is some cause for optimism that
progress towards formulating MHV rules for loop processes can still be
made \cite{Brandhuber:2004yw}.  One of the features that emerges at
loop order is that the new fields do not generate the same S-matrix as
the original ones because of the non-locality of the canonical
transformation. This effect accounts for the one-loop amplitudes for
gluons of purely positive helicity which would otherwise appear to be
absent from the theory. However, it is potentially damaging for the
efficiency of the MHV rules because it would seem to require an extra
ingredient in the calculation of amplitudes  to describe the
translation between the two sets of fields. It is the purpose of this
paper to show that although extra structure is required to translate
between the two sets of fields, these `translation kernels'
are required only for all plus amplitudes at
one-loop, so that for the calculation of general amplitudes
we are free to use Green functions for either set of fields, thus
partially regaining the simplicity of the CSW rules for the regulated
theory. In theories with exact supersymmetry these problems are absent
and it is known that four dimensional MHV vertices and MHV rules may
be used to recover all amplitudes at one loop
\cite{Brandhuber:2005kd}. 

We begin by describing the canonical transformation as it is
constructed in four dimensions.  Using light-cone co-ordinates in
Minkowski space \begin{equation} \hat x = \tfrac1{\sqrt 2}(t-x^3),
\quad \check x = \tfrac1{\sqrt 2}(t+x^3), \quad z = \tfrac1{\sqrt
2}(x^1+ix^2), \quad \bar z = \tfrac1{\sqrt 2}(x^1-ix^2).
\label{eq:lc-4d-coords} \end{equation} and the gauge condition $\hat
A=0$ allows the Yang-Mills action to be written in terms of positive
and negative helicity fields ${\gA}\equiv A_z$ and ${\bar \gA}\equiv
A_{\bar z}$ (after elimination of unphysical degrees of freedom) as
the light-cone action

\begin{equation}
\label{eq:ym-lc4d-action}
S = \frac 4{g^2} \int d\hat x\:\int_\Sigma d^3{\bf x}\:({\cal L}^{-+}+{\cal L}^{-++}+{\cal L}^{--+}+{\cal L}^{'--++}),
\end{equation}
where
\begin{align}
\label{eq:ym-4d-mp}
{\cal L}^{-+} &= \phantom{-}{\rm tr}\,
{\bar \gA}\,\left(\check\partial\hat\partial-
\partial\bar\partial\right)\,\gA\,,\\
\label{eq:ym-4d-mpp}
{\cal L}^{-++}&=-{\rm tr}\,
({\bar\partial}{\hat\partial}^{-1} {  \gA})\:
[{  \gA},{\hat\partial} {\bar\gA}]\,,\\
\label{eq:ym-4d-mmp}
{\cal L}^{--+}&=-{\rm tr}\, [{\bar
{\cal A}},{\hat\partial} {  \gA}]\:
({  \partial}{\hat\partial}^{-1} {\bar \gA})\,,\\
\label{eq:ym-4d-mmpp}
{\cal L}^{'--++}&=-{\rm tr}\, [{\bar \gA
},{\hat\partial} { \gA }]\:{\hat\partial}^{-2}\: [{ \gA
},{\hat\partial} {\bar \gA }]\,,
\end{align}
and $\Sigma$ is a constant-$\hat x$ quantisation surface and $d^3{\bf
x}=d\check x\,dz\,d\bar z$.
 
The combination ${\cal L}^{-+}+ {\cal L}^{-++}$ by itself describes
self-dual gauge theory \cite{Chalmers:1996rq}. At tree-level this is a
free theory because the only connected scattering amplitudes that can
be constructed involve one negative helicity particle and an arbitrary
number of positive helicity particles. The Feynman diagrams
contributing to this are the same as in the full Yang-Mills theory,
for which such amplitudes are known to vanish.  (Bizarrely, the
one-loop amplitudes for processes involving only positive helicity
particles are non-zero, and these are the only non-vanishing
amplitudes in the theory.) This encourages us to find a new field
$\gB$ that is a non-local functional of $\gA$ on the surface of
constant $\hat x$ such that ${\cal L}^{-+}+ {\cal L}^{-++}$ can be
written as a free theory, i.e. 
\begin{equation}
\label{eq:transform-4d}
{\cal L}^{-+}[\gA, {\bar\gA}] + {\cal L}^{-++}[\gA, {\bar\gA}] = {\cal L}^{-+}[\gB,{\bar\gB}]\,,
\end{equation}
where ${\bar\gB}$ is determined by the requirement that the
transformation be canonical:
\begin{gather}
  \hat\partial \bar\gA^a(\hat x,\vec x) = \int_\Sigma d^{3}\vec y \:
  \frac{\delta\gB^b(  \hat x,\vec y)}{\delta\gA^a(\hat x  ,\vec x)} \hat\partial
  \bar\gB^b(\hat x  ,\vec y)\Leftrightarrow \hat\partial
\bar\gB^a(\hat x,\vec x) = \int_\Sigma d^{3}\vec y \:
  \frac{\delta\gA^b(  \hat x,\vec y)}{\delta\gB^a(\hat x  ,\vec x)} \hat\partial
  \bar\gA^b(\hat x  ,\vec y).
  \label{eq:momentum-transform} 
\end{gather}
This transformation is readily expressed in terms of the fields after
taking the Fourier transform with respect to position within the
quantisation surface
\bq
&&\gB(\hat x,\vec p)=\gA(\hat x,\vec p)+\nonumber\\
&&
\sum_{n=2}^\infty \int {d^3k_1\over (2\pi)^3}\dots{d^3k_n\over (2\pi)^3} {{\hat p}^{n-1}\,(2\pi)^3\,\delta^3 (\vec p-\sum\vec k_i)\over
(p,k_1)\,(p,k_1+k_2)\dots(p,k_1+\dots+k_{n-1})} \gA(\hat x,\vec k_1)\dots
 \gA(\hat x,\vec k_n)\label{transf}
\eq
where 
\begin{equation}(k_1,k_2)\equiv {\hat k_1}{k_2}
-{\hat k_2}{k_1}\,.
\end{equation}
The transformation is therefore local in $\hat x$ and the coefficients
of the products $\gA\dots\gA$ are independent of both $\hat x$ and
$\bar k$.  (\ref{eq:momentum-transform}) shows that $\bar \gA$ is a
linear functional of $\bar \gB$, which we write as
\bq
&&\bar\gA(\hat x,\vec p)=\bar\gB(\hat x,\vec p)+\nonumber\\
&&
\sum_{m=3}^\infty \sum_{s=2}^m\int {d^3k_1\over (2\pi)^3}\dots{d^3k_n\over (2\pi)^3} {{\hat k_s}\over \hat p}\,\Xi^{s-1}(\vec p,
-\vec k_1,\dots,-\vec k_m)\times\nonumber\\
&&\hskip1in
(2\pi)^3 
\,\delta^3 (\vec p-\sum\vec k_i)  \gB(\hat x,\vec
k_1)\dots\bar\gB(\hat x,\vec k_s)\dots
 \gB(\hat x,\vec k_m)\label{bartransf}
\eq
so that when the remaining terms in the action are written in the new
variables we obtain an infinite series, each term of which contains
two powers of $\bar\gB$. Labelling these terms by their helicities
gives
\begin{equation}
 {\cal L}[\gA, {\bar\gA}]={\cal L}^{-+}[\gB,\bar\gB] + {\cal L}^{--+}[\gB,\bar\gB ] + {\cal L}^{--++}[\gB,\bar\gB ] + {\cal L} ^{--+++}[\gB,\bar\gB ] + \cdots.\label{lagexp}
\end{equation}
The coefficients of the fields in the interaction terms can be shown
\cite{Ettle:2006bw}, by explicit calculation, to consist of the
Parke-Taylor amplitudes \cite{Parke:1986gb} (continued off-shell).

\medskip

The LSZ procedure gives scattering amplitudes in terms of the momentum
space Green functions (suitably normalised) for $\gA$ and $\bar\gA$
fields by cancelling each external leg using a factor $p^2$ and then
taking each momentum on-shell, $p^2\rightarrow 0$. The equivalence
theorem for S-matrix elements seems to allow us to use Green functions
for the $\gB$ and $\bar\gB$  fields instead of the $\gA$ and
$\bar\gA$, provided we include a multiplicative wave-function
renormalisation.  This is because, to leading order in the fields,
$\gA$ is the same as $\gB$. In any Feynman diagram contributing to a
Green function these fields are attached to the rest of the diagram by
a propagator $\sim 1/p^2$ which cancels the LSZ factor of $p^2$ and so
survives the on-shell limit.  In the higher order terms in
(\ref{transf}) the momentum $p$ is shared between the $\gA$ fields, so
the propagators that attach these to diagrams cannot directly cancel
$p^2$. The cancellation can occur if the diagram forces just these
momenta to flow together through some internal line, because by
momentum conservation this line will contribute $\sim 1/p^2$. The
effect of such diagrams is to renormalise the field, and this will
cancel in the computation of scattering amplitudes. Another source of
$1/p^2$ could be the kernels in (\ref{transf}). These kernels are
non-local within the quantisation surface, and a requirement of the
equivalence theorem is that the transformation be local. However our
transformation is still local in light-cone `time' $\hat x$ which
means that the kernels are independent of $\check p$ (and also, for
other reasons, $\bar p$) so it is hard (but not impossible, as we will
see) to imagine how the kernels can generate the $1/p^2$ needed to
stop us generalising the theorem to the case in hand.

\medskip So it would seem safe to invoke the S-matrix equivalence
theorem and use the $\gB$ fields to calculate scattering amplitudes,
expecting to get physical gluon amplitudes.  It is clear that the new
Lagrangian would then generate the CSW (or MHV) rules of \cite{CSW},
and, once we have a Lagrangian we are much closer to being able to
generalise the rules beyond tree-level. However, this cannot be
correct as the rules cannot generate the one-loop amplitudes for
processes in which the gluons all have positive helicity. These
amplitudes have long been considered to be related to an anomaly
\cite{Bardeen:1995gk}. In the context of the change of variables from
$\gA$ to $\gB$ this anomaly could be related to the Jacobian which
ought to be unity since the transformation is canonical. However, in
\cite{us} it was shown instead that these amplitudes result from an
evasion of the equivalence theorem when the theory is formulated using
dimensional regularisation. This implies a flaw in the argument we
have just presented. Specifically, it was shown that in the case of
the four-point all-plus amplitude the change of variables can be
implemented with unit Jacobian by directly comparing both sides of:
\bq
&&
\lim_{p_i^2\rightarrow 0}
\int {\cal D}(\gA,\bar\gA)\,e^{iS_{lc}}\, p_1^2 \,\bar \gA^{a_1}(p_1)\dots p_4^2\, \bar \gA^{a_4}(p_4)
=\nonumber\\
&&
\lim_{p_i^2\rightarrow 0}
\int {\cal D}(\gB,\bar\gB)\,e^{iS_{MHV}}\, p_1^2 \,\{\bar \gB^{a_1}(p_1)+\dots \}\dots p_4^2\, \{\bar \gB^{a_4}(p_4)+\dots \}\label{funint}
\eq
where the dots in $\bar \gB(p_1)+\dots $ represent the extra terms
involving the $\Xi$ in (\ref{bartransf}). If we ignored these extra
terms, as the S-matrix equivalence theorem implies we should, then the
right-hand side would vanish because there are no interactions in
$S_{MHV}$ that would allow us to contract all the $\bar\gB$ together.
{Since it is known that this amplitude is in fact non-zero the extra
terms must contribute and the equivalence theorem is not directly
applicable.} These extra terms appear to spoil the efficiency of our
approach. If we have to include the details of the transformation in
computing scattering amplitudes then we are unlikely to be able to
profit from any gains resulting from the simplicity of the MHV
Lagrangian. It is the purpose of this paper to investigate just how
damaging this is. We will see that actually the problem is quite
contained and the equivalence theorem is only spoilt for a class of
known amplitudes.

To simplify our discussion we will regulate using
Four-Dimensional-Helicity regularisation \cite{Bern:1991aq} in which
the external helicity are in four dimensions and only the internal
momenta are in $D$ dimensions. It is not essential to use this scheme,
and in our earlier paper \cite{us} we used standard dimensional
regularisation, but it will simplify our expressions considerably. In
section \ref{sect:dim-reg}, we will describe this. Then in section
\ref{sect:canonial-trans}, we  examine the canonical transformation
using it. We will find that the effect of regularisation is to make
only minor changes to the recursion relations for the expansion
coefficients.  In order to avoid spurious poles in the recursion
expansion of $\Xi^s$, we also establish a new recursion relation of
$\Xi^s$ which involves only true singularities in each term of the
expansion. As a byproduct we also find a relation between the
tree-level light-cone diagrams and the canonical expansion
coefficients, which will facilitate the singularity analysis in the
translation kernel contribution later. We also review the tree-level
evasion of the equivalence theorem for the $(-++)$ amplitude in
section \ref{sect:tree-mpp}. 

After this preparation, in section \ref{sect:general-evasion}, we will
discuss systematically the different ways that the S-matrix
equivalence theorem can be evaded. We first argue that at tree-level
evasion will not occur in higher point amplitudes. Then we discuss the
three ways that the theorem can potentially be evaded at one-loop: by
dressing propagators, in tadpoles, and by infrared divergences. We
will conclude that only tadpoles can evade the equivalence theorem at
one-loop.  During this discussion, we find that there is a puzzle in
the $(++-)$ amplitude with an external leg dressed by a tadpole. By
examining the calculation of the $(+++-)$ amplitude in section
\ref{sect:mppp}, we find that the one-minus-helicity amplitudes should
come just from tadpoles made out of MHV vertices, but when we cut the
diagrams there appear to be additional contributions from equivalence
theorem evading tadpoles which can dress external legs.  In section
\ref{sect:oneloop-ppm}, we resolve this double-counting puzzle by
choosing a suitable limiting order in the LSZ procedure and show that
these extra terms do not contribute to the on-shell amplitude. Section
\ref{sect:conclusion} is the conclusion.

\section{Dimensional Regularisation}
\label{sect:dim-reg} We will regulate the ultra-violet divergences of
pure Yang-Mills by working in arbitrary space-time dimension, $D$, and
using co-ordinates which replace the pair $z, \,\bar z$ of complex
space-like co-ordinates by $D/2-1$ such pairs, $z_{(i)}, \,\bar
z_{(i)}$. In \cite{us} we used standard dimensional regularisation in
which the gauge-field $A_\mu$ has $D$ space-time components.  We could
instead use four-dimensional-helicity regularization (FDH)
\cite{Bern:1991aq} and keep $\mu$ four dimensional.  Consequently
polarisation vectors would remain four dimensional, so we retain just
two helicities, and the gauge invariance of the action is
four dimensional.  Just as in the usual dimensional regularisation the
momenta of `physical' gluons which appear in asymptotic states of
scattering processes also remain in four dimensions, but the momenta
of virtual gluons that appear as internal lines in Feynman diagrams
will be $D$ dimensional. The advantage of FDH is that the light-cone
gauge action is very similar to the four dimensional version, the only
change being in the free part which becomes
$${\cal L}^{-+} = \phantom{-}{\rm tr}\,
{\bar \gA}\,\left(\check\partial\hat\partial-
\sum_{i=1}^{D/2-1}\partial_{(i)}\bar\partial_{(i)} \right)\,\gA\,.
$$
Tree-level amplitudes are unchanged when the external legs all have
four dimensional momenta, however when the external legs are allowed
to have $D$ dimensional momenta then they are modified. In particular
the amplitudes in which all but one of the scattered gluons have the
same helicity no longer vanish. This is responsible for the
non-vanishing of the one-loop amplitude in which all the scattered
gluons have the same helicity because the optical theorem relates the
imaginary part of this latter amplitude to the product of tree-level
amplitudes of the former. The one-loop four-gluon all positive
helicity reduced amplitude is \cite{Bern:1995db}
\be
-{ig^4\over 48\pi^2}{\{p_1,p_2\}\,\{p_3,p_4\}\over (p_1,p_2)\,(p_3,p_4)}
\label{4pt}
\ee
where $p_1,\dots ,p_4$ are the momenta of the gluons and $\{p_1,p_2\}
\equiv {\hat p}_1 \bar p_{2}-{\hat p}_2 \bar p_{1}$.  The all-plus
one-loop amplitudes are missing from a na\"ive application of the MHV
rules at one-loop because if we are limited to vertices of
Parke-Taylor type then we cannot construct such amplitudes. (In
\cite{us} it was shown that such amplitudes originate in a failure of
the S-matrix equivalence of the $\gA$ and $B$ fields, we shall enlarge
on this later.)

The failure of the one minus rest plus helicity tree-level amplitudes
to vanish has significant consequences for the attempt to construct an
MHV Lagrangian in $D$ dimensions. Firstly it means that the theory
described by the truncated Lagrangian ${\cal L}^{-+}+ {\cal L}^{-++}$
that generates these amplitudes is not free. Secondly it means that
the Parke-Taylor vertices are likely to be much more complicated in
$D$ dimensions because their simplicity in four dimensions can be
explained within the BCFW recursion method \cite{Britto:2005fq} as
deriving from the vanishing of the one minus rest plus tree-level
amplitude. We will now investigate how damaging these facts are.

\section{Canonical transformation in $D$ dimensions}
\label{sect:canonial-trans}
\subsection{Recursion relations for the expansion coefficients}

Perhaps surprisingly we can still construct a canonical transformation
in $D$ dimensions so that (\ref{eq:transform-4d}) holds. Using FDH
regularization, and given (\ref{eq:momentum-transform}) we have to
solve 
\be
\omega\, \gA(x)+\gA(x)\,\left({\bar\partial \over\hat\partial}\gA(x)\right)-\left({\bar\partial \over\hat\partial}\gA(x)\right)\,\gA(x)
=\int_{\hat x={\it const.}}\omega^{\prime}\, \gB(x^\prime)\,{\delta\gA(x)\over\delta\gB(x^\prime)}\,d^{D-1}x^\prime
\ee
where
$$
\omega=\sum_{i=1}^{D/2-1}\partial_{(i)} \bar\partial_{(i)}/\hat\partial\,.
$$
Re-arranging:
\be
\omega\, \gA(x)-\int_{\hat x={\it const.}}\omega\, \gB(x^\prime)\,{\delta\gA(x)\over\delta\gB(x^\prime)}\,d^{D-1}x^\prime
=-\gA(x)\,\left({\bar\partial \over\hat\partial}\gA(x)\right)+\left({\bar\partial \over\hat\partial}\gA(x)\right)\,\gA(x)
\,.\label{bas}
\ee
We make the basic assumption, appropriate to perturbation theory, that
we can expand the Fourier transform of $\cal A$ in powers of the
transform of $\gB$, with kernels $\Upsilon$. (Note we use the same
symbol for the fields and their Fourier transforms) 
\be
\gA_p=\sum_{n=1}^\infty\int \Upsilon(p,p_1,\dots
,p_n)\,\delta(p+\sum_{i=1}^n p_i)\,\gB_{\bar1}\dots \gB_{\bar n}\, {\rm
d}^Dp_1\dots {\rm d}^Dp_n\,,
\label{upsi2}
\ee
where we adopt the notation that the subscripts of the fields label
the momenta: ${\cal A}_p\equiv {\cal A}(p)$ and ${\cal B}_{\bar
\imath}\equiv{\cal B}(-p_i)$. Then the first term on the
left-hand-side of (\ref{bas}) multiplies each term in the expansion by
the Fourier transform of $\omega$, $i\Omega_0\equiv i\Omega(p)$,
whereas the second replaces each $\gB_{\bar \imath}$ by
$i\Omega_i\gB_{\bar \imath}$, and the right-hand-side glues two
expansions together using what is essentially the three-point vertex
corresponding to helicities $++-$ (and which we are attempting to
eliminate from the theory by performing the canonical transformation
to new variables). This is most easily represented graphically. Let us
denote the expansion (\ref{upsi2}) by
\begin{eqnarray}
\begin{picture}(50,52)
        \SetOffset(30,24)\Text(0,-3)[bl]{$\gA=$}
\end{picture}
   \begin{picture}(50,52)
        \SetOffset(30,24)
        \Line(-25,0)(-2,0)\Vertex(-12,0){2}
        \Line(0,0)(25,0)
        \BCirc(0,0){8}
        \Text(3,-4)[br]{$1$}
 \Text(39,-3)[br]{$\gB$}
      \end{picture}
\begin{picture}(50,52)
        \SetOffset(30,24)\Text(0,0)[bl]{$+$}
\end{picture}
\begin{picture}(50,52)
        \SetOffset(30,24)
        \Line(-25,0)(-2,0)
        \Line(0,0)(25,15)
        \Line(0,0)(25,-15)
        \BCirc(0,0){8}\Vertex(-12,0){2}
        \Text(3,-4)[br]{$2$}
        \Text(38,13)[br]{$\gB$}
        \Text(38,-18)[br]{$\gB$}
      \end{picture}
\begin{picture}(50,52)
        \SetOffset(30,24)\Text(0,0)[bl]{$+$}
\end{picture}
\begin{picture}(50,52)
        \SetOffset(30,24)
        \Line(-25,0)(-2,0)
        \Line(0,0)(17,25)\Line(0,0)(25,0)\Text(38,-3)[br]{$\gB$}
        \Line(0,0)(17,-25)
        \BCirc(0,0){8}\Vertex(-12,0){2}
        \Text(3,-4)[br]{$3$}
        \Text(30,23)[br]{$\gB$}
        \Text(30,-28)[br]{$\gB$}
   \end{picture}
\begin{picture}(50,52)
        \SetOffset(30,24)\Text(0,0)[bl]{$+\,....$}
\end{picture}
\label{Agraph}
\end{eqnarray}
and the Fourier transform of the right-hand-side of (\ref{bas}) by
\\\\
\centerline{\begin{picture}(80,50)
        \SetOffset(30,25)
        \Line(-25,0)(-2,0)\Text(-25,0)[r]{$i$}
        \Vertex(-8,0){2}
        \Line(0,0)(34,20)
        \Line(0,0)(34,-20)
        \Vertex(0,0){4}
        \Text(36,20)[l]{$j$} \Text(36,-20)[l]{$k$}
\end{picture}
\begin{picture}(60,50)
        \SetOffset(0,25)
       \Text(0,0)[l]{$=\bar V^2(p_j,p_k,p_i)/\hat p_i\,,$} 
\end{picture}}
where $\bar V^2(p_1,p_2,p_3)= i(\bar 1/\hat1-\bar2/\hat 2)\hat 3$  is
the factor from the three-point $(++-)$ vertex of the lagrangian
(\ref{lagexp}).  The small black dots in the diagram denote the
minus-helicity end of the propagators. Then the terms in (\ref{bas})
with n $\gB$ fields give
\\
{\centerline{\begin{picture}(85,80)
        \SetOffset(30,40)\Text(0,-10)[bl]{$\Big(\sum_0^n \Omega_i\Big)$}
\end{picture}
\begin{picture}(60,80)
        \SetOffset(30,40)
        \Line(-25,0)(-2,0)
        \Line(0,0)(25,15)
        \Line(0,0)(25,-15)
        \BCirc(0,0){8}\Vertex(-12,0){2}
       \Text(3,-4)[br]{$n$}
        \Text(38,13)[br]{$\gB$}\Text(38,8)[br]{.}\Text(38,3)[br]{.}\Text(38,-2)[br]{.}\Text(38,-7)[br]{.}
        \Text(38,-18)[br]{$\gB$}
      \end{picture}
 \begin{picture}(60,80)
        \SetOffset(30,40)\Text(-25,-6)[bl]{$=-\sum_{r+s=n}$}
\end{picture}
\begin{picture}(60,80)
        \SetOffset(30,40)
        \Line(-25,0)(-2,0)\Vertex(-8,0){2}
        \Line(0,0)(34,20)
        \Line(0,0)(34,-20)
        \Vertex(0,0){4}
\Line(34,20)(59,35)
        \Line(34,20)(59,5)
        \BCirc(34,20){8}
        \Text(37,16)[br]{$r$}\Vertex(22.7,13.3){2}
        \Text(72,33)[br]{$\gB$}\Text(72,28)[br]{.}\Text(72,23)[br]{.}\Text(72,18)[br]{.}\Text(72,13)[br]{.}
        \Text(72,2)[br]{$\gB$}
\Line(34,-20)(59,-35)
        \Line(34,-20)(59,-5)
        \BCirc(34,-20){8}
        \Text(37,-23)[br]{$s$}\Vertex(22.7,-13.3){2}
        \Text(72,-40)[br]{$\gB$}\Text(72,-28)[br]{.}\Text(72,-23)[br]{.}\Text(72,-18)[br]{.}\Text(72,-13)[br]{.}
        \Text(72,-10)[br]{$\gB$}
      \end{picture}
}}
If we were to use usual dimensional regularisation rather than FDH,
we would have arrived at the same graphical equation, but with indices
attached to the lines and $\bar V^2(p_j,p_k,p_i)=i(\{p_i,p_j\}_K\delta_{IJ}/\hat p_k+
\{p_k,p_i\}_J\delta_{KI}/\hat p_j)$, in the notation of \cite{us}.
We can divide through by $\sum \Omega$ when it is
non-zero and obtain the recursion relation for $\Upsilon$ in momentum
space
\begin{eqnarray}
 \label{eq:-UpsilonR}
\Upsilon(\bar1\cdots \bar n) &=& {{ 1}\over \hat 1(\Omega_1
+\cdots+\Omega_n)}\sum^{n-1}_{j=2}\bar V^2(P_{2j},P_{j+1,n},1)
\Upsilon(- ,\bar 2,\dots,\bar j)\Upsilon(- ,\overline{j+1},\dots,\bar n)\,,
\end{eqnarray}
where we use the notation $P_{i,j}=p_i+p_{i+1}+\cdots+p_{j}$, for
$j>i$, $P_{i,j}=p_{i}+p_{i+1}+\cdots+p_{n}+p_1+\cdots +p_j$ for $j<i$,
$\bar n=-p_n$ and the $-$ in  the bracket of $\Upsilon$ denotes the
minus of the sum of all the other momenta in $\Upsilon$.  This can be
represented graphically 
\\
\centerline{
\begin{picture}(60,80)
        \SetOffset(30,40)
        \Line(-25,0)(-2,0)
        \Line(0,0)(25,15)
        \Line(0,0)(25,-15)
        \BCirc(0,0){8}\Vertex(-12,0){2}
       \Text(3,-4)[br]{$n$}
        \Text(38,13)[br]{$\gB$}\Text(38,8)[br]{.}\Text(38,3)[br]{.}\Text(38,-2)[br]{.}\Text(38,-7)[br]{.}
        \Text(38,-18)[br]{$\gB$}
      \end{picture}
\begin{picture}(20,80)
        \SetOffset(30,40)\Text(0,-6)[bl]{$=$}
\end{picture}
{\begin{picture}(60,80)
        \SetOffset(30,40)\Text(0,-10)[bl]{${1\over\sum_0^n \Omega_i}$}
\end{picture}}
{\begin{picture}(50,80)
        \SetOffset(30,40)\Text(-20,-6)[bl]{$\sum_{r+s=n}$}
\end{picture}}
{\begin{picture}(60,80)
        \SetOffset(30,40)
        \Line(-25,0)(-2,0)\Vertex(-8,0){2}
        \Line(0,0)(34,20)
        \Line(0,0)(34,-20)
        \Vertex(0,0){4}
\Line(34,20)(59,35)
        \Line(34,20)(59,5)
        \BCirc(34,20){8}
        \Text(37,16)[br]{$r$}\Vertex(22.7,13.3){2}
        \Text(72,33)[br]{$\gB$}\Text(72,28)[br]{.}\Text(72,23)[br]{.}\Text(72,18)[br]{.}\Text(72,13)[br]{.}
        \Text(72,2)[br]{$\gB$}
\Line(34,-20)(59,-35)
        \Line(34,-20)(59,-5)
        \BCirc(34,-20){8}
        \Text(37,-23)[br]{$s$}\Vertex(22.7,-13.3){2}
        \Text(72,-40)[br]{$\gB$}\Text(72,-28)[br]{.}\Text(72,-23)[br]{.}\Text(72,-18)[br]{.}\Text(72,-13)[br]{.}
        \Text(72,-10)[br]{$\gB$}
      \end{picture}}
}\\
We will encounter situations when $\sum \Omega$ vanishes, and then we
need a prescription for dealing with this singularity. We will address
this in the appendices.

If we denote ${-1/(\sum_0^n \Omega_i)}$ by a closed broken curve
cutting each line whose momentum appears in the sum, each order of the
expansion of $\cal A$ can be represented as  
\begin{eqnarray}\label{eq:UpsilonR}
\begin{picture}(68,90)
        \SetOffset(27,45)
        \Line(-25,0)(-2,0)
        \Line(0,0)(25,15)
        \Line(0,0)(25,-15)
        \BCirc(0,0){7}
\Vertex(-12,0){2}
       \Text(3,-4)[br]{$n$}
        \Text(38,13)[br]{${\cal B}_1$}\Text(36,2)[]{\vdots}
        \Text(38,-18)[br]{${\cal B}_n$}
      \end{picture}
  \begin{picture}(50,90)
        \SetOffset(30,45)\Text(-30,-5)[l]{$=
-\sum\limits_{r+s=n}$}
\end{picture}
\begin{picture}(60,90)
        \SetOffset(30,45)
        \DashCArc(0,0)(12,103,180){1}
        \DashLine(-3,12)(31,32){1}
\DashCArc(34,20)(12,13,103){1}
\DashCArc(0,0)(12,180,257){1}
\DashLine(-3,-12)(31,-32){1}
\DashCArc(34,-20)(12,257,347){1}
\DashLine(46,-22)(46,22){1}
        \Line(-25,0)(-2,0)
\Vertex(-8,0){2}
        \Line(0,0)(34,20)
        \Line(0,0)(34,-20)
        \Vertex(0,0){4}
\Line(34,20)(59,35)
        \Line(34,20)(59,5)
        \BCirc(34,20){7}
        \Text(37,16)[br]{$r$}\Vertex(22.7,13.3){2}
        \Text(72,33)[br]{${\cal B}_1$}\Text(57,22)[]{$\vdots$}
        \Text(72,2)[br]{$\cal B$}
\Line(34,-20)(59,-35)
        \Line(34,-20)(59,-5)
        \BCirc(34,-20){7}
        \Text(37,-23)[br]{$s$}\Vertex(22.7,-13.3){2}
        \Text(72,-40)[br]{${\cal B}_n$}\Text(57,-18)[]{$\vdots$}
        \Text(72,-10)[br]{${\cal B}$}
      \end{picture}
\end{eqnarray}
where
\\
\centerline{\begin{picture}(68,50)
        \SetOffset(27,25)
        \Line(-25,0)(-2,0) 
\Text(-26,0)[r]{${ }_{\bar p}$}
        \Line(0,0)(25,15)
        \Line(0,0)(25,-15)
        \BCirc(0,0){7}
\Vertex(-12,0){2}
       \Text(3,-4)[br]{$n$}
        \Text(38,13)[br]{${\cal B}_1$}\Text(36,2)[]{\vdots}
        \Text(38,-18)[br]{${\cal B}_n$}
      \end{picture}
\begin{picture}(130,50)
        \SetOffset(30,25)\Text(-30,0)[l]{$=\int_{1\cdots
n}\Upsilon(\bar p\bar 1\cdots \bar n){\cal B}_1 \cdots {\cal B}_n\,.$}
\end{picture}}
\\
This can be easily iterated, starting with the
leading term $\gA=\gB$:
\\
  \centerline{
\begin{picture}(20,70)
        \SetOffset(30,35)\Text(0,-3)[bl]{$\gA=$}
\end{picture}
   \begin{picture}(50,70)
        \SetOffset(30,35)
 \Text(39,-3)[br]{$\gB$}
      \end{picture}
\begin{picture}(50,70)
        \SetOffset(30,35)\Text(0,0)[bl]{$-$}
\end{picture}
\begin{picture}(50,70)
        \SetOffset(30,35)
        \Line(-25,0)(-2,0)\Vertex(-8,0){2}
        \Line(0,0)(25,15)
        \Line(0,0)(25,-15)
       \DashCArc(0,0)(12,0,360){1}
        \Vertex(0,0){4}
        \Text(38,13)[br]{$\gB$}
        \Text(38,-18)[br]{$\gB$}
      \end{picture}
\begin{picture}(50,70)
        \SetOffset(30,35)\Text(0,0)[bl]{$+$}
\end{picture}
\begin{picture}(60,70)
        \SetOffset(30,35)
        \DashCArc(0,0)(12,103,180){1}
        \DashLine(-3,12)(31,32){1}
\DashCArc(34,20)(12,283,103){1}
\DashCArc(0,0)(12,180,283){1}
        \Line(-25,0)(-2,0)
        \Line(0,0)(34,20)
        \Line(0,0)(34,-20)
        \Vertex(0,0){4}\Vertex(-8,0){2}
\Line(34,20)(59,35)
        \Line(34,20)(59,5)
        \Vertex(34,20){4}\Vertex(24.5,15){2}
\DashCArc(34,20)(8,283,103){1}\DashCArc(34,20)(8,283,103){1}
\DashCArc(24.5,15)(8,180,283){1}\DashCArc(24.5,15)(8,103,180){1}
\DashLine(23.5,24)(33,29){1}
\DashLine(25.5,6)(35,11){1}
        \Text(47,-23)[br]{$\gB$}
        \Text(72,2)[br]{$\gB$}
        \DashLine(3,-12)(37,8){1}
        \Text(72,32)[br]{$\gB$}
      \end{picture}
}\\
\centerline{\begin{picture}(50,70)
        \SetOffset(30,35)\Text(0,0)[bl]{$+$}
\end{picture}
\begin{picture}(60,70)
        \SetOffset(30,35)
        \DashCArc(0,0)(12,103,180){1}
        \DashLine(-3,-12)(31,-32){1}
\DashCArc(34,-20)(12,257,77){1}
\DashCArc(0,0)(12,77,253){1}
        \Line(-25,0)(-2,0)
        \Line(0,0)(34,20)
        \Line(0,0)(34,-20)\Vertex(24.5,-15){2}
        \Vertex(0,0){4}\Vertex(-8,0){2}
\Line(34,-20)(59,-35)
        \Line(34,-20)(59,-5)
        \Vertex(34,-20){4}
\DashCArc(34,-20)(8,257,77){1}
\DashCArc(24.5,-15)(8,103,180){1}\DashCArc(24.5,-15)(8,77,253){1}
\DashLine(23.5,-24)(33,-29){1}
\DashLine(25.5,-6)(35,-11){1}
        \Text(47,20)[br]{$\gB$}
        \Text(72,-6)[br]{$\gB$}
        \DashLine(3,12)(37,-8){1}
        \Text(72,-40)[br]{$\gB$}
      \end{picture}
\begin{picture}(50,70)
        \SetOffset(60,35)\Text(0,0)[bl]{$+\,\,\,\dots$}
\end{picture}
}
\\
Similarly we can expand $\bar\gA$ in terms of $\gB$, and $\bar\gB$ in
which it is linear.  It is more convenient to expand
$\hat\partial\bar\gA$ in terms of $\gB$, and $\hat\partial\bar\gB$,
and we denote this graphically by
\\
\centerline{
{\begin{picture}(50,52)
        \SetOffset(30,24)\Text(0,-3)[bl]{$\hat\partial\bar\gA=$}
\end{picture}}
   \begin{picture}(50,52)
        \SetOffset(30,24)
        \Line(-25,0)(-2,0)
        \Line(0,0)(25,0)
        \Vertex(13,0){2}
        \GCirc(0,0){8}{0.75}
 \Text(39,-3)[br]{$\hat\partial\bar\gB$}
      \end{picture}
{\begin{picture}(50,52)
        \SetOffset(30,24)\Text(0,0)[bl]{$+$}
\end{picture}}
{\begin{picture}(50,52)
        \SetOffset(30,24)
        \Line(-25,0)(-2,0)
        \Line(0,0)(25,15)
        \Line(0,0)(25,-15)
        \GCirc(0,0){8}{0.75}
        \Vertex(13,7){2}
        \Text(38,13)[br]{$\hat\partial\bar\gB$}
        \Text(38,-18)[br]{$\gB$}
      \end{picture}}
{\begin{picture}(50,52)
        \SetOffset(30,24)\Text(0,0)[bl]{$+$}
\end{picture}}
{\begin{picture}(50,52)
        \SetOffset(30,24)
        \Line(-25,0)(-2,0)
        \Line(0,0)(25,15)
        \Line(0,0)(25,-15)
        \GCirc(0,0){8}{0.75}
        \Vertex(13,-7){2}
        \Text(38,13)[br]{$\gB$}
        \Text(38,-18)[br]{$\hat\partial\bar\gB$}
      \end{picture}}
}
\\
\centerline{
{\begin{picture}(50,52)
        \SetOffset(30,24)\Text(0,0)[bl]{$+$}
\end{picture}}
{\begin{picture}(50,52)
        \SetOffset(30,24)
        \Line(-25,0)(-2,0)
        \Line(0,0)(17,25)\Line(0,0)(25,0)\Text(38,-3)[br]{$\hat\partial\bar\gB$}
        \Line(0,0)(17,-25)\Vertex(13,0){2}
        \GCirc(0,0){8}{0.75}
        \Text(30,23)[br]{$\gB$}
        \Text(30,-28)[br]{$\gB$}
   \end{picture}}
{\begin{picture}(50,52)
        \SetOffset(30,24)\Text(0,0)[bl]{$+$}
\end{picture}}
{\begin{picture}(50,52)
        \SetOffset(30,24)
        \Line(-25,0)(-2,0)
        \Line(0,0)(17,25)\Line(0,0)(25,0)\Text(38,-3)[br]{$\gB$}
        \Line(0,0)(17,-25)\Vertex(9,13){2}
        \GCirc(0,0){8}{0.75}
        \Text(30,23)[br]{$\hat\partial\bar\gB$}
        \Text(30,-28)[br]{$\gB$}
   \end{picture}}
{\begin{picture}(50,52)
        \SetOffset(30,24)\Text(0,0)[bl]{$+$}
\end{picture}}
{\begin{picture}(50,52)
        \SetOffset(30,24)
        \Line(-25,0)(-2,0)
        \Line(0,0)(17,25)\Line(0,0)(25,0)\Text(38,-3)[br]{$\gB$}
        \Line(0,0)(17,-25)
        \GCirc(0,0){8}{0.75}
        \Text(30,23)[br]{$\gB$}\Vertex(9,-13){2}
        \Text(30,-28)[br]{$\hat\partial\bar\gB$}
   \end{picture}}
{\begin{picture}(50,52)
        \SetOffset(30,24)\Text(0,0)[bl]{$+\,....$}
\end{picture}}
}
and in momentum space we use $\Xi$ to denote the expansion
coefficients
\\
\centerline{\begin{picture}(70,50)
        \SetOffset(25,25)
        \Line(-25,0)(-2,0)
\Text(-25,-2)[tl]{${ }_{\bar p}$}
        \Line(0,0)(25,15)
        \Line(0,0)(25,-15)
	\Line(0,0)(25,0) \Text (28,0)[l]{$\hat \imath\bar {\cal B}_i$}
        \GCirc(0,0){7}{.75} 
		\Vertex(10,0){2}
       \Text(0,0)[]{$n$}
        \Text(27,15)[l]{${\cal B}_1$}\Text(23,2)[b]{\vdots }\Text(23,2)[t]{\vdots }
        \Text(27,-15)[l]{${\cal B}_n$}
      \end{picture}
\begin{picture}(200,50)
        \SetOffset(50,25)
\Text(-50,0)[l]{$=\int_{1\cdots n} \hat \imath \,\Xi^{i}(\bar p\bar 1\cdots\bar
n){\cal B}_1\cdots\bar {\cal B}_i\cdots{\cal B}_n\,.$}
\end{picture}
}
\\
Using this and (\ref{Agraph}) allows us to depict the second of
(\ref{eq:momentum-transform}) as
\\
\centerline{\begin{picture}(80,80)
        \SetOffset(30,40)\Text(0,-6)[bl]{$\hat\partial\bar\gB=\sum$}
\end{picture}
{\begin{picture}(40,80)
        \SetOffset(30,40)
        \Line(25,0)(2,0)
        \Line(0,0)(-17,25)\Line(0,0)(-25,0)
        \Line(0,0)(-17,-25)
        \BCirc(0,0){8}\Vertex(12,0){2}
        \Text(-23,23)[br]{$\gB$}
        \Text(-23,-28)[br]{$\gB$}
      \Text(-23,16)[br]{.}\Text(-23,11)[br]{.}\Text(-23,-16)[br]{.}\Text(-23,-11)[br]{.}
         \end{picture}}
{\begin{picture}(100,80)
        \SetOffset(30,40)
        \Line(-25,0)(-2,0)
        \Line(0,0)(17,25)\Line(0,0)(25,0)\Text(38,-3)[br]{$\hat\partial\bar\gB$}
        \Line(0,0)(17,-25)\Vertex(13,0){2}
        \GCirc(0,0){8}{0.75}
        \Text(30,23)[br]{$\gB$}
        \Text(30,-28)[br]{$\gB$}
      \Text(30,16)[br]{.}\Text(30,11)[br]{.}\Text(30,-16)[br]{.}\Text(30,-11)[br]{.}
         \end{picture}}
}
\\
Since there are no $\gB$-fields on the left-hand-side we can equate to
zero the sum of terms on the right that contain precisely $n$
$\gB$-fields, when $n>0$:
\\\centerline{{\begin{picture}(80,80)
        \SetOffset(30,40)\Text(0,-6)[bl]{$0=\sum$}
\end{picture}}
{\begin{picture}(40,80)
        \SetOffset(30,40)
        \Line(25,0)(2,0)
        \Line(0,0)(-17,25)\Line(0,0)(-25,0)
        \Line(0,0)(-17,-25)
        \BCirc(0,0){8}\Vertex(12,0){2}
        \Text(-23,23)[br]{$\gB$}
        \Text(-23,-28)[br]{$\gB$}
      \Text(-23,16)[br]{.}\Text(-23,11)[br]{.}\Text(-23,-16)[br]{.}\Text(-23,-11)[br]{.}
         \end{picture}}
{\begin{picture}(100,80)
        \SetOffset(30,40)
        \Line(-25,0)(-2,0)
        \Line(0,0)(17,25)\Line(0,0)(25,0)\Text(38,-3)[br]{$\hat\partial\bar\gB$}
        \Line(0,0)(17,-25)\Vertex(13,0){2}
        \GCirc(0,0){8}{0.75}
        \Text(30,23)[br]{$\gB$}
        \Text(30,-28)[br]{$\gB$}
      \Text(30,16)[br]{.}\Text(30,11)[br]{.}\Text(30,-16)[br]{.}\Text(30,-11)[br]{.}
         \end{picture}}
}
\\
The term in which there are no $\gB$-fields in the left-hand factor is
the kernel we are looking for, so
\\
\begin{eqnarray}
\centerline{\begin{picture}(50,60)
        \SetOffset(30,30)
        \Line(-25,0)(-2,0)
        \Line(0,0)(17,25)\Line(0,0)(25,0)\Text(38,-3)[br]{$\hat\partial\bar\gB$}
        \Line(0,0)(17,-25)\Vertex(13,0){2}
        \GCirc(0,0){8}{0.75}
        \Text(30,23)[br]{$\gB$}
        \Text(30,-28)[br]{$\gB$}
      \Text(30,16)[br]{.}\Text(30,11)[br]{.}\Text(30,-16)[br]{.}\Text(30,-11)[br]{.}
         \end{picture}
\begin{picture}(80,60)
        \SetOffset(30,30)\Text(0,-6)[bl]{$=-\sum '$}
\end{picture}
\begin{picture}(40,60)
        \SetOffset(30,30)
        \Line(25,0)(2,0)
        \Line(0,0)(-17,25)\Line(0,0)(-25,0)
        \Line(0,0)(-17,-25)
        \BCirc(0,0){8}\Vertex(12,0){2}
        \Text(-23,23)[br]{$\gB$}
        \Text(-23,-28)[br]{$\gB$}
      \Text(-23,16)[br]{.}\Text(-23,11)[br]{.}\Text(-23,-16)[br]{.}\Text(-23,-11)[br]{.}
         \end{picture}
\begin{picture}(100,60)
        \SetOffset(30,30)
        \Line(-25,0)(-2,0)
        \Line(0,0)(17,25)\Line(0,0)(25,0)\Text(38,-3)[br]{$\hat\partial\bar\gB$}
        \Line(0,0)(17,-25)\Vertex(13,0){2}
        \GCirc(0,0){8}{0.75}
        \Text(30,23)[br]{$\gB$}
        \Text(30,-28)[br]{$\gB$}
      \Text(30,16)[br]{.}\Text(30,11)[br]{.}\Text(30,-16)[br]{.}\Text(30,-11)[br]{.}
         \end{picture}}
\label{eq:old-recursion}
\end{eqnarray}\\
where the prime on the sum indicates that we sum over terms in which
there is at least one $\gB$-field in the left-hand factor, and the
ordering of fields matches on both sides of the equation. This is
iterated to yield
\begin{eqnarray}
\begin{picture}(45,60)
        \SetOffset(15,30)\Text(0,-3)[bl]{$\hat\partial\bar\gA\,\,=\,\,\hat\partial\bar\gB$}
\end{picture}
\begin{picture}(45,60)
        \SetOffset(30,30)\Text(0,0)[bl]{$+$}
\end{picture}
\begin{picture}(50,60)
        \SetOffset(30,30)
        \Line(-25,0)(-2,0)\Vertex(7,-4){2}
        \Line(0,0)(25,15)
        \Line(0,0)(25,-15)
       \DashCArc(0,0)(12,0,360){1}
        \Vertex(0,0){4}
        \Text(38,13)[br]{$\gB$}
        \Text(38,-18)[br]{$\hat\partial\bar\gB$}
      \end{picture}
\begin{picture}(45,60)
        \SetOffset(30,30)\Text(0,0)[bl]{$+$}
\end{picture}
\begin{picture}(50,60)
        \SetOffset(30,30)
        \Line(-25,0)(-2,0)\Vertex(7,4){2}
        \Line(0,0)(25,15)
        \Line(0,0)(25,-15)
       \DashCArc(0,0)(12,0,360){1}
        \Vertex(0,0){4}
        \Text(38,13)[br]{$\hat\partial\bar\gB$}
        \Text(38,-18)[br]{$\gB$}
      \end{picture}
\begin{picture}(45,40)
        \SetOffset(30,24)\Text(0,0)[bl]{$-$}
\end{picture}
\begin{picture}(80,60)
        \SetOffset(30,30)
        \DashCArc(0,0)(12,103,180){1}
        \DashLine(-3,12)(31,32){1}
\DashCArc(34,20)(12,283,103){1}
\DashCArc(0,0)(12,180,283){1}
        \Line(-25,0)(-2,0)
        \Line(0,0)(34,20)
        \Line(0,0)(34,-20)\Vertex(8,-5){2}
        \Vertex(0,0){4}
\Line(34,20)(59,35)
        \Line(34,20)(59,5)
        \Vertex(34,20){4}\Vertex(24.5,15){2}
\DashCArc(34,20)(8,283,103){1}\DashCArc(34,20)(8,283,103){1}
\DashCArc(24.5,15)(8,180,283){1}\DashCArc(24.5,15)(8,103,180){1}
\DashLine(23.5,24)(33,29){1}
\DashLine(25.5,6)(35,11){1}
        \Text(47,-23)[br]{$\hat\partial\bar\gB$}
        \Text(72,2)[br]{$\gB$}
        \DashLine(3,-12)(37,8){1}
        \Text(72,32)[br]{$\gB$}
      \end{picture}
\nonumber \\
\begin{picture}(45,60)
        \SetOffset(30,30)\Text(0,0)[bl]{$-$}
\end{picture}
\begin{picture}(80,60)
        \SetOffset(30,30)
        \DashCArc(0,0)(12,103,180){1}
        \DashLine(-3,-12)(31,-32){1}
\DashCArc(34,-20)(12,257,77){1}
\DashCArc(0,0)(12,77,253){1}
        \Line(-25,0)(-2,0)
        \Line(0,0)(34,20)
        \Line(0,0)(34,-20)\Vertex(24.5,-15){4}
        \Vertex(0,0){4}\Vertex(8,-5){2}
\Line(34,-20)(59,-35)
        \Line(24.5,-15)(58.5,5)
        \Vertex(34,-20){2}
\DashCArc(9.5,-5)(8,257,77){1}
\DashCArc(0,0)(8,103,180){1}\DashCArc(0,0)(8,77,253){1}
\DashLine(-1,-9)(8.5,-14){1}
\DashLine(1,9)(10.5,4){1}
\Text(47,20)[br]{$\gB$}
        \Text(72,3)[br]{$\gB$}
        \DashLine(3,12)(37,-8){1}
        \Text(72,-40)[br]{$\hat\partial\bar\gB$}
      \end{picture}
\begin{picture}(45,40)
        \SetOffset(30,24)\Text(0,0)[bl]{$+$}
\end{picture}
\begin{picture}(60,60)
        \SetOffset(30,30)
        \Line(-25,0)(-2,0)
        \Line(0,0)(34,20)
        \Line(0,0)(34,-20)\Vertex(24.5,-15){4}
        \Vertex(0,0){4}\Vertex(8,-5){2}
\DashCArc(8,-5)(8,257,77){1}
\DashCArc(0,0)(8,103,180){1}\DashCArc(0,0)(8,77,253){1}
\DashLine(-0.5,-9)(8.5,-14){1}
\DashLine(1.5,9)(10.5,4){1}
\Line(34,-20)(59,-35)
        \Line(24.5,-15)(58.5,5)
        \Vertex(34,-20){2}
\DashCArc(34,-20)(8,257,77){1}
\DashCArc(24.5,-15)(8,103,180){1}\DashCArc(24.5,-15)(8,77,253){1}
\DashLine(23.5,-24)(33,-29){1}
\DashLine(25.5,-6)(35,-11){1}
 \Text(47,20)[br]{$\gB$}
        \Text(72,3)[br]{$\gB$}
        \Text(72,-40)[br]{$\hat\partial\bar\gB$}
      \end{picture}
\begin{picture}(70,60)
        \SetOffset(60,30)\Text(0,0)[l]{$+\,\,\,...$}
\end{picture}
\label{eq:barA-exp}
\end{eqnarray}
However, the broken curves in the above diagrams do not denote the
real singularities in the expansion of $\bar{\cal A}$. Some
singularities are cancelled out. For example, by explicit calculation,
one finds that  the singularity represented by the inner broken curves
around the left big black dot in the fifth and sixth terms are
cancelled out.  In fact, by induction, one can prove another recursion
relation of $\Xi^s$:
\begin{eqnarray}
\Xi^{i-1}_{\bar 1,\dots ,\bar n}&=&-\frac
1{\Omega_1+\cdots+\Omega_n}\Big(\sum_{l=2}^{i-1}
\frac 1{\hat P_{l+1,n}}\bar V^2(p_1,P_{2,l},P_{l+1,n})\Upsilon(
-,\bar 2,\dots, \bar l)\Xi^{i-l}(-,\overline{l+1},\dots,\bar n)
\nonumber \\&&
+\sum_{l=i}^{n-1}\frac 1{\hat P_{2,l}}\bar V^2(P_{l+1,n},p_1,P_{2,l})\Xi^{i-1}(
-,\bar 2,\dots,\bar l)\Upsilon(-,\overline{l+1},\dots, \bar n)
\Big)\,.\label{eq:NewXiR-}
\end{eqnarray}
Using this we can represent each order of the expansion of
$\hat\partial\bar {\cal A}$ by diagrams:
\begin{align}
\begin{picture}(84,90)
        \SetOffset(25,45)
        \Line(-25,0)(-2,0)
        \Line(0,0)(25,15)
        \Line(0,0)(25,-15)
	\Line(0,0)(25,0) \Text (28,0)[l]{$\hat \partial\bar {\cal B}$}
        \GCirc(0,0){7}{.75} 
		\Vertex(10,0){2}
       \Text(0,0)[]{$n$}
        \Text(27,15)[l]{${\cal B}$}\Text(23,2)[b]{\vdots }\Text(23,2)[t]{\vdots }
        \Text(27,-15)[l]{${\cal B}$}
      \end{picture}
\begin{picture}(50,90)
        \SetOffset(27,45)\Text(-25,0)[l]{$=
\sum\limits_{r+s=n
}\Bigg($}
\end{picture}
&
\begin{picture}(100,90)
        \SetOffset(30,45)
        \DashCArc(0,0)(12,103,180){1}
        \DashLine(-3,12)(31,32){1}
\DashCArc(34,20)(12,13,103){1}
\DashCArc(0,0)(12,180,257){1}
\DashLine(-3,-12)(31,-32){1}
\DashCArc(34,-20)(12,257,347){1}
\DashLine(46,-22)(46,22){1}
        \Line(-25,0)(-2,0)
        \Line(0,0)(34,20)
        \Line(0,0)(34,-20)\Vertex(6.8,-4){2}
        \Vertex(0,0){4}
\Line(34,20)(59,35)
        \Line(34,20)(59,5)
        \BCirc(34,20){7}
        \Text(34,20)[]{$r$}\Vertex(22.7,13.3){2}
        \Text(72,33)[br]{${\cal B}$}\Text(59,22)[]{$\vdots$}
        \Text(72,2)[br]{$\cal B$}
\Line(34,-20)(59,-35)
        \Line(34,-20)(59,-5)
        \Line(34,-20)(59,-20)\Text(59,-18)[b]{$\vdots$}\Text(61,-20)[l]{$\hat \partial \bar {\cal
B}$}\Text(59,-18)[t]{$\vdots$}
	\GCirc(34,-20){7}{.75}
\Vertex(45,-20){2}
        \Text(34,-20)[]{$s$}
        \Text(72,-40)[br]{${\cal B}$}
        \Text(72,-10)[br]{$\cal B$}
      \end{picture}
\nonumber \\
&\hskip -1cm
\begin{picture}(40,90)
        \SetOffset(30,45)\Text(-25,-4)[l]{$+
$}
\end{picture}
\begin{picture}(100,90)
        \SetOffset(30,45)
        \DashCArc(0,0)(12,103,180){1}
        \DashLine(-3,12)(31,32){1}
\DashCArc(34,20)(12,13,103){1}
\DashCArc(0,0)(12,180,257){1}
\DashLine(-3,-12)(31,-32){1}
\DashCArc(34,-20)(12,257,347){1}
\DashLine(46,-22)(46,22){1}
        \Line(-25,0)(-2,0)
        \Line(0,0)(34,20)\Vertex(6.8,4){2}
        \Line(0,0)(34,-20)
        \Vertex(0,0){4}
\Line(34,20)(59,35)
        \Line(34,20)(59,5)
	\Line(34,20)(59,20)\Text(59,22)[b]{$\vdots$}\Text(61,20)[l]{$\hat \partial \bar {\cal
		B}$}\Text(59,22)[t]{$\vdots$}
        \GCirc(34,20){7}{.75}
\Vertex(45,20){2}
        \Text(34,20)[]{$r$}
        \Text(72,33)[br]{${\cal B}$}
        \Text(72,2)[br]{$\cal B$}
\Line(34,-20)(59,-35)
        \Line(34,-20)(59,-5)
       	\BCirc(34,-20){7}
	\Text(59,-18)[]{$\vdots$}
        \Text(34,-20)[]{$s$}\Vertex(22.7,-13.3){2}
        \Text(72,-40)[br]{${\cal B}$}
        \Text(72,-10)[br]{$\cal B$}
	\Text(85,0)[]{$\Bigg)$}
\end{picture}\label{eq:dia-barA}
\end{align}
The proof of the new recursion relation starts with the old one
(\ref{eq:old-recursion}) and uses relation \\
\begin{equation}
\begin{picture}(132,90)
        \SetOffset(52,45)
        \DashCArc(0,0)(12,103,180){1}
        \DashLine(-3,12)(31,32){1}
\DashCArc(34,20)(12,13,103){1}
\DashCArc(0,0)(12,180,257){1}
\DashLine(-3,-12)(31,-32){1}
\DashCArc(34,-20)(12,257,347){1}
\DashLine(46,-22)(46,22){1}
        \Line(-30,0)(-50,20)\Text(-51,20)[r]{${ }_l$}
	\Line(-30,0)(-50,-20)\Text(-25,-25)[r]{${ }_{l+r+s+1}$}
	\Line(-30,0)(-50,0)\Text(-51,0)[r]{${ }_1$}\Vertex(-50,0){1}\Vertex(-20,0){2}
	\Text(-48,2)[b]{$\vdots$}\Text(-48,3)[t]{$\vdots$}
	\Line(-30,0)(-2,0)
        \BCirc(-30,0){7}
 	\Line(0,0)(34,20)
        \Line(0,0)(34,-20)\Vertex(6.8,-4){2}
        \Vertex(0,0){4}
\Line(34,20)(59,35)
        \Line(34,20)(59,5)
        \BCirc(34,20){7}
        \Text(37,16)[br]{$r$}\Vertex(22.7,13.3){2}
        \Text(59,35)[l]{${ }_{l\!+\!1}$}\Text(57,22)[]{$\vdots$}
        \Text(59,5)[l]{${ }_{l+r}$}
\Line(34,-20)(59,-35)
        \Line(34,-20)(59,-20) \Text(60,-20)[l]{ ${ }_i$}
	\Line(34,-20)(59,-5)
        \GCirc(34,-20){7}{.75}\Vertex(45,-20){2}
        \Text(37,-23)[br]{$s$}
        \Text(59,-35)[l]{${ { }_{l\!+r\!+s}}$}\Text(57,-18)[b]{$\vdots$}\Text(57,-17)[t]{$\vdots$}
        \Text(59,-5)[l]{${ }_{l\!+r\!+1}$}
\end{picture}
\begin{picture}(10,90)
        \SetOffset(5,45)
       \Text(0,0)[]{$-$} 
\end{picture}
\begin{picture}(110,90)
        \SetOffset(51,45)
\DashCArc(34,20)(12,13,103){1}
\DashCArc(34,-20)(12,257,347){1}
        \Line(-30,0)(-50,20)
	\Line(-30,0)(-50,-20)
	\Line(-30,0)(-50,0)\Vertex(-50,0){1}\Vertex(-20,0){2}
	\Text(-48,2)[b]{$\vdots$}\Text(-48,3)[t]{$\vdots$}
	\Line(-30,0)(-2,0)
        \BCirc(-30,0){7}
 	\DashCArc(-30,0)(12,103,257){1}
	\DashLine(-33,-12)(31,-32){1}	
	\DashLine(-33,12)(31,32){1}\DashLine(46,-22)(46,22){1}
	\Line(0,0)(34,20)
        \Line(0,0)(34,-20)\Vertex(6.8,-4){2}
        \Vertex(0,0){4}
\Line(34,20)(59,35)
        \Line(34,20)(59,5)
        \BCirc(34,20){7}
        \Text(37,16)[br]{$r$}\Vertex(22.7,13.3){2}
	\Text(57,22)[]{$\vdots$}
\Line(34,-20)(59,-35)
        \Line(34,-20)(59,-20) 
	\Line(34,-20)(59,-5)
        \GCirc(34,-20){7}{.75}\Vertex(45,-20){2}
        \Text(37,-23)[br]{$s$}
	\Text(57,-18)[b]{$\vdots$}
	\Text(57,-17)[t]{$\vdots$}
\end{picture}
\begin{picture}(24,90)
        \SetOffset(12,45)
       \Text(0,0)[]{$=-$} 
\end{picture}
\begin{picture}(120,90)
        \SetOffset(51,45)
        \DashCArc(0,0)(10,103,257){1}
        \DashLine(-3,10)(31,29){1}
\DashCArc(34,20)(10,13,103){1}
\DashLine(-3,-10)(31,-29){1}
\DashCArc(34,-20)(10,257,347){1}
\DashLine(44,-21)(44,21){1}
        \Line(-30,0)(-50,20)
	\Line(-30,0)(-50,-20)
	\Line(-30,0)(-50,0)\Vertex(-50,0){1}\Vertex(-20,0){2}
	\Text(-48,2)[b]{$\vdots$}\Text(-48,3)[t]{$\vdots$}
	\Line(-30,0)(-2,0)
        \Vertex(-30,0){7}%
 	\DashCArc(34,20)(12,13,103){1}\DashCArc(34,-20)(12,257,347){1}\DashCArc(-30,0)(12,103,257){1}
	\DashLine(46,-22)(46,22){1}\DashLine(-33,-12)(31,-32){1}\DashLine(-33,12)(31,32){1}
	\Line(0,0)(34,20)
        \Line(0,0)(34,-20)\Vertex(6.8,-4){2}
        \Vertex(0,0){4}
\Line(34,20)(59,35)
        \Line(34,20)(59,5)
        \BCirc(34,20){6}
        \Text(37,16)[br]{$r$}\Vertex(22.7,13.3){2}
	\Text(57,22)[]{$\vdots$}
\Line(34,-20)(59,-35)
        \Line(34,-20)(59,-20) 
	\Line(34,-20)(59,-5)
        \GCirc(34,-20){6}{.75}\Vertex(45,-20){2}
        \Text(37,-23)[br]{$s$}
	\Text(57,-18)[b]{$\vdots$}\Text(57,-17)[t]{$\vdots$}
\end{picture}
\label{eq:relation}
\end{equation}
\\
repeatedly (see appendix \ref{sect:proof} for a sketch of the proof).
The relation above is simply a result of the equation
\begin{eqnarray}
\frac1{\Omega_{l+r+s+1,l}+\sum_{i=l+1}^{l+r+s}\Omega_{i}}-\frac1{\sum_{i=1}^{n}\Omega_{i}}
=\frac{\Omega_{l+1,l+r+s}+\sum_{i=l+r+s+1}^{l}\Omega_{i}}{(\Omega_{l+r+s+1,l}+\sum_{i=1}^{l+1}\Omega_{l+r+s})\sum_{i=1}^{n}\Omega_{i}}
\label{eq-flipping} 
\end{eqnarray}
where $\Omega_{i,j}= P_{i,j}\bar P_{i,j}/\hat P_{i,j}$.  The numerator
on the right-hand-side of (\ref{eq-flipping}) will cancel the
denominator of the left $\Upsilon$ blob in the diagrams.  We denote
this cancellation by filling in the left-hand blob. In fact, there is
an easy way to prove this recursion relation in four dimensions where
we  do not care about regularization: If we use the relation obtained
in \cite{Ettle:2006bw}
\begin{eqnarray}\label{eq:UpsilonXi}
\Xi^{i-1}(1\cdots n)=-\frac{\hat \imath}{\hat 1}\Upsilon(1\cdots n)\,,
\end{eqnarray}
this recursion relation recovers that of (\ref{eq:UpsilonR}) for
$\Upsilon$.

\subsection{Reconstructing the expansion coefficients from tree-level
light-cone diagrams}\label{sect:lc-Cononical}
From (\ref{eq:-UpsilonR}) we observe that the expansion terms of $\cal
A$ can be constructed as follows: for each term of the expansion, draw
all the tree-level Feynman diagrams with an $\cal A$ as one end of an
external propagator and all $\cal B$s in the term as amputated
external lines using only $(++-)$ vertices; then calculate this
diagram using $\bar V^2$ as vertices and $1/(\hat
p(\Omega_p+\sum\Omega))$ as corresponding propagators.  Notice that
the light-cone  Feynman rule for vertex $(++-)$ is 
\begin{eqnarray}
\bar V(1,2,3)=i\frac
4{g^2} \bar V^2(\bar 1,\bar 2,\bar 3)=-i\frac
4{g^2}  \bar V^2( 1, 2, 3)
\end{eqnarray}
 and the light-cone propagator is 
\begin{eqnarray}
\langle{\cal A}_p{\cal A}_{\bar p}\rangle
=-i\frac{g^2}{2p^2}\,.
\end{eqnarray}
So 
\begin{eqnarray}
\langle{\cal A}_3{\cal A}_{\bar 3}\rangle\bar
V(1,2,3)=-\frac2{p_3^2}\bar V^2( 1, 2, 3)
\end{eqnarray}
is consistent with the coefficient of each term in the recursion
relation if we make the replacement $-2/{p_3^2}\to 1/(\hat
p_3(\Omega_3+\Omega_1+\Omega_2))$.  As a result, we can reconstruct
the terms of $\cal A$ from light-cone tree-level calculations by
replacing the light-cone propagators using
\begin{eqnarray}
\frac 1{P^2_{ij}}
\to -{{ 1}\over 2\hat P_{j+1,i-1}(\Omega_{j+1,i-1}+\Omega_{i}+\Omega_{i+1}
+\cdots+\Omega_j)}.
\label{eq:replace}
\end{eqnarray}
Here $P_{j+1,i-1}$ should be understood as the sum of all momenta
except those labelled from $i$ to $j$. The momentum in each term in
the bracket of the denominators corresponds to the outgoing momentum
of the external line of the sub-tree diagram not involving $\cal A$
when the propagator is cut. For example, for terms with ${\cal
B}_2{\cal B}_3{\cal B}_4$:
\begin{eqnarray}
{\cal A}_{\bar 1}&\sim& \Upsilon(\bar 1\bar2\bar3\bar 4){\cal B}_2{\cal B}_3{\cal B}_4 
\nonumber\\
&=& \frac 1{\hat 1(\Omega_1+\cdots+\Omega_4)}\bigg(\frac{\bar V^2(2,34,1)\bar
V^2(3,4,12)}{\hat P_{12}(\Omega_{12}+\Omega_3+\Omega_4)}+\frac{\bar
V^2(23,4,1)\bar V^2(2,3,41)}{\hat P_{41}(\Omega_{41}+\Omega_2+\Omega_3)}\bigg)
\nonumber \\
&&\times{\cal B}_2{\cal B}_3{\cal B}_4 
\,.\label{eq:A-4}
\end{eqnarray}
The corresponding diagrams are:
\\
\centerline{
\includegraphics[height=2cm]{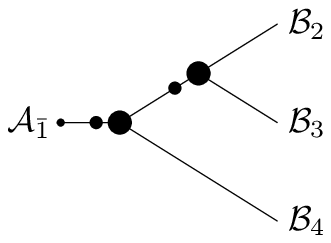}
\hspace{1cm}
\includegraphics[height=2cm]{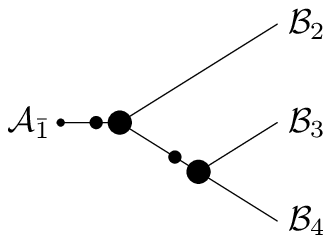}
}\\
From the light-cone calculation of these Feynman diagrams, we have:
\begin{eqnarray}
2^2\frac 1{p_1^2}\bigg(\frac{\bar V^2(2,34,1)\bar
V^2(3,4,12)}{P^2_{12}}+\frac{\bar
V^2(23,4,1)\bar V^2(2,3,41)}{ P^2_{41}}\bigg)
\label{eq:A-4-light-cone}
\end{eqnarray}
in which the two terms correspond to the two tree-level Feynman
diagrams.  We can see that (\ref{eq:A-4}) and
(\ref{eq:A-4-light-cone}) only differ by the change 
\begin{eqnarray}
1/p_1^2& \rightarrow& -1/(2\hat 1(\Omega_1+\cdots+\Omega_4))
\,,\label{eq:replace-1}\\
1/ P^2_{12}&\rightarrow& -1/(2\hat
P_{12}(\Omega_{12}+\Omega_3+\Omega_4)) 
\,,\label{eq:replace-2}\\
1/ P^2_{41}&\rightarrow& -1/(2\hat
P_{41}(\Omega_{41}+\Omega_2+\Omega_3)).
\label{eq:replace-3} 
\end{eqnarray}
If we put $p_2, p_3, p_4$ on shell, the $\rightarrow$ in the above
equations can be replaced by $=$, thus (\ref{eq:A-4}) is equal to
(\ref{eq:A-4-light-cone}) which gives the translation kernel
contribution to the amplitude as it should.

For $\bar{\cal A}$, the same rule also holds allowing us to
reconstruct the expansion of $\bar {\cal A}$ from light-cone
calculations: one needs to first draw the tree-level diagrams with one
${\bar {\cal A} }$ as an external propagator, all the ${\cal B}$,
$\bar {\cal B}$ in the term as amputated legs using $(++-)$ vertices,
and then calculate the diagram using the light-cone Feynman rules with
the replacement (\ref{eq:replace}).  This can be justified from the
recursion relation (\ref{eq:NewXiR-}) with a similar discussion to
that for $\Upsilon$: First, in (\ref{eq:NewXiR-}) all the $\Upsilon$'s
already obey this rule. The $-1/\hat P_{l+1,n}=1/\hat P_{1,l}$ in the
first term in the bracket will combine with the $1/\sum\Omega$ factor
in the expansion of the next $\Xi$ in this term to be
$1/(P_{1,l}(\Omega_{1,l}+\sum_{i=l+1}^{n}\Omega_i))$ which is just
what we need to be consistent with the rule. It is the same for the
second term in the bracket.  We only need to consider the factor of
the $\Xi$ in the first iteration and the last iteration.  We should
divide the expansion of $\hat \partial \bar {\cal A}$ by the
corresponding $i\hat p$ in the momentum space, to obtain the expansion
of $\bar {\cal A}$.  This factor $1/\hat p_1$ will combine with the
factor of the first $\Xi$ to be $1/(\hat p_1(\sum_{i=1}^{n}\Omega_i))$
in (\ref{eq:NewXiR-}). The last iteration corresponds to the
right-most grey blob adjacent to $\hat \partial {\cal B}$ in each term
of the full iteratively expanded diagrams in (\ref{eq:dia-barA}).  The
extra factor $1/\hat \imath$ in the $\Xi$ of the last step of the
iteration will cancel the $\hat \imath$ in the $\hat \imath {\cal
B}_i$ from $\hat \partial {\cal B}$. So just as in the case of
$\Upsilon$, the expansion of $\bar {\cal A}$ requires calculating
tree-level diagrams using $\bar V^2$ as vertices and $1/(\hat
p(\Omega_p+\sum\Omega))$ as the propagators, and so obeys the same
rule.

\section{`Missing' amplitudes from equivalence theorem evasion reviewed}
\label{sect:tree-mpp}
In \cite{us} we explained how the tree-level $(-++)$ and the one-loop
$(++\dots ++)$ amplitudes are obtained from the $\gB$, $\bar\gB$
theory, despite there being no vertices in this theory that could
contribute. The amplitudes are non-zero because the equivalence
theorem is not directly applicable to our non-local transformation.
Thus $\gA$ and $\bar\gA$ do not create the same particles as $\gB$,
$\bar\gB$. This would appear to drastically complicate the calculation
of amplitudes within the $\gB$, $\bar\gB$ theory. It is the main
purpose of this paper to show that only certain amplitudes are
affected by this, and that in the general case we can use either set
of fields to generate amplitudes. In this section we briefly review
the `missing' tree-level amplitude. 

In light-cone gauge Yang-Mills theory the tree-level contribution to
the Green function $\langle \,\gA(p_1)\,\bar\gA(p_2)\,\bar\gA(p_3)
\,\rangle$ comes from the vertex in ${\cal L}^{-++}$, so to this
order, and taking account of the $i\epsilon$-prescription in
propagators (and suppressing Lie algebra indices on the understanding
that we deal with colour-ordered amplitudes)
\\
\centerline{\begin{picture}(60,52)
         \SetOffset(60,0) \Text(47,19)[br]{$(p_1^2+i\epsilon)\,(p_2^2+i\epsilon)\,(p_3^2+i\epsilon)\,\langle \,\gA(p_1)\,\bar\gA(p_2)\,\bar\gA(p_3)
\,\rangle\,\,\,\,=$}
           \end{picture}
\begin{picture}(60,52)
        \SetOffset(100,24)
        \Line(-25,0)(-2,0)
        \Vertex(-8,0){2}
        \Line(0,0)(34,20)
        \Line(0,0)(34,-20)
        \Vertex(0,0){4}
        \Text(-40,-15)[br]{$\Bigg($}\Text(65,-15)[br]{$\Bigg)$}\Text(80,-3)[br]{$\hat p_1$,}
        \Text(47,17)[br]{$p_2$} \Text(47,-23)[br]{$p_3$}
         \Text(-28,-3)[br]{$p_1$}
\end{picture}
}
\\
and as all three momenta go on-shell this becomes the three-point
amplitude (which vanishes in four dimensional Minkowski space, but is
non-zero in other signatures and dimensions.) Clearly $\langle
\,\gB(p_1)\,\bar\gB(p_2)\,\bar\gB(p_3) \,\rangle=0$ at tree-level due
to the helicity assignment of the Parke-Taylor vertices.  To compute
the Green function in the $\gB$, $\bar\gB$ theory we must use the
translation kernels:
\\
\centerline{\begin{picture}(70,52)
          \Text(47,19)[br]{$\langle \,\gA(p_1)\,\bar\gA(p_2)\,\bar\gA(p_3)
\,\rangle\,\,\,\,=$}
           \end{picture}
   \begin{picture}(40,40)
        \SetOffset(10,24)\Text(10,-15)[br]{$-\langle\,\,\,\Bigg($}
 \Text(39,-3)[br]{$\gB (p_1)$}
      \end{picture}
\begin{picture}(30,40)
        \SetOffset(15,24)\Text(0,0)[bl]{$+$}
\end{picture}
\begin{picture}(125,40)
        \SetOffset(35,24)\Text(-28,-3)[br]{$p_1$}
        \Line(-25,0)(-2,0)\Vertex(-8,0){2}
        \Line(0,0)(25,15)
        \Line(0,0)(25,-15)
       \DashCArc(0,0)(12,0,360){1}
        \Vertex(0,0){4}
        \Text(38,13)[br]{$\gB$}
        \Text(38,-18)[br]{$\gB$}\Text(90,-15)[br]{$+\,...\,\,\Bigg)\,\,\times$}
\end{picture}
}
\\
\centerline{
\begin{picture}(50,40)
        \SetOffset(30,24)\Text(0,-15)[br]{${1\over \hat p_2}\,\Bigg($}
 \Text(39,-3)[br]{$\hat\partial\bar\gB(p_2)$}
      \end{picture}
\begin{picture}(50,40)
        \SetOffset(30,24)\Text(0,0)[bl]{$-$}
\end{picture}
\begin{picture}(50,40)
        \SetOffset(30,24)\Text(-28,-3)[br]{$p_2$}
        \Line(-25,0)(-2,0)\Vertex(7,-4){2}
        \Line(0,0)(25,15)
        \Line(0,0)(25,-15)
       \DashCArc(0,0)(12,0,360){1}
        \Vertex(0,0){4}
        \Text(38,13)[br]{$\gB$}
        \Text(38,-18)[br]{$\hat\partial\bar\gB$}
      \end{picture}
\begin{picture}(50,40)
        \SetOffset(30,24)\Text(0,0)[bl]{$-$}
\end{picture}
\begin{picture}(50,40)
        \SetOffset(30,24)\Text(-28,-3)[br]{$p_2$}
        \Line(-25,0)(-2,0)\Vertex(7,4){2}
        \Line(0,0)(25,15)
        \Line(0,0)(25,-15)
       \DashCArc(0,0)(12,0,360){1}
        \Vertex(0,0){4}
        \Text(38,13)[br]{$\hat\partial\bar\gB$}
        \Text(38,-18)[br]{$\gB$}
      \end{picture}
\begin{picture}(50,40)
        \SetOffset(30,24)\Text(0,0)[bl]{$+\,...$}\Text(40,-15)[br]{$\Bigg)\,\,\times$}
\end{picture}
}\\
\centerline{
\begin{picture}(50,40)
        \SetOffset(30,24)\Text(0,-15)[br]{${1\over \hat p_3}\,\Bigg($}
 \Text(39,-3)[br]{$\hat\partial\bar\gB(p_3)$}
      \end{picture}
\begin{picture}(50,40)
        \SetOffset(30,24)\Text(0,0)[bl]{$-$}
\end{picture}
\begin{picture}(50,40)
        \SetOffset(30,24)\Text(-28,-3)[br]{$p_3$}
        \Line(-25,0)(-2,0)\Vertex(7,-4){2}
        \Line(0,0)(25,15)
        \Line(0,0)(25,-15)
       \DashCArc(0,0)(12,0,360){1}
        \Vertex(0,0){4}
        \Text(38,13)[br]{$\gB$}
        \Text(38,-18)[br]{$\hat\partial\bar\gB$}
      \end{picture}
\begin{picture}(50,40)
        \SetOffset(30,24)\Text(0,0)[bl]{$-$}
\end{picture}
\begin{picture}(50,40)
        \SetOffset(30,24)\Text(-28,-3)[br]{$p_3$}
        \Line(-25,0)(-2,0)\Vertex(7,4){2}
        \Line(0,0)(25,15)
        \Line(0,0)(25,-15)
       \DashCArc(0,0)(12,0,360){1}
        \Vertex(0,0){4}
        \Text(38,13)[br]{$\hat\partial\bar\gB$}
        \Text(38,-18)[br]{$\gB$}
      \end{picture}
\begin{picture}(50,40)
        \SetOffset(30,24)\Text(0,0)[bl]{$+\,...$}\Text(40,-15)[br]{$\Bigg)\,\,\rangle$}
\end{picture}
}
since no vertices contribute to leading order this can be computed by
contracting the $\gB$, $\bar\gB$ fields using the free propagator,
which we denote by $\Photon(0,0)(70,0){1.5}{8}$ 
\\
\begin{picture}(70,52)
          \Text(5,19)[l]{$\langle \,\gA(p_1)\,\bar\gA(p_2)\,\bar\gA(p_3)
\,\rangle\,\,\,\,=$}
           \end{picture}
\\
\centerline{
\begin{picture}(130,52)
        \SetOffset(30,24)
        \Line(-38,0)(0,0)
        \Vertex(-8,0){2}    \DashCArc(0,0)(12,0,360){1}
        \Photon(0,0)(34,20){1.5}{8}
        \Photon(0,0)(34,-20){1.5}{8}
        \Vertex(0,0){4}
        \Text(70,-3)[br]{$ +$}
        \Text(47,17)[br]{$p_2$} \Text(47,-23)[br]{$p_3$}
         \Text(-42,-3)[br]{$p_1$}
\end{picture}
\begin{picture}(160,52)
        \SetOffset(30,24)
        \Photon(-38,0)(0,0){1.5}{8}
        \Vertex(-8,0){2}    \DashCArc(0,0)(12,0,360){1}
        \Line(0,0)(34,20)
        \Photon(0,0)(34,-20){1.5}{8}
        \Vertex(0,0){4}
        \Text(-50,-15)[br]{$\Bigg($}\Text(65,-15)[br]{$\Bigg)$}\Text(80,-8)[br]{${\hat p_1
\over\hat p_2}$} \Text(100,-3)[br]{$ +$}
        \Text(47,17)[br]{$p_2$} \Text(47,-23)[br]{$p_3$}
         \Text(-42,-3)[br]{$p_1$}
\end{picture}
\begin{picture}(100,52)
        \SetOffset(30,24)
        \Photon(-38,0)(0,0){1.5}{8}
        \Vertex(-8,0){2}    \DashCArc(0,0)(12,0,360){1}
        \Photon(0,0)(34,20){1.5}{8}
        \Line(0,0)(34,-20)
        \Vertex(0,0){4}
        \Text(-50,-15)[br]{$\Bigg($}\Text(65,-15)[br]{$\Bigg)$}\Text(80,-8)[br]{${\hat p_1
\over \hat p_3}$}
        \Text(47,17)[br]{$p_2$} \Text(47,-23)[br]{$p_3$}
         \Text(-42,-3)[br]{$p_1$}
\end{picture}
}\\
\centerline{
\begin{picture}(200,52)
        \SetOffset(-50,24)\Text(-55,-3)[br]{$ =$}
        \Line(-25,0)(-2,0)
        \Vertex(-8,0){2}
        \Line(0,0)(34,20)
        \Line(0,0)(34,-20)
        \Vertex(0,0){4}\DashCArc(0,0)(12,0,360){1}
        \Text(-40,-15)[br]{$\Bigg($}\Text(65,-15)[br]{$\Bigg)$}\Text(270,-3)[br]{${\hat p_1\over(p_1^2+i\epsilon)\,(p_2^2+i\epsilon)\,(p_3^2+i\epsilon) }\Big( {p_1^2+i\epsilon\over \hat p_1}+{p_2^2+i\epsilon\over \hat p_2}+{p_3^2+i\epsilon\over \hat p_3}\Big)$,}
        \Text(47,17)[br]{$p_2$} \Text(47,-23)[br]{$p_3$}
         \Text(-28,-3)[br]{$p_1$}
\end{picture}
}
\\The broken line cutting the three lines denotes division by 
$$
-{\sum_j\Omega(p_j)}=-{\sum_j\sum_{i=1}^{D/2-1}p_{j(i)} \bar
p_{j(i)}/\hat p_j}\,,  $$
which does not depend on the $\check p_j$.  However, if we add $\sum_j
\check p_j$, which vanishes by momentum conservation, this becomes
${\sum_j p_j^2/\hat p_j}$. If we also include $i\epsilon$ terms to
match the last factor then we reproduce the light-cone Yang-Mills
amplitude. This tells us how to treat $1/\sum\Omega$ when the
denominator is singular, so in general the broken lines in our
diagrams will denote
$$
{1\over\sum_j {p_j^2+i\epsilon\over\hat p_j}}\,.
$$
It is of course not surprising that we reproduce the usual Green
function, as all we have done is transform to new variables to do the
calculation.  It will be useful, for what comes later, to examine how
the equivalence theorem has been evaded. Note that the combined limit
$p_1^2+i\epsilon,p_2^2+i\epsilon,p_3^2+i\epsilon\to0$ is not valid for each term
separately, because the value of  
\[\lim_{p_1+i\epsilon^2,p_2^2+i\epsilon,p_3^2+i\epsilon\to0}
\frac{p_1^2+i\epsilon}{\frac{p_1^2+i\epsilon}{\hat 1}
+\frac{p_2^2+i\epsilon}{\hat 2}+\frac{p_3^2+i\epsilon}{\hat 3}}\]
depends on the order in which the limits are taken, but it is valid to
take the limit of the sum of the three terms because the factor
$(p_1^2+i\epsilon)/{\hat 1} +(p_2^2+i\epsilon)/{\hat
2}+(p_3^2+i\epsilon)/{\hat 3}$ in the denominator is cancelled out.
Consequently we can take the limit of the sum in any order. Suppose we
take the legs on-shell one after another, beginning with $p_2$ and
$p_3$. We include $\epsilon$ in the mass-shell condition because it
enters the propagators for external legs that have to be cancelled by
the LSZ factors.  Since $p_2^2+i\epsilon$ and $p_3^2+i\epsilon$ cancel
the propagators in the first diagram, but not in the other two, it is
clear that for general $p_1$ the contributions from the last two
diagrams are wiped out in the limit leaving 
\\
\centerline{\begin{picture}(-30,52)
         \SetOffset(60,0) \Text(47,19)[br]{$\lim_{p_2^2+i\epsilon\rightarrow 0}\,\,\lim_{p_3^2+i\epsilon\rightarrow 0}\,(p_2^2+i\epsilon)\,(p_3^2+i\epsilon)\,\langle \,\gA(p_1)\,\bar\gA(p_2)\,\bar\gA(p_3)
\,\rangle\,\,\,\,=$}
           \end{picture}
}\\
\centerline{\begin{picture}(-40,52)
        \SetOffset(60,24) \Text(-60,-7)[br]{$\lim_{p_2^2+i\epsilon\rightarrow 0}\,\,\lim_{p_3^2+i\epsilon\rightarrow 0}\,(p_2^2+i\epsilon)\,(p_3^2+i\epsilon)\,\,$}
        \Line(-38,0)(0,0)
        \Vertex(-8,0){2}    \DashCArc(0,0)(12,0,360){1}
        \Photon(0,0)(34,20){1.5}{8}
        \Photon(0,0)(34,-20){1.5}{8}
        \Vertex(0,0){4}
        \Text(47,17)[br]{$p_2$} \Text(47,-23)[br]{$p_3$}
         \Text(-42,-3)[br]{$p_1$}\Text(-55,-15)[br]{$\Bigg($}\Text(75,-15)[br]{$\Bigg)\,=$}
\end{picture}
}\\
\centerline{\begin{picture}(-40,52)
        \SetOffset(60,24) \Text(-60,-7)[br]{$\lim_{p_2^2+i\epsilon\rightarrow 0}\,\,\lim_{p_3^2+i\epsilon\rightarrow 0}\,\,{1\over{{p_1^2+i\epsilon\over \hat p_1}+{p_2^2+i\epsilon\over \hat p_2}+{p_3^2+i\epsilon\over \hat p_3}}}\,\,$}
        \Line(-38,0)(0,0)
        \Vertex(-8,0){2}    
        \Photon(0,0)(34,20){1.5}{8}
        \Photon(0,0)(34,-20){1.5}{8}
        \Vertex(0,0){4}
        \Text(47,17)[br]{$p_2$} \Text(47,-23)[br]{$p_3$}
         \Text(-42,-3)[br]{$p_1$}\Text(-55,-15)[br]{$\Bigg($}\Text(75,-15)[br]{$\Bigg)\,=$}
\end{picture}
}
\centerline{\begin{picture}(-40,52)
        \SetOffset(60,24) \Text(-60,-7)[br]{${\hat p_1\over p_1^2+i\epsilon}\,\,\,$}
        \Line(-38,0)(0,0)
        \Vertex(-8,0){2}    
        \Photon(0,0)(34,20){1.5}{8}
        \Photon(0,0)(34,-20){1.5}{8}
        \Vertex(0,0){4}
        \Text(47,17)[br]{$p_2$} \Text(47,-23)[br]{$p_3$}
         \Text(-42,-3)[br]{$p_1$}\Text(-55,-15)[br]{$\Bigg($}\Text(75,-15)[br]{$\Bigg)\,.$}
\end{picture}
}
\\
So the $1/(p_1^2+i\epsilon)$ needed to cancel the $p_1^2+i\epsilon$
coming from the LSZ prescription is generated as part of the
translation kernels, even though these appeared to be independent of
the $\check p$ components of momenta. We should point out that
`missing amplitudes' can be generated in different ways if the theory
is formulated differently such as in the gauge fixing of the twistor
action \cite{Wen} or in the light-cone friendly regularisation of
\cite{Brandhuber:2007vm}.

\section{Equivalence theorem evasion in general.}
\label{sect:general-evasion} When the equivalence theorem holds we can
ignore all except the leading translation kernels.  However the
theorem will be evaded whenever the translation kernels that express
$\gA (p)$ or $\bar \gA (p)$ in terms of $\gB$ and $\bar \gB$ produce a
$1/(p^2+i\epsilon)$ that can cancel the LSZ factors. We will now list
all the types of process in which this can occur. The singular terms
originate in the $1/\sum \Omega$ represented by the broken lines in
our diagrams. These must cut the line with momentum $p$ if we are to
end up with $1/(p^2+i\epsilon)$. Suppose that the other lines cut
carry momenta $p_1,\dots ,p_n$, then $$ {1\over \sum \Omega}={1\over
{p^2+i\epsilon\over \hat p}+\sum_{j=1}^n{p_j^2+i\epsilon\over \hat
p_j}}\,,$$ so we have to examine the conditions under which
$\sum_{j=1}^n{(p_j^2+i\epsilon)/\hat p_j}=0$. Notice that here we
actually take $\sum_{j=1}^n{(p_j^2+i\epsilon)/\hat p_j}\to0$ limit
first and then $p^2\to 0$ in the LSZ procedure. We will see that this
is valid in a similar way to the three point case.

\subsection{Tree-level}

In the absence of loops there are two ways that
$\sum_{j=1}^n{(p_j^2+i\epsilon)/\hat p_j}=0$.  The first is that each
of the legs cut by the broken line are external and so their momenta
must be put on shell.  For example, in the four-particle process with
one $-$ helicity and three $+$ helicity gluons we need the Green
function $\langle \,\gA(p_1)\,\bar\gA(p_2)\,\bar\gA(p_3)\,\bar\gA(p_4)
\,\rangle $. Contributing to this are translation kernels for
$\gA(p_1)$, $\bar\gA(p_2)$, $\bar\gA(p_3)$, and $\bar\gA(p_4)$ which
give rise to diagrams like
\\
\centerline{\begin{picture}(150,80)
        \SetOffset(30,24)
        \DashCArc(0,0)(12,103,180){1}
        \DashLine(-3,12)(31,32){1}
\DashCArc(34,20)(12,283,103){1}
\DashCArc(0,0)(12,180,283){1}
        \Line(-25,0)(-2,0)\Text(-30,0)[br]{$p_1$}
        \Line(0,0)(34,20)
        \Photon(0,0)(34,-20){1.5}{8}
        \Vertex(0,0){4}\Vertex(-8,0){2}
\Photon(34,20)(59,35){1.5}{8}
        \Photon(34,20)(59,5){1.5}{8}
        \Vertex(34,20){4}\Vertex(24.5,15){2}
\DashCArc(34,20)(8,283,103){1}\DashCArc(34,20)(8,283,103){1}
\DashCArc(24.5,15)(8,180,283){1}\DashCArc(24.5,15)(8,103,180){1}
\DashLine(23.5,24)(33,29){1}
\DashLine(25.5,6)(35,11){1}
        \Text(47,-23)[br]{$p_4$}
        \Text(72,2)[br]{$p_3$}
        \DashLine(3,-12)(37,8){1}
        \Text(72,32)[br]{$p_2$}
      \end{picture}}
\\
in which four external legs are cut by the broken line. Since all the
external lines will be cut by the broken curve  we cannot include any
Parke-Taylor vertices.  Consequently, in the general case we can only
ever have a contribution to a tree-level amplitude with one $-$
helicity external gluon and $n$ $+$ helicity external gluons. For each
light-cone tree-level diagram of such an amplitude, there are terms
from the translation kernels that contribute. For example, for the
four-point diagram:
\\
\centerline{
\begin{picture}(150,80)
        \SetOffset(30,24)
        \Line(-25,0)(-2,0)
	\Text(-30,0)[br]{$p_1$}
        \Line(0,0)(34,20)
        \Line(0,0)(34,-20)
        \Vertex(0,0){4}\Vertex(-8,0){2}
\Line(34,20)(59,35)
        \Line(34,20)(59,5)
        \Vertex(34,20){4}\Vertex(24.5,15){2}
        \Text(47,-23)[br]{$p_4$}
        \Text(72,2)[br]{$p_3$}
        \Text(72,32)[br]{$p_2$}
      \end{picture}}
\\
using the method in section (\ref{sect:lc-Cononical}), we can
construct the translation kernel contribution to this diagram from the
canonical expansion of ${\cal A}_1$, $\bar{\cal A}_2$, $\bar{\cal
A}_3$, $\bar{\cal A}_4$ which can be represented graphically:
\\
\centerline{
\begin{picture}(150,80)
        \SetOffset(30,24)
        \DashCArc(0,0)(12,103,180){1}
        \DashLine(-3,12)(31,32){1}
\DashCArc(34,20)(12,283,103){1}
\DashCArc(0,0)(12,180,283){1}
        \Line(-25,0)(-2,0)\Text(-30,0)[br]{$p_1$}
        \Line(0,0)(34,20)
        \Photon(0,0)(34,-20){1.5}{8}
        \Vertex(0,0){4}\Vertex(-8,0){2}
\Photon(34,20)(59,35){1.5}{8}
        \Photon(34,20)(59,5){1.5}{8}
        \Vertex(34,20){4}\Vertex(24.5,15){2}
        \DashCArc(34,20)(8,0,360){1}
        \Text(47,-23)[br]{$p_4$}
        \Text(72,2)[br]{$p_3$}
        \DashLine(3,-12)(37,8){1}
        \Text(72,32)[br]{$p_2$}
      \end{picture}
\begin{picture}(150,80)
        \SetOffset(30,24)
        \DashCArc(0,0)(12,103,180){1}
        \DashLine(-3,12)(31,32){1}
\DashCArc(34,20)(12,283,103){1}
\DashCArc(0,0)(12,180,283){1}
        \Photon(-25,0)(-2,0){1.5}{6}\Text(-30,0)[br]{$p_1$}
        \Line(0,0)(34,20)
        \Photon(0,0)(34,-20){1.5}{8}
        \Vertex(0,0){4}\Vertex(-8,0){2}
\Line(34,20)(59,35)
        \Photon(34,20)(59,5){1.5}{8}
        \Vertex(34,20){4}\Vertex(24.5,15){2}
        \DashCArc(0,0)(8,0,360){1}
        \Text(47,-23)[br]{$p_4$}
        \Text(72,2)[br]{$p_3$}
        \DashLine(3,-12)(37,8){1}
        \Text(72,32)[br]{$p_2$}
\end{picture}
}\\
\centerline{\begin{picture}(150,80)
        \SetOffset(30,24)
        \DashCArc(0,0)(12,103,180){1}
        \DashLine(-3,12)(31,32){1}
\DashCArc(34,20)(12,283,103){1}
\DashCArc(0,0)(12,180,283){1}
        \Photon(-25,0)(-2,0){1.5}{6}\Text(-30,0)[br]{$p_1$}
        \Line(0,0)(34,20)
        \Photon(0,0)(34,-20){1.5}{8}
        \Vertex(0,0){4}\Vertex(-8,0){2}
\Photon(34,20)(59,35){1.5}{8}
        \Line(34,20)(59,5)
        \Vertex(34,20){4}\Vertex(24.5,15){2}
        \DashCArc(0,0)(8,0,360){1}
        \Text(47,-23)[br]{$p_4$}
        \Text(72,2)[br]{$p_3$}
        \DashLine(3,-12)(37,8){1}
        \Text(72,32)[br]{$p_2$}
      \end{picture}
\begin{picture}(150,80)
        \SetOffset(30,24)
        \DashCArc(0,0)(12,103,180){1}
        \DashLine(-3,12)(31,32){1}
\DashCArc(34,20)(12,283,103){1}
\DashCArc(0,0)(12,180,283){1}
        \Photon(-25,0)(-2,0){1.5}{6}\Text(-30,0)[br]{$p_1$}
        \Line(0,0)(34,20)
        \Line(0,0)(34,-20)
        \Vertex(0,0){4}\Vertex(-8,0){2}
\Photon(34,20)(59,35){1.5}{8}
        \Photon(34,20)(59,5){1.5}{8}
        \Vertex(34,20){4}\Vertex(24.5,15){2}
        \DashCArc(34,20)(8,0,360){1}
        \Text(47,-23)[br]{$p_4$}
        \Text(72,2)[br]{$p_3$}
        \DashLine(3,-12)(37,8){1}
        \Text(72,32)[br]{$p_2$}
      \end{picture}}
\\ 
The difference between these translation kernels and the light-cone
contribution is only in the denominators. Examining these:
\begin{align}
\lim_{p_1^2,p_2^2,p_3^2,p_4^2\to0}p_1^2\,p_2^2\,p_3^2\,p_4^2
&
\left\{\frac1{\hat 1\left(\frac{p_1^2}{\hat 1}+\frac{p_2^2}{\hat 2}
+\frac{p_3^2}{\hat 3}+\frac{p_4^2}{\hat 4}\right)(\hat 1+\hat
4)\left(\frac{p_{41}^2}{\hat 1+\hat 4}+\frac{p_2^2}{\hat 2}
+\frac{p_3^2}{\hat 3}\right)p_2^2p_3^2p_4^2}
\right.
\nonumber\\
&+
\frac1{\hat 2\left(\frac{p_1^2}{\hat 1}+\frac{p_2^2}{\hat 2}
+\frac{p_3^2}{\hat 3}+\frac{p_4^2}{\hat 4}\right)(\hat 2+\hat
3)\left(\frac{p_{23}^2}{\hat 2+\hat 3}+\frac{p_1^2}{\hat 1}
+\frac{p_4^2}{\hat 4}\right)p_1^2p_3^2p_4^2}
\nonumber\\
&+\frac1{\hat 3\left(\frac{p_1^2}{\hat 1}+\frac{p_2^2}{\hat 2}
+\frac{p_3^2}{\hat 3}+\frac{p_4^2}{\hat 4}\right)(\hat 2+\hat
3)\left(\frac{p_{23}^2}{\hat 2+\hat 3}+\frac{p_1^2}{\hat 1}
+\frac{p_4^2}{\hat 4}\right)p_1^2p_2^2p_4^2}
\nonumber\\
&\left.+\frac1{\hat 4\left(\frac{p_1^2}{\hat 1}+\frac{p_2^2}{\hat 2}
+\frac{p_3^2}{\hat 3}+\frac{p_4^2}{\hat 4}\right)(\hat 1+\hat
4)\left(\frac{p_{41}^2}{\hat 1+\hat 4}+\frac{p_2^2}{\hat 2}
+\frac{p_3^2}{\hat 3}\right)p_1^2p_2^2p_3^2}
\right\}
\nonumber \\=&
\lim_{p_1^2,p_2^2,p_3^2,p_4^2\to0}\frac{-\frac{p_{41}^2}{\hat 1+\hat 4}+\frac{p_1^2}{\hat 1}+\frac{p_4^2}{\hat 4}
-\frac{p_2^2}{\hat 2}-\frac{p_3^2}{\hat 3} }
{(\hat 1+\hat 4)\left(\frac{p_{41}^2}{\hat 1+\hat 4}+\frac{p_2^2}{\hat 2}
+\frac{p_3^2}{\hat 3}\right)\left(-\frac{p_{41}^2}{\hat 1+\hat 4}+\frac{p_1^2}{\hat 1}
+\frac{p_4^2}{\hat 4}\right)}
\nonumber 
\\=&\frac{1}
{p_{41}^2}\,,
\end{align}
(We omit the $+i\epsilon$ accompanying each $p^2$ here since it is not
important in our discussion.) we see that the factor ${p_1^2}/{\hat
1}+{p_2^2}/{\hat 2} +{p_3^2}/{\hat 3}+{p_4^2}/{\hat 4}$ in the
denominator is cancelled out and the limit procedure is valid at last.
The denominator provides the propagator needed in the light-cone
computation. Since the combined limit is valid, like in the $(++-)$
case, we could take the limit in any order, for example take the
$p_1^2,p_2^2,p_3^2\to0$ first and then $p_4^2\to 0$ at last. Then one
finds the first three diagrams vanish and the contribution comes only
from last diagram and the factor $\hat 4({p_1^2}/{\hat
1}+{p_2^2}/{\hat 2} +{p_3^2}/{\hat 3}+{p_4^2}/{\hat 4})$ becomes
$p_4^2$ to be cancelled with $p_4^2$ in the numerator from the LSZ
procedure. This reproduces the light-cone computation of the
amplitude. One can imagine that the same thing happens for general
multileg one-minus-helicity amplitudes.  Fortunately these amplitudes
vanish at tree-level in four dimensional Minkowski space, (and for
$n>2$ in arbitrary signature) which means that the translation kernel
contributions add up to zero.

The other way that $\sum_{j=1}^n{(p_j^2+i\epsilon)/\hat p_j}=0$
without all of the $p_j$ being external legs is if some of the terms
in the sum cancel against each other, or if the translation kernel is
connected to a vertex by a momentum that is on-shell. At tree-level
this can only occur for special choices of the momenta of the external
particles, and cannot contribute to an amplitude with generic values
of external momenta. So at tree-level the equivalence theorem can be
used for non-trivial generic amplitudes, which is why the MHV rules
correctly reproduce tree-level amplitudes without having to take
account of the translation between $\gA$, $\bar \gA$ and $\gB$,
$\bar\gB$ fields.

\subsection{One loop}
There are several processes that can occur at one-loop order that give
rise to evasions of S-matrix equivalence. The first is that loops can
dress the propagators that occur in tree-level diagrams. Secondly, we
can have tadpole diagrams in which two legs of a translation kernel
are contracted with each other. These diagrams are responsible for the
all positive helicity amplitudes `missing' from a straightforward
application of the MHV rules. Thirdly we can have more general
processes in which the loop integration has an infra-red divergence
that might cancel the LSZ factor.

\subsubsection{Dressing propagators}\label{sect:dressing-prop}
Loops can dress propagators, so, at one-loop, as for tree-level
$\sum_{j=1}^n{(p_j^2+i\epsilon)/\hat p_j}$ can vanish when each of the
$p_j$ is the momentum of an on-shell gluon. For example, the first
interaction in (\ref{lagexp}), ${\cal L}^{--+}$ which we denote by
\\
\centerline{\begin{picture}(60,60)
        \SetOffset(30,30)
        \Line(-39.5,0)(-2,0)\Text(-47,-3)[br]{$\gB$}
        \Vertex(7,4){2}\Vertex(7,-4){2}
        \Line(0,0)(34,20)
        \Line(0,0)(34,-20)
        \Vertex(0,0){4}
        \Text(47,17)[br]{$\bar\gB$} \Text(47,-23)[br]{$\bar\gB$}
\end{picture}}
\\
can be contracted with the fifth term in the expansion of
$\hat\partial \bar\gA$ equation (\ref{eq:barA-exp})
\\
\centerline{\begin{picture}(80,80)
        \SetOffset(30,40)
        \DashCArc(0,0)(12,103,180){1}
        \DashLine(-3,-12)(31,-32){1}
\DashCArc(34,-20)(12,257,77){1}
\DashCArc(0,0)(12,77,253){1}
        \Line(-25,0)(-2,0)
        \Line(0,0)(34,20)
        \Line(0,0)(34,-20)\Vertex(24.5,-15){4}
        \Vertex(0,0){4}\Vertex(8,-5){2}
\Line(34,-20)(59,-35)
        \Line(24.5,-15)(58.5,5)
        \Vertex(34,-20){2}
\DashCArc(9.5,-5)(8,257,77){1}
\DashCArc(0,0)(8,103,180){1}\DashCArc(0,0)(8,77,253){1}
\DashLine(-1,-9)(8.5,-14){1}
\DashLine(1,9)(10.5,4){1}
\Text(47,20)[br]{$\gB$}
        \Text(72,3)[br]{$\gB$}
        \DashLine(3,12)(37,-8){1}
        \Text(72,-40)[br]{$\hat\partial\bar\gB$}
      \end{picture}}
\\
to give
\\
\centerline{\begin{picture}(80,80)
        \SetOffset(30,40)
        \DashCArc(0,0)(12,103,180){1}
        \DashLine(-3,-12)(31,-32){1}
\DashCArc(34,-20)(12,257,77){1}
\DashCArc(0,0)(12,77,253){1}
        \Line(-25,0)(-2,0)
        \Line(0,0)(34,20)
        \Line(0,0)(24.5,-15)\Vertex(24.5,-15){4}
        \Vertex(0,0){4}\Vertex(8,-5){2}
        \PhotonArc(48.5,-30)(30.4,30,150){1.5}{8}
        \PhotonArc(48.5,0)(30.4,210,330){1.5}{8}
\Vertex(30,-25){2}\Vertex(67,-5){2}
\DashCArc(9.5,-5)(8,257,77){1}
\DashCArc(0,0)(8,103,180){1}\DashCArc(0,0)(8,77,253){1}
\DashLine(-1,-9)(8.5,-14){1}
\DashLine(1,9)(10.5,4){1}
\Text(47,20)[br]{$\gB$}
        \Vertex(72.5,-15){4}\Line(72.5,-15)(97.5,-15)\Text(108.5,-20)[br]{$\bar\gB$}\Vertex(82,-15){2}
        \DashLine(3,12)(37,-8){1}
      \end{picture}}
\\
This will contribute to the Green function $\langle
\,\gA(p_1)\,\bar\gA(p_2)\,\bar\gA(p_3) \,\rangle$, for example by
contracting $\gB$ with the leading term in the expansion of
$\bar\gA(p_2)$ and $\bar\gB$ with that of $\gA(p_1)$.  The propagators
cancel two of the LSZ factors for the $++-$ amplitude.  Taking
$p_2^2+i\epsilon=0$ and $p_1^2+i\epsilon=0$ causes the $1/\sum\Omega$
factor denoted by the inner broken curve to reduce to
$1/(p_3^2+i\epsilon)$, which will cancel the remaining LSZ factor,
thus evading the equivalence theorem and producing a contribution to
the three-point amplitude that is the same as the tree-level diagram
with a self-energy insertion on the $p_1$ leg. In Minkowski space the
three-point amplitude vanishes on-shell anyway, so this evasion
appears inconsequential. For complex on-shell momenta, of the kind
used in the BCFW rules, this amplitude does not vanish, so it is
worthwhile considering this further. We noted earlier that the
relations (\ref{eq:NewXiR-}) enable us to re-write the series for
$\bar\gA$ in a way that moves the position of the dotted lines so that
the singularity $1/\sum_i\Omega_i$ corresponding to the dotted lines
around the left big black dot is cancelled out after we sum the fifth
and sixth term in (\ref{eq:barA-exp}). These combine to give 
\\\centerline{\begin{picture}(140,80)
        \SetOffset(30,40)
        \DashCArc(0,0)(12,103,180){1}
        \DashLine(-3,-12)(31,-32){1}
\DashCArc(34,-20)(12,257,77){1}
\DashCArc(0,0)(12,77,253){1}
        \Line(-25,0)(-2,0)\Text(-25,-3)[t]{$\bar{\cal A}$}
	\Vertex(-25,0){1}
        \Line(0,0)(34,20)
        \Line(0,0)(24.5,-15)\Vertex(24.5,-15){4}
        \Vertex(0,0){4}\Vertex(8,-5){2}
	\DashCArc(24.5,-15)(8,0,360){1}
        \PhotonArc(48.5,-30)(30.4,30,150){1.5}{8}
        \PhotonArc(48.5,0)(30.4,210,330){1.5}{8}
\Vertex(30,-25){2}\Vertex(67,-5){2}
\Text(47,20)[br]{$\cal B$}
        \Vertex(72.5,-15){4}\Line(72.5,-15)(97.5,-15)\Text(108.5,-20)[br]{$\bar{\cal
B}$}\Vertex(82,-15){2}
        \DashLine(3,12)(37,-8){1}
\end{picture}}
\\
Since the contribution of the internal line to the denominator $\sum
p_i^2/\hat p_i$ represented by the inner broken curve can not be zero,
it is obvious that there is no $1/p^2$ generated in this diagram. So
this diagram can not contribute to the amplitude. The same is true for
the case of a dressed propagator on a $\cal B$ leg:
\\\centerline{\begin{picture}(140,80)
        \SetOffset(30,40)
        \DashCArc(0,0)(12,103,180){1}
        \DashLine(-3,-12)(31,-32){1}
\DashCArc(34,-20)(12,257,77){1}
\DashCArc(0,0)(12,77,253){1}
        \Line(-25,0)(-2,0)\Text(-25,-3)[t]{$\bar{\cal A}$}
	\Vertex(-25,0){1}
        \Line(0,0)(34,20)
        \Line(0,0)(24.5,-15)\Vertex(24.5,-15){4}
        \Vertex(0,0){4}\Vertex(8,5){2}
	\DashCArc(24.5,-15)(8,0,360){1}
        \PhotonArc(48.5,-30)(30.4,30,150){1.5}{8}
        \PhotonArc(48.5,0)(30.4,210,330){1.5}{8}
\Vertex(16.5,-10){2}\Vertex(67,-5){2}\Vertex(67,-25){2}
\Text(47,20)[br]{$\bar{\cal B}$}
        \Vertex(72.5,-15){4}\Line(72.5,-15)(97.5,-15)\Text(108.5,-20)[br]{${\cal
B}$}
        \DashLine(3,12)(37,-8){1}
\end{picture}} 

The three-point interaction can dress a propagator either in the way
just described, or, potentially by two such vertices being glued
together 
\\\centerline{\begin{picture}(80,80)
        \SetOffset(30,40)
        \Line(0,-15)(24.5,-15)\Vertex(24.5,-15){4}\Vertex(15,-15){2}
        \PhotonArc(48.5,-30)(30.4,30,150){1.5}{8}
        \PhotonArc(48.5,0)(30.4,210,330){1.5}{8}
\Vertex(30,-25){2}\Vertex(67,-5){2}
\Text(-7,-20)[br]{$\bar\gB$}
        \Vertex(72.5,-15){4}\Line(72.5,-15)(97.5,-15)\Text(108.5,-20)[br]{$\bar\gB$}\Vertex(82,-15){2}
      \end{picture}}
\\
An insertion of this kind into a diagram effectively changes a
$\gB$-field into a $\bar\gB$-field, however explicit calculation shows
that this vanishes.  At one-loop the only other vertices that can
contribute to dressing propagators are contained in ${\cal L}^{--++}$:
\\
\centerline{\begin{picture}(60,52)
        \SetOffset(30,24)
         \Line(-34,20)(0,0)\Text(-41,17)[br]{$\bar\gB$}
        \Line(-34,-20)(0,0)\Text(-41,-23)[br]{$\gB$}
        \Vertex(-7,4){2}\Vertex(7,-4){2}
        \Line(0,0)(34,20)
        \Line(0,0)(34,-20)
        \Vertex(0,0){4}
        \Text(47,17)[br]{$\gB$} \Text(47,-23)[br]{$\bar\gB$}
\end{picture}
}\\
and these produce insertions that connect $\gB$ with $\bar\gB$
\\
\centerline{\begin{picture}(60,75)
        \SetOffset(30,24)
\PhotonArc(0,20)(20,0,360){1.5}{16}
         \Line(-34,-20)(0,0)
       \Text(-41,-23)[br]{$\gB$}
        \Vertex(-7,2){2}\Vertex(7,-4){2}
        \Line(0,0)(34,-20)
        \Vertex(0,0){4}
 \Text(47,-23)[br]{$\bar\gB$}
\end{picture}}
\\
Dressing propagators can produce diagrams that evade the equivalence
theorem, but only if the corresponding tree-level diagrams do so
already, in which case the result is proportional to the tree-level
amplitude.  As we have seen this only happens for amplitudes that
vanish in the physical dimension, so this source of equivalence
theorem evasion has no physical consequence. However there is a
subtlety involved in the one-loop $(++-)$ amplitude in $(++--)$
signature. In section \ref{sect:mppp}, we will find that the tadpoles
formed from MHV vertices already include the diagrams with external
leg corrections. Including translation kernel contributions in the
amplitude would appear to count the diagrams with external leg
corrections twice.  We will solve  this  puzzle in section
\ref{sect:oneloop-ppm}.

\subsubsection{Tadpoles}
At tree-level we dismissed the second way that
$\sum_{j=1}^n{(p_j^2+i\epsilon)/\hat p_j}$ could vanish because it
could only apply to special configurations of external momenta.  When
we integrate over loop momenta such special configurations can easily
arise, and so we must analyse them.

The simplest way that two of the terms in
$\sum_{j=1}^n{(p_j^2+i\epsilon)/\hat p_j}$ could cancel without each
being on-shell occurs in the translation kernel for $\bar\gA$ when a
$\gB$ and $\bar\gB$ field are contracted, because then their lines
carry equal and opposite momenta. These are `tadpoles' when drawn in
terms of the translation kernels, e.g.  
\\ \centerline{
\begin{picture}(50,60)
        \SetOffset(30,30)
        \Line(-25,0)(-2,0)
        \Line(0,0)(17,25)
\Line(0,0)(-17,25)\Text(-22,23)[br]{$\gB$}
        \PhotonArc(8,-12)(14.5,0,360){1.5}{16}
        \GCirc(0,0){8}{0.75}
        \Text(30,23)[br]{$\gB$}\Vertex(-7.5,-11){2}
   \end{picture}
}
\\
but are rather more complicated when drawn in terms of the graphical
solution. For example, one of the terms contributing to this tadpole
originates in the following term which appears in the expansion of
$\bar\gA$:
\\
\centerline{\begin{picture}(80,90)
        \SetOffset(30,45)
        \DashCArc(0,0)(12,103,180){1}\DashCArc(-34,-20)(16,103,180){1}
        \DashLine(-3,12)(31,32){1}
\DashCArc(34,20)(12,283,103){1}
\DashCArc(0,0)(12,180,283){1}\DashCArc(-34,-20)(16,180,310){1}
        \Line(0,0)(34,20)
        \Line(0,0)(-25,15)\Vertex(-8,-5){2}
        \Vertex(0,0){4}
\Line(34,20)(59,35)
        \Line(34,20)(9,35)
        \Vertex(34,20){4}\Vertex(24.5,15){2}
            \Vertex(-34,-20){4}\Line(-34,-20)(0,-40)\Vertex(-26,-25){2}
\Line(-34,-20)(0,0)\Line(-34,-20)(-57,-20)
\DashCArc(34,20)(8,283,103){1}\DashCArc(34,20)(16,310,103){1}
\DashCArc(24.5,15)(8,180,310){1}\DashCArc(24.5,15)(8,103,180){1}
\DashLine(23.5,24)(33,29){1}
\DashLine(25.5,6)(35,11){1}
        \Text(13,-43)[br]{$\hat\partial\bar\gB$}
        \Text(72,32)[br]{$\gB$}
        \DashLine(3,-12)(37,8){1}\DashLine(-39,-5)(31,36){1}
        \DashLine(-29,-35)(43,7){1}
        \Text(6,32)[br]{$\gB$}\Text(-28,12)[br]{$\gB$}
      \end{picture}}
\\
Contracting $\bar\gB$ with a $\gB$ and the remaining fields with
external gluons gives
\\
\centerline{\begin{picture}(80,90)
        \SetOffset(30,45)
        \DashCArc(0,0)(12,103,180){1}\DashCArc(-34,-20)(16,103,180){1}
        \DashLine(-3,12)(31,32){1}
\DashCArc(34,20)(12,283,103){1}
\DashCArc(0,0)(12,180,283){1}\DashCArc(-34,-20)(16,180,310){1}
        \Line(0,0)(34,20)
        \Line(0,0)(-34,-20)\Vertex(-8,-5){2}
        \Vertex(0,0){4}
        \Photon(34,20)(9,35){1.5}{8}
        \Vertex(34,20){4}\Vertex(24.5,15){2}
            \Vertex(-34,-20){4}\Line(-34,-20)(-57,-20)\Vertex(-29,-27){2}
\Photon(-25,15)(0,0){1.5}{8}
\DashCArc(34,20)(8,283,103){1}\DashCArc(34,20)(16,310,103){1}
\DashCArc(24.5,15)(8,180,310){1}\DashCArc(24.5,15)(8,103,180){1}
\DashLine(23.5,24)(33,29){1}
\DashLine(25.5,6)(35,11){1}
        \DashLine(3,-12)(37,8){1}\DashLine(-39,-5)(31,36){1}
        \DashLine(-29,-35)(43,7){1}
 \PhotonArc(0,0)(40,215,35){1.5}{16}\Text(6,32)[br]{$p_3$}\Text(-28,12)[br]{$p_2$}\Text(-61,-22)[br]{$p_1$}
\end{picture}}
\\
When the $p_2$ and $p_3$ are put on-shell, $p_2^2+i\epsilon=0$ and
$p_3^2+i\epsilon=0$, so the dotted line cutting the three external
momenta and the gluon propagator reduces to
$\sum_{j=1}^n{(p_j^2+i\epsilon)/\hat p_j}=(p_1^2+i\epsilon)/\hat p_1$
resulting in an evasion of S-matrix equivalence.  Because the
contraction used to make a tadpole removes a $\gB$ and a $\bar\gB$
field from the translation kernels for $\bar\gA$ they can contribute
to one-loop amplitudes involving only positive helicity gluons. In
\cite{us} it was found that it is this mechanism that is responsible
for generating the one-loop all plus four-point amplitude (\ref{4pt})
that a na\"ive application of the MHV rules cannot account for.

\subsubsection{Infra-red divergent loop integration}
\label{sec:infra-red}
Evasion of S-matrix equivalence might arise in a more general
situation when a vertex is attached to a translation kernel. For
illustration we focus on one term in the expansion of $\bar\gA$ and
contract two of the legs with those of some arbitrary subgraph denoted
by the open circle:
\\
\centerline{\begin{picture}(80,80)
        \SetOffset(30,40)
        \DashCArc(0,0)(12,103,180){1}
        \DashLine(-3,-12)(31,-32){1}
\DashCArc(34,-20)(12,257,77){1}
\DashCArc(0,0)(12,77,253){1}
        \Line(-25,0)(-2,0)\Text(-32,-3)[br]{$p$}
        \Photon(0,0)(34,20){1.5}{8}
        \Line(0,0)(24.5,-15)\Vertex(24.5,-15){4}\DashCArc(24.5,-15)(9,0,360){1} 
        \Vertex(0,0){4}\Vertex(8,-5){2}
        \PhotonArc(48.5,-30)(30.4,30,150){1.5}{8}\Text(50,5)[br]{$q$}
        \PhotonArc(48.5,0)(30.4,210,330){1.5}{8}\Text(70,-45)[br]{$-j-q$}
\Vertex(30,-25){2}\Vertex(67,-5){2}
\Text(47,20)[br]{$p_1$}\Text(109,2)[br]{$\bar\gB$}\Text(109,-38)[br]{$\gB$}
       \BCirc(72.5,-15){4}\Line(75,-12.5)(97.5,5)\Line(75,-17.5)(97.5,-35)\Vertex(82,-8){2}
        \DashLine(3,12)(37,-8){1}\Text(100,0)[br]{$.$}\Text(100,-5)[br]{$.$}
        \Text(100,-10)[br]{$.$}\Text(100,-15)[br]{$.$}\Text(100,-20)[br]{$.$}
        \Text(100,-25)[br]{$.$}\Text(100,-30)[br]{$.$}
      \end{picture}
}
\\
If we take $p_1^2+i\epsilon=0$, (having cancelled the corresponding
LSZ factor with the propagator,) the loop integration is
\be
\int d^Dq \,\frac{1}{\frac{p^{2}+i\epsilon}{\hat{p}}+\frac{q^{2}+i\epsilon}{\hat{q}}-\frac{(q+j)^{2}+i\epsilon}{\hat q+\hat j}}
\frac{1}{\frac{j^{2}+i\epsilon}{\hat \jmath}+\frac{q^{2}+i\epsilon}{\hat{q}}-\frac{(q+j)^{2}+i\epsilon}{\hat q+\hat j}}
\,\,{1\over q^2+i\epsilon}\,\,
{1\over (j+q)^2+i\epsilon}f(j,q)
\label{kernelloop1} \ee
with $j=p+p_1$. 

We need to investigate whether this integral can generate a factor of
$1/(p^{2}+i\ \epsilon)$. To do so it would have to be divergent as $p$
goes on-shell.  The integrand has a number of singularities as a
function of the components of loop momentum $q^\mu$ but by deforming
the integration contours into the complex $q^\mu$-planes the surfaces
where the integrand diverges can typically be avoided so that the
integral is well-defined. We are aided in identifying the directions
in which to deform the contours by the $i \epsilon$ prescription.  (We
can ignore what happens as $q\rightarrow\infty$ as the ultra-violet
behaviour is regulated). However, as we vary $p$ the positions of
these singularities move, and it is possible that our integration
surface may lie between several singularity surfaces that approach
each other for some values of $p$ and pinch the contours so that they
can no longer be deformed to avoid the singularity. As this happens
the value of the integral itself diverges as a function of $p$. Prior
to taking the on-shell limit we can deform the integration surface so
that it consists of a piece surrounding the singularities and a piece
that we can move well away from either singularity. In the on-shell
limit we can ignore this last piece because of the LSZ factor,
$p^2+i\epsilon$. We now focus on the contribution from the piece
surrounding the singularity, which means that in the loop integral we
take $f(j,q)$ as constant.  We begin by  integrating out the
$\check{q}$ component. The first two factors of the integrand come
from the translation kernels and so do not depend on $\check{q}$. As
we close the $\check{q}$ contour in the complex plane we pick up
singularities from the propagators. Using conservation of momentum the
residue can be put into a form similar to that of the kernel, but
without $(p^{2}+i\epsilon)/\hat{p}$:
\be
\frac{\theta(\hat{q})\theta(-\hat q-\hat j)-\theta(-\hat{q})\theta(\hat q+\hat j)}{\hat q\,(\hat q+\hat j)}\frac{2\pi i}{\frac{q^{2}+i\epsilon}{\hat{q}}-\frac{(q+j)^{2}+i\epsilon}{\hat q+\hat j}}
\ee
which, of course, does not depend on $\check q$.  This allows us to
extract $1/(p^{2}+i\epsilon)$ explicitly from the integral
(\ref{kernelloop1}) which becomes

\bq
&&- \frac{2\pi i\,\hat p}{p^{2}+i\epsilon}\int\left(\prod\limits_{i=1}^{D/2-1} dq_{(i)}d\bar q_{(i)}\right)\,d\hat{q} \, \frac{\theta(\hat{q})\theta(-\hat q-\hat j)-\theta(-\hat{q})\theta(\hat q+\hat j)}{\hat q\,(\hat q+\hat j)}
\\
&& \hspace{2.1cm} \times
\left[\frac{1}{\frac{p^{2}+i\epsilon}{\hat{p}}+\frac{q^{2}+i\epsilon}{\hat{q}}-\frac{(q+j)^{2}+i\epsilon}{\hat
q+\hat
j}}-\frac{1}{\frac{q^{2}+i\epsilon}{\hat{q}}-\frac{(q+j)^{2}+i\epsilon}{\hat
q+\hat j}}\right]f_1 (j,q)\label{kernelloop2}
\,.\eq
Since the second factor in the integrand of (\ref{kernelloop1}) is
finite when the first factor is singular, it is irrelevant to our
discussion and we absorb it into $f_1$.

The LSZ factor is cancelled by the $1/(p^2+i\epsilon)$. If we now take
the on-shell limit $p^2+i\epsilon\rightarrow 0$ then the two terms in
square brackets cancel and the integral actually vanishes, provided
that no singularity is encountered as we integrate over $\hat q$.
However, for certain values of $j$ and $q_{(i)}$ both terms in the
square brackets are divergent close to the real axis, so we have to
investigate the location of these singularities. The first diverges
for
\be
\hat q=\frac{-b\pm\sqrt{b^2-4a(c+i\epsilon\hat j)}}{2a}
\label{eq:q-hat solution} \ee
with
\be
a=\phi-2\check j\,,\quad
\phi=\frac{p^{2}+i\epsilon}{\hat{p}}\,,\quad
b=\hat j\phi-j^2+2\sum_i \left(q_{(i)}\bar j_{(i)}+\bar q_{(i)} j_{(i)}\right),
\quad c=-2\hat j\sum_i q_{(i)}\bar q_{(i)}\,,
\ee
whilst the location of the pole in the second term is given by the
above expression with $\phi$ set to zero.  For the moment treat $\phi$
as being real. Then for $b^2>4ac$ the poles are close to the real
axis, with an imaginary piece
\be
\mp \frac{\epsilon\hat j}{\sqrt{b^2-4ac}}\,.
\ee
Since these are on the same side of the real axis for both terms in
square brackets it is clear that the contribution to the integral of
these two terms cancels even when the singularities are close to the
real axis. Consequently there is no S-matrix equivalence evasion in
this case, provided that we keep $\phi$ real as we take the on-shell
limit for external legs.

\section{One-loop $(-+++)$ amplitude}
\label{sect:mppp}
In \cite{us} we described how the one-loop $(++++)$ amplitude arises
in this approach as a tadpole-like diagram constructed from
translation kernel.  By contrast the one-minus helicity amplitude is
constructed from the tadpole diagram of a MHV vertex.


Let us now look at a box diagram, $A(1^-2^+3^+4^+)$ with $1^-$
attached to MHV $(--+)$ vertex. The integrand of the light-cone
amplitude is
\begin{eqnarray}
A^{(1)}(1^-2^+3^+4^+)=2^4 {V^2(\bar a234,1,a)\bar V^2(4,1a,\bar a 23)\bar V^2(3,41a,\bar a
2)\bar V^2(2,341a,\bar a ) \over p_{\bar a }^2p_{1a}^2p_{\bar a
23}^2p_{\bar a 2}^2} 
\,.\label{eq:A-box-1}
\end{eqnarray}
It must come from the tadpole diagram in the CSW method by connecting
two lines of six-point MHV vertices.  We can identify the tadpole
contributions to this amplitude in the following way. First, we can
cut any one of the four internal lines and get four tree-level MHV
diagrams. 
\\
\centerline{\footnotesize
\begin{picture}(60,60)
\SetOffset(30,30)
       \Line(-20,-20)(-10,-10)
	\Line(-20,20)(-10,10)
	\Line(20,20)(10,10) 
	\Line(20,-20)(10,-10) 
       \ArrowLine(-10,-10)(-10,10)
	\Line(-10,10)(10,10)
	\Line(10,10)(10,-10)
	\Line(10,-10)(-10,-10)
	\Vertex(-10,-10){2}
	\Vertex(-10,10){2}
	\Vertex(10,-10){2}
	\Vertex(10,10){2}
	\Vertex(-10,6){1.3}
	\Vertex(6,10){1.3}
	\Vertex(-6,-10){1.3}
	\Vertex(10,-6){1.3}
	\Vertex(-13,-13){1.3}
	\Text(-21,-21)[tr]{$1^-$}
	\Text(-21,21)[br]{$2^+$}
	\Text(21,21)[bl]{$3^+$}
	\Text(21,-21)[tl]{$4^+$}
	\Text(-11,0)[r]{$p_a$}
\end{picture}
\begin{picture}(30,60)
\SetOffset(15,30)
\Text(0,0)[ ]{$\longrightarrow$}
\end{picture}
\begin{picture}(60,60)
\SetOffset(30,30)
       \Line(-20,-20)(-10,-10)
	\Line(-20,20)(-10,10)
	\Line(20,20)(10,10) 
	\Line(20,-20)(10,-10) 
        \Photon(-10,-10)(-10,-2){1}{2}
	\Photon(-10,2)(-10,10){1}{2}
	\Line(-10,10)(10,10)
	\Line(10,10)(10,-10)
	\Line(10,-10)(-10,-10)
	\Vertex(-10,-10){2}
	\Vertex(-10,10){2}
	\Vertex(10,-10){2}
	\Vertex(10,10){2}
	\Vertex(-10,6){1.3}
	\Vertex(6,10){1.3}
	\Vertex(-6,-10){1.3}
	\Vertex(10,-6){1.3}
	\Vertex(-13,-13){1.3}
	\Text(-21,-21)[tr]{$1^-$}
	\Text(-21,21)[br]{$2^+$}
	\Text(21,21)[bl]{$3^+$}
	\Text(21,-21)[tl]{$4^+$}
\end{picture}
\hspace{0.5cm}
\begin{picture}(60,60)
\SetOffset(30,30)
       \Line(-20,-20)(-10,-10)
	\Line(-20,20)(-10,10)
	\Line(20,20)(10,10) 
	\Line(20,-20)(10,-10) 
        \Line(-10,-10)(-10,10)
	\Line(-10,10)(10,10)
	\Line(10,10)(10,-10)
	\Photon(-2,-10)(-10,-10){1}{2}
	\Photon(2,-10)(10,-10){1}{2}
	\Vertex(-10,-10){2}
	\Vertex(-10,10){2}
	\Vertex(10,-10){2}
	\Vertex(10,10){2}
	\Vertex(-10,6){1.3}
	\Vertex(6,10){1.3}
	\Vertex(-6,-10){1.3}
	\Vertex(10,-6){1.3}
	\Vertex(-13,-13){1.3}	\Text(-21,-21)[tr]{$1^-$}
	\Text(-21,21)[br]{$2^+$}
	\Text(21,21)[bl]{$3^+$}
	\Text(21,-21)[tl]{$4^+$}
\end{picture}
\hspace{0.5cm}
\begin{picture}(60,60)
\SetOffset(30,30)
       \Line(-20,-20)(-10,-10)
	\Line(-20,20)(-10,10)
	\Line(20,20)(10,10) 
	\Line(20,-20)(10,-10) 
       \Line(-10,-10)(-10,10)
	\Line(-10,10)(10,10)
	\Photon(10,10)(10,2){1}{2}
	\Photon(10,-10)(10,-2){1}{2}
	\Line(10,-10)(-10,-10)
	\Vertex(-10,-10){2}
	\Vertex(-10,10){2}
	\Vertex(10,-10){2}
	\Vertex(10,10){2}
	\Vertex(-10,6){1.3}
	\Vertex(6,10){1.3}
	\Vertex(-6,-10){1.3}
	\Vertex(10,-6){1.3}
	\Vertex(-13,-13){1.3}
	\Text(-21,-21)[tr]{$1^-$}
	\Text(-21,21)[br]{$2^+$}
	\Text(21,21)[bl]{$3^+$}
	\Text(21,-21)[tl]{$4^+$}
\end{picture}
\hspace{0.5cm}
\begin{picture}(60,60)
\SetOffset(30,30)
       \Line(-20,-20)(-10,-10)
	\Line(-20,20)(-10,10)
	\Line(20,20)(10,10) 
	\Line(20,-20)(10,-10) 
       \Line(-10,-10)(-10,10)
	\Photon(2,10)(10,10){1}{2}
	\Photon(-10,10)(-2,10){1}{2}
	\Line(10,10)(10,-10)
	\Line(10,-10)(-10,-10)
	\Vertex(-10,-10){2}
	\Vertex(-10,10){2}
	\Vertex(10,-10){2}
	\Vertex(10,10){2}
	\Vertex(-10,6){1.3}
	\Vertex(6,10){1.3}
	\Vertex(-6,-10){1.3}
	\Vertex(10,-6){1.3}
	\Vertex(-13,-13){1.3}
	\Text(-21,-21)[tr]{$1^-$}
	\Text(-21,21)[br]{$2^+$}
	\Text(21,21)[bl]{$3^+$}
	\Text(21,-21)[tl]{$4^+$}
\end{picture}
}
\\
MHV vertices are generated by expanding the $\cal A$ and $\bar {\cal
A}$ in the lagrangian ${\cal L}^{--+}$. We first identify the three
point MHV $(--+)$ vertex in the tree-level diagrams and the three
parts in the diagrams corresponding to the expansion of $\cal A$ and
$\bar{\cal A}$ in  ${\cal L}^{--+}$. By comparing with the three parts
of the diagram, we can find out the corresponding three parts in the
light cone amplitude (\ref{eq:A-box-1}). Then by replacing the
propagators in the light-cone amplitude using (\ref{eq:replace}) we
can reconstruct the contribution to the one-loop box diagram of the
tadpole.  The four tree-level diagram contributions are (we label the
internal line between leg $1$ and $2$ as $a$):
\begin{eqnarray}
A^{(1,1)}=2^4 {V^2(\bar a234,1,a)\bar V^2(4,1a,\bar a 23)\bar V^2(3,41a,\bar a
2)\bar V^2(2,341a,\bar a )\over p_{\bar a}^2\hat P_{\bar a 23} \hat P_{\bar
a 2}\hat P_{1a} (\frac {P_{1a}^2}{\hat P_{1a}}+\frac {P_{\bar a}^2}{\hat P_{\bar
a}}) (\frac {P_{41a}^2}{\hat P_{41a}}+\frac {P_{\bar a}^2}{\hat P_{\bar
a}}) (\frac {P_{341a}^2}{\hat P_{341a}}+\frac {P_{\bar a}^2}{\hat P_{\bar
a}})}
\,,\\
A^{(1,2)}=2^4 {V^2(\bar a234,1,a)\bar V^2(4,1a,\bar a 23)\bar V^2(3,41a,\bar a
2)\bar V^2(2,341a,\bar a )\over P_{1a}^2\hat P_{\bar a 23} \hat P_{\bar
a 2}\hat P_{\bar a} (\frac {P_{1a}^2}{\hat P_{1a}}+\frac {P_{\bar a}^2}{\hat P_{\bar
a}}) (\frac {P_{\bar a2}^2}{\hat P_{\bar a2}}+\frac {P_{1 a}^2}{\hat
P_{1a}}) (\frac {P_{\bar a 23}^2}{\hat P_{\bar a 23}}+\frac {P_{1 a}^2}{\hat P_{ 1a}})}
\,,\\
A^{(1,3)}=2^4 {V^2(\bar a234,1,a)\bar V^2(4,1a,\bar a 23)\bar V^2(3,41a,\bar a
2)\bar V^2(2,341a,\bar a )\over P_{\bar a23}^2\hat P_{\bar a} \hat P_{\bar
a 2}\hat P_{1a} (\frac {P_{41a}^2}{\hat P_{41a}}+\frac {P_{\bar a}^2}{\hat P_{\bar
a}}) (\frac {P_{41a}^2}{\hat P_{41a}}+\frac {P_{\bar a2}^2}{\hat P_{\bar
a2}}) (\frac {P_{1a}^2}{\hat P_{1a}}+\frac {P_{\bar a23}^2}{\hat P_{\bar
a23}})}
\,,\\
A^{(1,4)}=2^4 {V^2(\bar a234,1,a)\bar V^2(4,1a,\bar a 23)\bar V^2(3,41a,\bar a
2)\bar V^2(2,341a,\bar a )\over P_{341a}^2\hat P_{41a} \hat P_{\bar
a}\hat P_{1a} (\frac {P_{1a}^2}{\hat P_{1a}}+\frac {P_{\bar a2}^2}{\hat P_{\bar
a2}}) (\frac {P_{41a}^2}{\hat P_{41a}}+\frac {P_{\bar a2}^2}{\hat P_{\bar
a2}}) (\frac {P_{341a}^2}{\hat P_{341a}}+\frac {P_{\bar a}^2}{\hat P_{\bar
a}})}
\,.
\end{eqnarray}
We have already set the $p_i^2$ in the denominator of the external
particles to zero, since there is no singularity when we put the
external particles on-shell. It makes no difference if we take the
on-shell limit before or after the LSZ procedure. It is easy to check
that these four terms add up to the integrand of the light-cone
amplitude for the box diagram (\ref{eq:A-box-1}).  The other box
diagrams, triangle, bubble diagrams of light-cone amplitude can be
checked in the same way. There are some subtle problems with diagrams
including corrections to external propagators  which we will address
in next section. So, in general, one can believe that the one loop
one-minus helicity amplitudes should all come just from the tadpoles
of MHV vertices.

\section{One-loop $(++-)$ amplitude with external tadpole dressing
propagators}
\label{sect:oneloop-ppm}

At first sight, for three point $(++-)$ amplitude, there could also be
contributions to the cut diagrams considered in the previous section
from translation kernels with dressed external propagators, since they
yield contributions proportional to tree-level amplitudes which are
not zero in $(++--)$ signature. This seems to count diagrams with
corrections to external propagators twice.  This problem arises from
the order of limits in LSZ procedure. In the example of the  previous
section it does not matter when we take the on shell limit because no
singularities are encountered in this limit.  But we must be more
careful with the diagrams with dressed propagators on external legs
because there will then be singularities from $1/\sum \Omega$.  We
should first calculate the off-shell Green function and then apply the
LSZ procedure. Also from the discussion in section
(\ref{sect:dressing-prop}) the Green function receives contributions
should not just from tadpoles of MHV five point vertices, but also
from translation kernels with dressed propagators. Let us look at the
example of a light-cone diagram for $\langle \bar{\cal
A}_{\bar1}\bar{\cal A}_{\bar 2}{\cal A}_{\bar 3}\rangle$:
\\\centerline{
\begin{picture}(140,80)
        \SetOffset(30,40)
        \Line(-25,0)(-2,0)\Text(-25,-3)[t]{$\bar{\cal A}_{\bar 1}$}
	\Vertex(-25,0){1}
        \Line(0,0)(34,20)\Vertex(34,20){1}
        \Line(0,0)(24.5,-15)\Vertex(24.5,-15){4}
        \Vertex(0,0){4}\Vertex(8,-5){2}
        \CArc(48.5,-30)(30.4,30,150)
        \CArc(48.5,0)(30.4,210,330)
	\Text(48.5,-33)[tc]{$l$}\Text(48.5,3)[bc]{$l+p_3$}
\Vertex(30,-25){2}\Vertex(67,-5){2}
\Text(47,20)[br]{$\bar{\cal A}_{\bar 2}$}
        \Vertex(72.5,-15){4}\Line(72.5,-15)(97.5,-15)\Text(97.5,-20)[lb]{${\cal
A}_{\bar3}$}
	\Vertex(82.5,-15){2}
	\Vertex(97.5,-15){1}
\end{picture}
}\\ 
The integrand of the light-cone computation of the diagram for the
Green function is
\begin{eqnarray}
A(1^+2^+3^-)=ig^6\frac{\bar V^2(1,2,3)\bar V^2(\bar3,l+p_3,\bar
l)V^2(-l-p_3,3,l)}{p_1^2
p_2^2(p_3^2)^2 l^2(l+p_3)^2}\,.
\end{eqnarray}
We have omitted the $i\epsilon$ in the propagators. According to the
method of the last section, the contribution from the tadpole of MHV
five-point vertices to this diagram can be constructed by replacing
the corresponding $1/p^2\to -1/(2\hat p\sum\Omega)$:
\\
\centerline{
\vbox{\hbox{
\begin{picture}(140,70)
        \SetOffset(30,35)
        \DashCArc(0,0)(12,103,180){1}
        \DashLine(-3,-12)(31,-32){1}
\DashCArc(34,-20)(12,257,77){1}
\DashCArc(0,0)(12,77,253){1}
        \Line(-25,0)(-2,0)
	\Vertex(-25,0){1}
	\Text(-25,-3)[lt]{$\bar {\cal A}_{\bar 1}$}
        \Line(0,0)(34,20)
	\Vertex(34,20){1}
        \Line(0,0)(24.5,-15)\Vertex(24.5,-15){4}
        \Vertex(0,0){4}\Vertex(8,-5){2}
        \CArc(48.5,-30)(30.4,30,150)
        \PhotonArc(48.5,0)(30.4,210,260){1.5}{8}
	\PhotonArc(48.5,0)(30.4,280,330){1.5}{8}
\Vertex(30,-25){2}\Vertex(67,-5){2}
\DashCArc(9.5,-5)(8,257,77){1}
\DashCArc(0,0)(8,103,180){1}\DashCArc(0,0)(8,77,253){1}
\DashLine(-1,-9)(8.5,-14){1}
\DashLine(1,9)(10.5,4){1}
\Text(47,20)[br]{$\bar {\cal A}_{\bar2}$}
        \Vertex(72.5,-15){4}
	\Line(72.5,-15)(97.5,-15)
	\Vertex(97.5,-15){1}\Text(97.5,-15)[l]{${\cal A}_{\bar 3}$}\Vertex(82,-15){2}
        \DashLine(3,12)(37,-8){1}
\end{picture}
}
\hbox{
\begin{picture}(140,35)
        \SetOffset(70,20)
	\Text(0,0)[]{$A^{(1)}$}
\end{picture}
}}
\vbox{\hbox{
\begin{picture}(140,70)
        \SetOffset(30,35)
        \DashCArc(0,0)(12,103,180){1}
        \DashLine(-3,-12)(31,-32){1}
\DashCArc(34,-20)(12,257,77){1}
\DashCArc(0,0)(12,77,253){1}
        \Line(-25,0)(-2,0)\Vertex(-25,0){1}
	\Text(-25,-3)[lt]{$\bar {\cal A}_{\bar 1}$}
        \Line(0,0)(34,20)\Vertex(34,20){1}
        \Line(0,0)(24.5,-15)\Vertex(24.5,-15){4}
        \Vertex(0,0){4}\Vertex(8,-5){2}
        \CArc(48.5,0)(30.4,210,330)
        \PhotonArc(48.5,-30)(30.4,30,80){1.5}{8}
	\PhotonArc(48.5,-30)(30.4,100,150){1.5}{8}
\Vertex(30,-25){2}\Vertex(67,-5){2}
\DashCArc(9.5,-5)(8,257,77){1}
\DashCArc(0,0)(8,103,180){1}\DashCArc(0,0)(8,77,253){1}
\DashLine(-1,-9)(8.5,-14){1}
\DashLine(1,9)(10.5,4){1}
\Text(47,20)[br]{$\bar {\cal A}_{\bar2}$}
        \Vertex(72.5,-15){4}
	\Line(72.5,-15)(97.5,-15)\Vertex(97.5,-15){1}\Text(97.5,-15)[l]{${\cal
A}_{\bar 3}$}\Vertex(82,-15){2}
        \DashLine(3,12)(37,-8){1}
\end{picture}
}
\hbox{
\begin{picture}(140,35)
        \SetOffset(70,20)
	\Text(0,0)[]{$A^{(2)}$}
\end{picture}
}}
}
\begin{eqnarray}
A^{(1)}&=&ig^6\frac{\bar V^2(1,2,3)\bar V^2(\bar3,l+p_3,\bar
l)V^2(-l-p_3,3,l)}{p_1^2
p_2^2p_3^2 \hat 3\left(\frac {p_1^2}{\hat 1}+\frac {p_2^2}{\hat 2}+\frac
{p_3^2}{\hat 3}\right)(\hat l+\hat 3)\left(\frac{(l+p_3)^2}{\hat l+\hat
3}-\frac{ l^2}{\hat l}+\frac{p_1^2}{\hat 1}+\frac{p_2^2}{\hat
2}\right)  l^2}\,,
\label{eq:tadpole-MHV1}\\
A^{(2)}&=&ig^6\frac{\bar V^2(1,2,3)\bar V^2(\bar3,l+p_3,\bar
l)V^2(-l-p_3,3,l)}{p_1^2
p_2^2p_3^2 \hat 3\left(\frac {p_1^2}{\hat 1}+\frac {p_2^2}{\hat 2}+\frac
{p_3^2}{\hat 3}\right)(-\hat l)\left(\frac{(l+p_3)^2}{\hat l+\hat
3}-\frac{ l^2}{\hat l}+\frac{p_1^2}{\hat 1}+\frac{p_2^2}{\hat
2}\right)  (l+p_3)^2}\,.
\label{eq:tadpole-MHV2}
\end{eqnarray}
The contribution from translation kernels with dressed propagators can
be represented as four diagrams:
\\
\centerline{
\vbox{\hbox{
\begin{picture}(140,60)
        \SetOffset(30,30)
        \Line(-25,0)(-2,0)\Vertex(-25,0){1}
	\Text(-25,-3)[lt]{$\bar {\cal A}_{\bar 1}$}
        \Photon(0,0)(34,20){1.5}{8}
	\Vertex(34,20){1}
        \Photon(0,0)(24.5,-15){1.5}{6}\Vertex(24.5,-15){4}
        \Vertex(0,0){4}\Vertex(8,-5){2}
        \CArc(48.5,-30)(30.4,30,150)
        \PhotonArc(48.5,0)(30.4,210,260){1.5}{8}
	\PhotonArc(48.5,0)(30.4,280,330){1.5}{8}
\Vertex(30,-25){2}\Vertex(67,-5){2}
        \DashCArc(0,0)(8,0,360){1}
\DashCArc(24.5,-15)(8,0,360){1}
\Text(47,20)[br]{$\bar {\cal A}_{\bar2}$}
        \Vertex(72.5,-15){4}
	\Photon(72.5,-15)(97.5,-15){1.5}{6}\Vertex(97.5,-15){1}\Text(97.5,-15)[l]{${\cal
A}_{\bar 3}$}\Vertex(82,-15){2}
\end{picture}
}
\hbox{\begin{picture}(140,35)
        \SetOffset(70,20)
	\Text(0,0)[]{$A^{(3)}$}
\end{picture}
}}
\vbox{\hbox{
\begin{picture}(140,60)
        \SetOffset(30,30)
        \DashCArc(0,0)(8,0,360){1}
	\DashCArc(24.5,-15)(8,0,360){1}
        \Line(-25,0)(-2,0)\Vertex(-25,0){1}
	\Text(-25,-3)[lt]{$\bar {\cal A}_{\bar 1}$}
        \Photon(0,0)(34,20){1.5}{8}\Vertex(34,20){1}
        \Photon(0,0)(24.5,-15){1.5}{6}\Vertex(24.5,-15){4}
        \Vertex(0,0){4}\Vertex(8,-5){2}
        \CArc(48.5,0)(30.4,210,330)
        \PhotonArc(48.5,-30)(30.4,30,80){1.5}{8}
	\PhotonArc(48.5,-30)(30.4,100,150){1.5}{8}
\Vertex(30,-25){2}\Vertex(67,-5){2}
\Text(47,20)[br]{$\bar {\cal A}_{\bar2}$}
        \Vertex(72.5,-15){4}
	\Photon(72.5,-15)(97.5,-15){1.5}{6}\Vertex(97.5,-15){1}\Text(97.5,-15)[l]{${\cal
A}_{\bar 3}$}\Vertex(82,-15){2}
\end{picture}
}
\hbox{
\begin{picture}(140,35)
        \SetOffset(70,20)
	\Text(0,0)[]{$A^{(4)}$}
\end{picture}
}}
}\\
\centerline{\vbox{\hbox{
\begin{picture}(140,60)
        \SetOffset(30,30)
        \Photon(-25,0)(-2,0){1.5}{6}\Vertex(-25,0){1}
	\Text(-25,-3)[lt]{$\bar {\cal A}_{\bar 1}$}
        \Line(0,0)(34,20)
	\Vertex(34,20){1}
        \Photon(0,0)(24.5,-15){1.5}{6}\Vertex(24.5,-15){4}
        \Vertex(0,0){4}\Vertex(8,-5){2}
        \CArc(48.5,-30)(30.4,30,150)
        \PhotonArc(48.5,0)(30.4,210,260){1.5}{8}
	\PhotonArc(48.5,0)(30.4,280,330){1.5}{8}
\Vertex(30,-25){2}\Vertex(67,-5){2}
        \DashCArc(0,0)(8,0,360){1}
\DashCArc(24.5,-15)(8,0,360){1}
\Text(47,20)[br]{$\bar {\cal A}_{\bar2}$}
        \Vertex(72.5,-15){4}
	\Photon(72.5,-15)(97.5,-15){1.5}{6}\Vertex(97.5,-15){1}\Text(97.5,-15)[l]{${\cal
A}_{\bar 3}$}\Vertex(82,-15){2}
\end{picture}
}
\hbox{
\begin{picture}(140,35)
        \SetOffset(70,20)
	\Text(0,0)[]{$A^{(5)}$}
\end{picture}
}}
\vbox{\hbox{
\begin{picture}(140,60)
        \SetOffset(30,30)
        \DashCArc(0,0)(8,0,360){1}
	\DashCArc(24.5,-15)(8,0,360){1}
        \Photon(-25,0)(-2,0){1.5}{6}\Vertex(-25,0){1}
	\Text(-25,-3)[lt]{$\bar {\cal A}_{\bar 1}$}
        \Line(0,0)(34,20)
	\Vertex(34,20){1}
        \Photon(0,0)(24.5,-15){1.5}{6}\Vertex(24.5,-15){4}
        \Vertex(0,0){4}\Vertex(8,-5){2}
        \CArc(48.5,0)(30.4,210,330)
        \PhotonArc(48.5,-30)(30.4,30,80){1.5}{8}
	\PhotonArc(48.5,-30)(30.4,100,150){1.5}{8}
\Vertex(30,-25){2}\Vertex(67,-5){2}
\Text(47,20)[br]{$\bar {\cal A}_{\bar2}$}
        \Vertex(72.5,-15){4}
	\Photon(72.5,-15)(97.5,-15){1.5}{6}\Vertex(97.5,-15){1}\Text(97.5,-15)[l]{${\cal
A}_{\bar 3}$}\Vertex(82,-15){2}
\end{picture}
}
\hbox{\begin{picture}(140,35)
        \SetOffset(70,20)
	\Text(0,0)[]{$A^{(6)}$}
\end{picture}
}}
}
\begin{eqnarray}
\label{eq:tadpole-completion1}
A^{(3)}&=&ig^6\frac{\bar V^2(1,2,3)\bar V^2(\bar3,l+p_3,\bar
l)V^2(-l-p_3,3,l)}{
p_2^2(p_3^2)^2 \hat 1\left(\frac {p_1^2}{\hat 1}+\frac {p_2^2}{\hat 2}+\frac
{p_3^2}{\hat 3}\right)(\hat l+\hat 3)\left(\frac{(l+p_3)^2}{\hat l+\hat
3}-\frac{ l^2}{\hat l}-\frac{p_3^2}{\hat 3}\right)  l^2}
\,,\\
A^{(4)}&=&ig^6\frac{\bar V^2(1,2,3)\bar V^2(\bar3,l+p_3,\bar
l)V^2(-l-p_3,3,l)}{
p_2^2(p_3^2)^2 \hat 1\left(\frac {p_1^2}{\hat 1}+\frac {p_2^2}{\hat 2}+\frac
{p_3^2}{\hat 3}\right)(-\hat l)\left(\frac{(l+p_3)^2}{\hat l+\hat
3}-\frac{ l^2}{\hat l}-\frac{p_3^2}{\hat 3}\right)  (l+p_3)^2}
\,,\\
A^{(5)}&=&ig^6\frac{\bar V^2(1,2,3)\bar V^2(\bar3,l+p_3,\bar
l)V^2(-l-p_3,3,l)}{
p_1^2(p_3^2)^2 \hat 2\left(\frac {p_1^2}{\hat 1}+\frac {p_2^2}{\hat 2}+\frac
{p_3^2}{\hat 3}\right)(\hat l+\hat 3)\left(\frac{(l+p_3)^2}{\hat l+\hat
3}-\frac{ l^2}{\hat l}-\frac{p_3^2}{\hat 3}\right)  l^2}
\,,\\
A^{(6)}&=&ig^6\frac{\bar V^2(1,2,3)\bar V^2(\bar3,l+p_3,\bar
l)V^2(-l-p_3,3,l)}{
p_1^2(p_3^2)^2 \hat 2\left(\frac {p_1^2}{\hat 1}+\frac {p_2^2}{\hat 2}+\frac
{p_3^2}{\hat 3}\right)(-\hat l)\left(\frac{(l+p_3)^2}{\hat l+\hat
3}-\frac{ l^2}{\hat l}-\frac{p_3^2}{\hat 3}\right)
(l+p_3)^2}\label{eq:tadpole-completion4}
\,.
\end{eqnarray}
After summing over $A^{(1)}$ to $A^{(6)}$ one finds that the factor
${p_1^2}/{\hat 1}+{p_2^2}/{\hat 2}+ {p_2^2}/{\hat 3}$ in the
denominator is cancelled and we can apply the LSZ procedure:
\begin{align}
\lim_{p_1^2,p_2^2,p_3^2\to
0}&(p_1^2p_2^2p_3^2)\int {\rm
d}^4l\sum_{i=1}^6A^{(i)}
\nonumber \\&
\begin{aligned}
=\lim_{p_1^2,p_2^2,p_3^2\to 0}ig^6\frac1{p_3^2}\int {\rm
d}^4l&\frac{\bar V^2(1,2,3)\bar V^2(\bar3,l+p_3,\bar
l)V^2(-l-p_3,3,l)}{l^2  (l+p_3)^2}
\nonumber \\
&\times\frac{\left(\frac{(l+p_3)^2}{\hat l+\hat
3}-\frac{ l^2}{\hat l}+\frac {p_1^2}{\hat 1}+\frac {p_2^2}{\hat 2}-\frac{p_3^2}{\hat 3}\right)\left(\frac{(l+p_3)^2}{\hat l+\hat
3}-\frac{ l^2}{\hat l}\right)  }{\left(\frac{(l+p_3)^2}{\hat l+\hat
3}-\frac{ l^2}{\hat l}-\frac{p_3^2}{\hat 3}\right)\left(\frac{(l+p_3)^2}{\hat l+\hat
3}-\frac{ l^2}{\hat l}+\frac{p_1^2}{\hat 1}+\frac{p_2^2}{\hat
2}\right) }
\end{aligned}
\nonumber \\
&=\lim_{p_1^2,p_2^2,p_3^2\to 0}ig^6\frac{f(p_1^2,p_2^2,p_3^2)}{p_3^2}
\nonumber\\
&=\lim_{p_1^2,p_2^2,p_3^2\to 0}ig^6\frac{\partial
f(p_1^2,p_2^2,p_3^2)}{\partial p_3^2}
\,.
\end{align}
Since the integration is uniformly convergent after regularization, we
can take the limit before integration and differentiation which will
give the same on-shell integral as the light-cone calculation. So we
have reproduced the light-cone computation. For the other diagrams
with dressed propagators a similar situation happens and it can be
checked that they give the same amplitudes as light-cone calculations.

From this example, we see that we should first collect the diagrams
with the same internal helicity configurations and with tadpoles on
the same legs  and then impose the limit $p_1^2,p_2^2,p_3^2\to 0$ in
the LSZ procedure. Just like at the tree-level, we can also change the
order of limits. Because we should take the $p_3^2$ at the last step
after integration, we choose the limit $p_1^2, p_2^2\to 0$ first.
Then we find that after we multiply $p_1^2 p_2^2p_3^2$ and take
$p_1^2, p_2^2\to 0$, the translation kernel contributions from
(\ref{eq:tadpole-completion1})--(\ref{eq:tadpole-completion4}) vanish
and the whole contribution to the amplitude comes from the tadpole of
MHV vertices (\ref{eq:tadpole-MHV1})--(\ref{eq:tadpole-MHV2}). The
${p_1^2}/{\hat 1}+{p_2^2}/{\hat 2}+{p_3^2}/{\hat 3}$ simply
contributes to the propagator $i/p_3^2$ needed in the amplitude. So
the result is that we do not need to consider the translation kernel
contribution in this case. 

 In section \ref{sect:dressing-prop}, we have argued that since the
sum of one-loop diagrams in which the external legs are dressed are
proportional to tree-level amplitudes, their contributions to higher
point one-minus-helicity amplitudes vanish. But it is also instructive
to apply the above arguments to these higher point amplitudes. In
fact, a similar situation occurs. For these amplitudes there are also
$1/(\sum_i p_i^2/\hat \imath)$ factors both from the tadpoles of MHV
vertices and the translation kernels, where $i$ enumerates all the
external momenta. If we collect the diagrams with the same internal
helicity configuration and with tadpoles on the same legs first,
(including tadpoles of MHV vertices and translation kernels,) then the
$\sum_i p_i^2/\hat \imath$ in the denominator is cancelled and we can
take the on-shell limits in any order. If we first set all the
external legs on-shell except that with the tadpole then the
translation kernel contributions vanish leaving just the tadpoles of
MHV vertices.  So we come to the conclusion that we do not need to
consider external propagators dressed by tadpoles from translation
kernel.

\section{Conclusion and higher loops}
\label{sect:conclusion}
We have seen that the S-matrix equivalence theorem is not immediately
applicable to the change of variables from $\gA$ and $\bar\gA$ to
$\gB$ and $\bar\gB$ because of the non-locality of the translation
kernels, and this accounts for the one-loop all plus helicity
amplitudes apparently missing from the CSW rules.  However, by
analysing the mechanisms that generate singularities in the external
momenta that are able to cancel the LSZ factors we have seen that the
types of amplitude in which S-matrix equivalence is violated are very
restricted. At tree-level the amplitudes that might have displayed
this violation actually vanish.  At one-loop the equivalence violating
amplitudes that do not vanish are ones in which all the gluons have
positive helicity, and these have a known form, e.g. (\ref{4pt}).
Because the only non-zero one-loop amplitudes that show S-matrix
equivalence violation are given by the tadpole diagrams in which the
single $\bar\gB$ field of an $\bar\gA$ translation kernel is
contracted with a $\gB$ field it follows that higher loops can only
contribute to violating processes by dressing the legs of these
one-loop diagrams.  So, apart from this class of known amplitudes we
are free to calculate S-matrix elements using the $\gB$ and $\bar\gB$
fields directly. 

Since one-minus-helicity diagrams can not be constructed from more
than one MHV vertex or from completion vertices, they can only arise
as tadpoles of MHV vertices. By analysing an example we saw how the
light-cone amplitudes really can be reconstructed from tadpoles of MHV
vertices.

We found a new recursion relation for the expansion coefficients
$\Xi^{s}$ of $\bar {\cal A}$, which encoded a cancellation of certain
singularities that would otherwise have contributed to further evasion
of the S-matrix equivalence theorem.  Using this recursion relation
for $\Xi$ together with the one for $\Upsilon$ we were led to a better
understanding of the canonical transformation: they can be
reconstructed from the light-cone tree level diagrams built with only
$(++-)$ vertices by replacing the propagator using (\ref{eq:replace}).
This was useful in discussing the relationship between light-cone and
MHV methods.

A few remarks about the rational parts of one-loop diagrams is in
order.  The CSW or MHV rules, although initially conjectured and
proven at tree-level have been studied at one-loop level. It has been
shown that they give supersymmetric amplitudes correctly
\cite{Brandhuber:2004yw,Quigley:2004pw,Bedford:2004py,Brandhuber:2005kd},
but when applied to non-supersymmetric amplitudes, the rational
parts can not be correctly reproduced \cite{Bedford:2004nh}, not only
in all-plus diagrams.  Our discussion in the present paper has focussed on
the (limited) breakdown of the equivalence theorem that is
responsible, in our approach, for the rational one-loop
all plus amplitude in non-supersymmetric Yang-Mills. Our conclusion is
that only these amplitudes require the use of the translation kernels,
and so all other one-loop amplitudes can be calculated directly from
the Green functions of the $B$ fields. One may then ask where the
missing rational parts of the other diagrams might come from. Here, we
should point out that we have formulated the transformation from
light-cone Yang-Mills to the new MHV Lagrangian in $D$ dimensions.
Consequently our canonical transformation coefficients $\Upsilon$ and
$\Xi$ are formulated in $D$ dimensions ($D=4-2\epsilon$) and the MHV
vertices derived from these coefficients are also in $D$ dimensions,
whether one uses standard dimensional regularisation as in \cite{us}
or FDH.  This is different from the usual analyses of MHV one-loop
calculations in \cite{Brandhuber:2004yw,
Quigley:2004pw,Bedford:2004py,Brandhuber:2005kd,Bedford:2004nh} which
use four dimensional MHV vertices. In the FDH procedure the $\epsilon$
dependence enters the transformation coefficients only through $\sum
\Omega=\sum p^2/\hat p$ where $p^2$ is the D-dimensional momentum,
rather than the four dimensional momentum, in recursion relations
(\ref{eq:-UpsilonR}) and (\ref{eq:NewXiR-}), but this is enough to
make the vertices of our Lagrangian different from the ordinary four
dimensional Parke-Taylor vertices.  In ordinary dimensional
regularisation the vertices would, in addition, acquire indices
relating to the extra dimensions.  In either formalism the vertices
differ from the four dimensional ones.  One would expect that, in
general, these modifications would produce terms proportional to
$\epsilon$ which would cancel the divergence $1/\epsilon$ from the
loop integration resulting in rational pieces missing in the ordinary
MHV calculation. 

Our arguments can easily be extended to super Yang-Mills theory using the
supersymmetry transformation in \cite{Morris:2008uc}.  We expect the
supersymmety transformation is not affected in $D$ dimensions, and
the results in \cite{Morris:2008uc} can be directly used here after
setting the chiral fields to zero. The $\cal{A}$ transformation is not
changed. From equation (B.7) and (C.14) in \cite{Morris:2008uc},
$\bar{\cal A}$ has an additional term which involves gluino $\Pi$:
\begin{eqnarray}
\bar{\cal A}^{B\Pi}_q&=&-\frac 1 {\sqrt2 \hat q}\sum_{n=2}^\infty\int_{1\cdots
n}\sum_{s=1}^{n}\bigg[\Xi^{s}_{q,\bar1\cdots\bar
n}\sum_{l=1,l\neq s}^{n}(-1)^{\delta_{ls}}{\cal
B}_{1}\cdots\Pi_l\cdots\bar \Pi_s\cdots
{\cal B}_{n}\bigg]\delta_{q\bar1\cdots\bar n}\,,
\label{eq:canon-trans-barA-3} 
\end{eqnarray} 
$\Xi^{s}$ is just the coefficient appearing in the pure bosonic
expansion.  The fermion propagator and bosonic propagators are 
\begin{eqnarray}
\langle \Pi\bar\Pi\rangle=\frac { i\,g^2 \hat
p}{\sqrt 2\,p^2}\,, \quad
\langle {\cal A} \bar {\cal A}\rangle=
\frac { i\,g^2 }{2p^2 }\,. 
\end{eqnarray}
Considering the $\sqrt 2\hat p$ factor, when connected to a gluino
propagator the coefficient is the same as the one in the pure bosonic
expansion up to a sign. So all the foregoing discussion can be applied
to diagrams with inner gluinos.  Therefore one would expect that only
the tadpole would evade the equivalence theorem. One can easily check
that the all-plus translation kernel contribution to the  amplitude is
cancelled using above expansion (\ref{eq:canon-trans-barA-3}). We also
expect that our MHV calculation should reproduce the light-cone super
Yang-Mills calculation, so the one-minus-helicity amplitude in
supersymmetric Yang-Mills should also be zero.  As is well known
\cite{Bern:1994cg}, in supersymmetric theories the rational parts of
amplitudes are determined uniquely by their (four dimensional)
cut-constructible parts. It follows that all the remaining rational
parts discussed in the previous paragraph should be cancelled in the
supersymmetric theory.

\section*{Acknowledgements}
Tim, Xiao, and Paul thank STFC for support under the rolling grants
ST/G000557/1 and ST/G000433/1, and Jonathan thanks the Richard Newitt
bursary scheme, for financial support. It is a pleasure to acknowledge
useful conversations with James Ettle.

\appendix
\section{Some remaining thoughts on translation kernels}\label{a1}

The careful reader may have noticed that the translation kernel can
become ill-defined due to the symmetry of graphs. For instance in the
tadpole graph below arising from self-contraction of the
$\Xi^{2}\bar{\mathcal{B}}\mathcal{B}\mathcal{B}$ term the gluons
flowing in and out of the kernel must carry equal and opposite momenta
as required by conservation of momentum. As a result the factors
$(p_{j}^{2}+i\epsilon)/\hat{p}_{j}$ which appear in the denominator of
the kernel cancel in pairs. The same cancellation can also occur for
special values of momentum. Note that in this case the standard
$i\epsilon$ prescription fails to prevent
$\sum(p_{j}^{2}+i\epsilon)/\hat{p}_{j}$ from vanishing.

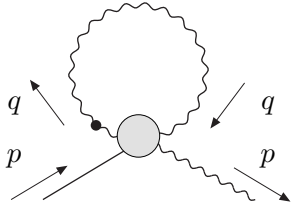
\begin{figure}[!h]
\centering
\subfigure{  \begin{picture}(126,75) (12,-9)
    \PhotonArc(64,39)(25.3,-18,342){1}{22}
    \Photon(64,15)(108,-9){1}{8}
    \Line(28,-9)(68,15)
    \SetWidth{0.5}
    \Vertex(48,19){2}
    \SetWidth{0.375}
    \LongArrow(100,3)(120,-9)
    \LongArrow(16,-9)(36,3)
    \GOval(64,15)(8,8)(0){0.882}
    \LongArrow(36,19)(24,35)
    \LongArrow(104,35)(92,19)
    \Text(20,23)[br]{$q$}
    \Text(116,23)[br]{$q$}
    \Text(20,3)[br]{$p$}
    \Text(116,3)[br]{$p$}
  \end{picture}
}
\caption{Translation kernels diverge in symmetrical graphs}
\label{symmetricloop}
\end{figure}


This problem can be fixed by adding a small correction to the
definition of translation kernels. To break symmetry we distinguish
the $i\epsilon$ associated with $\mathcal{A}$ fields and $\mathcal{B}$
fields. $\Upsilon(123)$ is now modified as 

\be
\Upsilon (123) =\frac{i\left(\frac{\bar{p}_{2}}{\hat{p}_{2}}-\frac{\bar{p}_{3}}{\hat{p}_{3}}\right)}{\frac{p_{1}^{2}+i\epsilon_{A}}{\hat{p_{1}}}+\frac{p_{2}^{2}+i\epsilon_{B}}{\hat{p_{2}}}+\frac{p_{3}^{2}+i\epsilon_{B}}{\hat{p_{3}}}}\ee

Higher order terms in the $\mathcal{A}$ field expansion can all be
redefined following the same spirit and the coefficients for the
$\bar{\mathcal{A}}$ expansion are in turn determined from the
canonical transformation condition (\ref{eq:barA-exp}). However a
small price is to be paid for getting around the divergences. By
substituting the modified kernels back into (\ref{eq:transform-4d})
which used to define $\Upsilon$ we find two sides of the equation
slightly mismatch. The differences generate new vertices carrying
infinitesimal corrections.

\bq
&& \frac{p^{2}+i\epsilon_{A}}{\hat{p}}A(p)+i\int d^{D-1}q \left[ \frac{\bar{q}}{\hat{q}}A(q),A(p-q) \right] \\
&& = \int d^{D-1}q  \frac{q^{2}+i\epsilon_{A}}{\hat{q}}B(q)\frac{\delta A(p)}{\delta B(q)} \nonumber \\ 
&& \hspace{1cm} +  \sum\limits_{n=2}^{\infty} \int \left( \prod \limits_{i=2}^{n}d^{D-1}q_{(i)} \right) \left(\sum\limits_{j=2}^{n}\frac{i\left(\epsilon_{A}-\epsilon_{B}\right)}{\hat{q_{j}}}\right)\Upsilon_{12\dots n}B(q_{2})\dots B(q_{n})  \\
&& =\int d^{D-1}q\frac{q^{2}+i\epsilon_{A}}{\hat{q}}B(q)\frac{\delta A(p)}{\delta B(q)}+\frac{i\left(\epsilon_{A}-\epsilon_{B}\right)}{\hat{q}}B(q)\frac{\delta A(p)}{\delta B(q)}-\frac{i\left(\epsilon_{A}-\epsilon_{B}\right)}{\hat{p}}B(p) \label{eq5} \eq

Equivalently this can be written as

\be
\mathcal{L}^{-+}[\gA ,\bar{\gA}] + \mathcal{L}^{-++}[\gA ,\bar{\gA}]
= \mathcal{L}^{-+}[\gB ,\bar{\gB}] + \mathcal{L}_{\epsilon}[\gB, \bar{\gB}] \label{eq6} \ee

where $\mathcal{L}_{\epsilon}$ represents the new vertex terms.

\bq
\mathcal{L}_{\epsilon}[\gB ,\bar{\gB}] 
&& = - \bar{\gA}i\left(\epsilon_{A}-\epsilon_{B}\right)\gB + \bar{\gB}\left(\epsilon_{A}-\epsilon_{B}\right)\gB  \nonumber \\
&& =\left(\sum\limits_{m=2}^{\infty} \sum\limits_{s=2}^{m} \int_{2\dots m} \frac{\hat{s}}{\hat{p}}\Xi^{s-1} \gB \dots \bar{\gB} \dots \gB \right)i\left(\epsilon_{A}-\epsilon_{B}\right) \gB \label{eq7} \eq

Introducing double circles to denote the factor $ \frac{\hat{s}}{\hat{p}}i\left(\epsilon_{A}-\epsilon_{B}\right) $, these terms are expressed graphically as
%
%
%
\begin{figure}[!h]
  \centering 
\subfigure{
  \begin{picture}(100,66) (10,20)
    \SetWidth{0.375}
    \Line(27,43)(58,33)
    \Line(25,25)(58,32)
    \SetWidth{0.5}
    \Vertex(45,37){2}
    \Text(11,20)[br]{$\mathcal{B}$}
    \Text(10,38)[br]{$\bar{\mathcal{B}}$}
    \Text(123,29)[br]{$\mathcal{B}$}
    \SetWidth{0.375}
    \Line(57,33)(85,33)
    \Line(86,33)(114,33)
    \GCirc(58,33){8}{0.75}
    \BCirc(86,33){4}
    \BCirc(86,33){2}
  \end{picture}
}
\subfigure{
\begin{picture}(10,10)
\end{picture}
}
\subfigure{
  \begin{picture}(100,66) (10,20)
    \SetWidth{0.375}
    \Line(27,43)(58,33)
    \Line(25,25)(58,32)
    \SetWidth{0.5}
    \Vertex(45,29){2}
    \Text(11,20)[br]{$\bar{\mathcal{B}}$}
    \Text(10,38)[br]{$\mathcal{B}$}
    \Text(123,29)[br]{$\mathcal{B}$}
    \SetWidth{0.375}
    \Line(57,33)(85,33)
    \Line(86,33)(114,33)
    \GCirc(58,33){8}{0.75}
    \BCirc(86,33){4}
    \BCirc(86,33){2}
  \end{picture}
}
\\

\subfigure{
  \begin{picture}(100,73) (11,20)
    \SetWidth{0.5}
    \Text(123,36)[br]{$\mathcal{B}$}
    \Text(11,50)[br]{$\bar{\mathcal{B}}$}
    \SetWidth{0.375}
    \Line(27,54)(58,40)
    \Line(27,23)(58,40)
    \SetWidth{0.5}
    \Vertex(46,46){2}
    \SetWidth{0.375}
    \Line(58,40)(28,40)
    \Text(11,36)[br]{$\mathcal{B}$}
    \Text(12,20)[br]{$\mathcal{B}$}
    \Line(57,40)(85,40)
    \Line(86,40)(114,40)
    \GCirc(58,40){8}{0.75}
    \BCirc(86,40){4}
    \BCirc(86,40){2}
  \end{picture}
}
\subfigure{
\begin{picture}(10,10)
\end{picture}
}
\subfigure{
  \begin{picture}(100,73) (11,20)
    \SetWidth{0.5}
    \Text(123,36)[br]{$\mathcal{B}$}
    \Text(11,50)[br]{$\bar{\mathcal{B}}$}
    \SetWidth{0.375}
    \Line(27,54)(58,40)
    \Line(27,23)(58,40)
    \SetWidth{0.5}
    \Vertex(46,46){2}
    \SetWidth{0.375}
    \Text(11,33)[br]{$.$}
    \Text(11,36)[br]{$.$}
    \Text(11,39)[br]{$.$}
    \Text(12,20)[br]{$\mathcal{B}$}
    \Line(57,40)(85,40)
    \Line(86,40)(114,40)
    \GCirc(58,40){8}{0.75}
    \BCirc(86,40){4}
    \BCirc(86,40){2}
  \end{picture}
} 
 \label{newvertex}
 \caption{Infinitesimal vertex terms}
\end{figure}
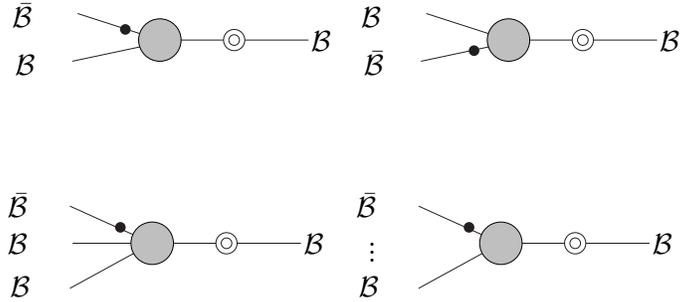

%
%
%

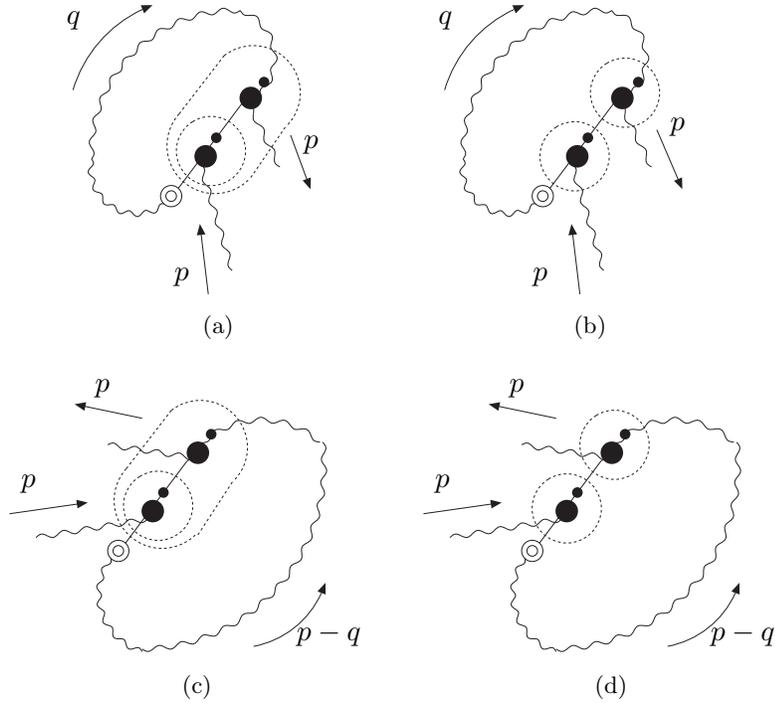
\begin{figure}[!h]
\centering
  \subfigure[]{
    \begin{picture}(126,116) (21,-35)
    \SetWidth{0.375}
    \Line(68,4)(92,36)
    \PhotonArc(86,54)(18.97,-71.57,71.57){-1}{5.5}
    \PhotonArc(90,18)(54.04,87.88,182.12){-1}{10.5}
    \PhotonArc(52.97,12.58)(17.31,168.61,330.28){-1}{7.5}
    \SetWidth{0.5}
    \Vertex(79,17){4.24}
    \Vertex(83,24){2}
    \Vertex(101,45){2}
    \Vertex(96,39){4.24}
    \SetWidth{0.375}
    \Photon(95,37)(107,13){1}{3}
    \Photon(77,14)(89,-26){1}{5}
    \LongArrow(80,-35)(77,-10)
    \LongArrow(111,25)(118,6)
    \Text(122,18)[br]{$p$}
    \Text(73,-32)[br]{$p$}
    \LongArrowArcn(76,28)(49.04,163.41,110.28)
    \Text(32,65)[br]{$q$}
    \DashCArc(96.92,41)(17.75,-42.53,127.95){1}
    \DashCArc(82.48,21)(18.05,138.33,313.92){1}
    \DashLine(96,9)(109,28){1}
    \DashLine(69,33)(86,55){1}
    \DashCArc(81,19)(13.04,122,482){1}
    \BCirc(66,2){4}
    \BCirc(66,2){2}
  \end{picture}
  \label{sheep1}
}     
\subfigure[]{
  \begin{picture}(126,116) (21,-35)
    \SetWidth{0.375}
    \Line(68,4)(92,36)
    \PhotonArc(86,54)(18.97,-71.57,71.57){-1}{5.5}
    \PhotonArc(90,18)(54.04,87.88,182.12){-1}{10.5}
    \PhotonArc(52.97,12.58)(17.31,168.61,330.28){-1}{7.5}
    \DashCArc(78,17)(13.04,122,482){1}
    \SetWidth{0.5}
    \Vertex(79,17){4.24}
    \Vertex(83,24){2}
    \SetWidth{0.375}
    \DashCArc(97,41)(13.04,122,482){1}
    \SetWidth{0.5}
    \Vertex(102,45){2}
    \Vertex(96,39){4.24}
    \SetWidth{0.375}
    \Photon(95,37)(107,13){1}{3}
    \BCirc(66,2){4}
    \BCirc(66,2){2}
    \Photon(77,14)(89,-26){1}{5}
    \LongArrow(80,-35)(77,-10)
    \LongArrow(109,27)(118,6)
    \Text(120,25)[br]{$p$}
    \Text(68,-32)[br]{$p$}
    \LongArrowArcn(76,28)(49.04,163.41,110.28)
    \Text(32,65)[br]{$q$}
  \end{picture}
    \label{sheep2}
   }   
   \subfigure[]{
  \begin{picture}(142,113) (25,-40)
    \SetWidth{0.375}
    \Line(68,1)(92,33)
    \SetWidth{0.5}
    \Vertex(79,14){4.24}
    \Vertex(83,21){2}
    \Vertex(101,43){2}
    \Vertex(96,36){4.24}
    \SetWidth{0.375}
    \DashCArc(96.92,38)(17.75,-42.53,127.95){1}
    \DashCArc(82.48,18)(18.05,138.33,313.92){1}
    \DashLine(96,5)(109,25){1}
    \DashLine(69,30)(86,52){1}
    \DashCArc(81,16)(13.04,122,482){1}
    \PhotonArc(81.31,-17.28)(22.61,126.04,253.81){-1}{7.5}
    \PhotonArc(119.97,15.26)(33.13,48.32,147.61){-1}{6.5}
    \Photon(95,34)(62,39){1}{4}
    \Photon(77,11)(35,4){1}{5}
    \LongArrow(75,49)(51,54)
    \LongArrow(25,13)(53,17)
    \Text(35,21)[br]{$p$}
    \Text(63,57)[br]{$p$}
    \LongArrowArc(110.5,-2.5)(35.11,-79.33,-17.4)
    \Text(158,-37)[br]{$p-q$}
    \PhotonArc(74.32,31.23)(69.24,-89.44,7.27){1}{13.5}
    \BCirc(66,-1){4}
    \BCirc(66,-1){2}
  \end{picture}
    \label{sheep3}
  }   
  \subfigure[]{
    \begin{picture}(142,113) (25,-40)
    \SetWidth{0.375}
    \Line(68,1)(92,33)
    \SetWidth{0.5}
    \Vertex(79,14){4.24}
    \Vertex(83,21){2}
    \Vertex(101,43){2}
    \Vertex(96,36){4.24}
    \SetWidth{0.375}
    \PhotonArc(81.31,-17.28)(22.61,126.04,253.81){-1}{7.5}
    \PhotonArc(119.97,15.26)(33.13,48.32,147.61){-1}{6.5}
    \Photon(95,34)(62,39){1}{4}
    \Photon(77,11)(35,4){1}{5}
    \LongArrow(75,49)(51,54)
    \LongArrow(25,13)(53,17)
    \Text(35,21)[br]{$p$}
    \Text(63,57)[br]{$p$}
    \LongArrowArc(110.5,-2.5)(35.11,-79.33,-17.4)
    \Text(158,-37)[br]{$p-q$}
    \PhotonArc(74.32,31.23)(69.24,-89.44,7.27){1}{13.5}
    \BCirc(66,-1){4}
    \BCirc(66,-1){2}
    \DashCArc(79,15)(13.04,122,482){1}
    \DashCArc(97,39)(13.04,122,482){1}
  \end{picture}
    \label{sheep4}
  }  \caption{Contributions to the $\left\langle \bar{\mathcal{A}}\bar{\mathcal{A}}\right\rangle$  symmetric loop graph }
\end{figure}

\begin{figure}[!h]
\centering
  \subfigure{
    \begin{picture}(135,43) (3,-15)
    \SetWidth{0.375}
    \CArc(60,4)(17.03,130,490)
    \Line(14,4)(42,4)
    \Line(78,4)(106,4)
    \Text(55,9)[br]{$+$}
    \Text(55,-5)[br]{$-$}
    \Text(72,9)[br]{$-$}
    \Text(72,-4)[br]{$+$}
    \Text(105,7)[br]{$-$}
    \Text(24,7)[br]{$-$}
    \Text(89,7)[br]{$+$}
    \Text(40,7)[br]{$+$}
    \Text(12,3)[br]{$\bar{\mathcal{A}}$}
    \Text(117,3)[br]{$\bar{\mathcal{A}}$}
  \end{picture}
\label{abarabar}  } 
\caption{ $\langle \bar{\mathcal{A}}\bar{\mathcal{A}} \rangle $
self-energy graph in the LCYM theory}
   \end{figure}
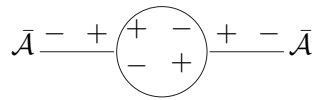

In most cases these corrections do not really enter into our
calculations because of the infinitesimal nature of the vertices,
except for extremely divergent graphs such as
(Fig.\ref{symmetricloop}). Because of the asymmetry treatment the
factor $\sum(p_{j}^{2}+i\epsilon)/\hat{p}_{j}$ in the denominator of
the kernel do not cancel completely. A factor of
$i(\epsilon_{A}-\epsilon_{B})/\hat{p}$ in the translation kernel is
left to cancel the infinitesimal factor brought by the new vertex,
resulting a finite contribution to the loop integral. It is
straightforward to show the following four graphs (Fig.\ref{sheep1} to
Fig.\ref{sheep4}) constructed from the new vertex restore the $\langle
\bar{\mathcal{A}}\bar{\mathcal{A}} \rangle $ self-energy bubble
integral in the LCYM theory (Fig.\ref{abarabar}).


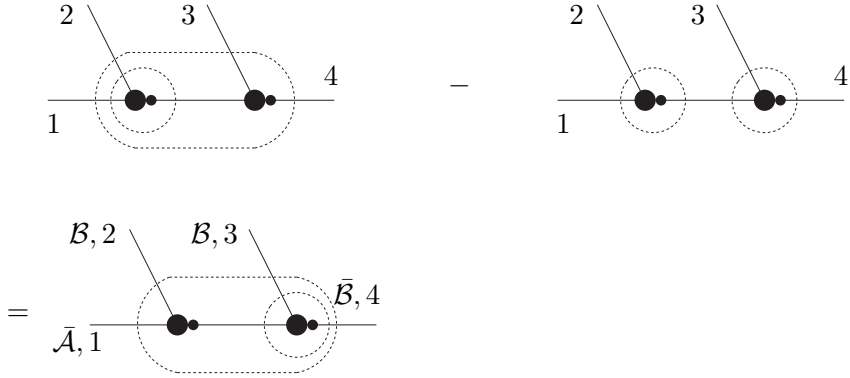
\begin{figure}[!h]
  \subfigure{  \begin{picture}(156,79) (6,7)
    \SetWidth{0.5}
    \Vertex(48,25){4}
    \Vertex(54,25){2}
    \SetWidth{0.375}
    \DashCArc(51,25)(12.21,145,505){1}
    \SetWidth{0.5}
    \Vertex(93,25){4}
    \SetWidth{0.375}
    \Line(15,25)(123,25)
    \Line(75,61)(93,25)
    \Line(30,61)(48,25)
    \SetWidth{0.5}
    \Vertex(99,25){2}
    \SetWidth{0.375}
    \DashCArc(51.83,25.44)(18.84,111.27,258.26){1}
    \DashCArc(89.7,25)(18.3,-79.61,79.61){1}
    \DashLine(48,7)(93,7){1}
    \DashLine(45,43)(93,43){1}
   \Text(20,13)[br]{$1$}
    \Text(25,54)[br]{$2$}
    \Text(71,54)[br]{$3$}
    \Text(125,31)[br]{$4$}
  \end{picture}
  }
    \subfigure{
  \begin{picture}(15,70)
  \Text(2,20)[br]{$-$}
  \end{picture}
  } 
  \subfigure{
    \begin{picture}(156,79) (6,7)
    \SetWidth{0.5}
    \Vertex(48,25){4}
    \Vertex(54,25){2}
    \SetWidth{0.375}
    \DashCArc(51,25)(12.21,145,505){1}
    \SetWidth{0.5}
    \Vertex(93,25){4}
    \SetWidth{0.375}
    \Line(15,25)(123,25)
    \Line(75,61)(93,25)
    \Line(30,61)(48,25)
    \SetWidth{0.5}
    \Vertex(99,25){2}
    \SetWidth{0.375}
    \DashCArc(93,25)(12.21,145,505){1}
    \Text(20,13)[br]{$1$}
    \Text(25,54)[br]{$2$}
    \Text(71,54)[br]{$3$}
    \Text(125,31)[br]{$4$}
  \end{picture}
  }
  
  \subfigure{
  \begin{picture}(5,70)
  \Text(2,20)[br]{$=$}
  \end{picture}
  }  
  \subfigure{
    \begin{picture}(156,79) (6,7)
    \SetWidth{0.5}
    \Vertex(48,25){4}
    \Vertex(54,25){2}
    \SetWidth{0.375}
    \SetWidth{0.5}
    \Vertex(93,25){4}
    \SetWidth{0.375}
    \Line(15,25)(123,25)
    \Line(75,61)(93,25)
    \Line(30,61)(48,25)
    \SetWidth{0.5}
    \Vertex(99,25){2}
    \SetWidth{0.375}
    \DashCArc(93,25)(12.21,145,505){1}
    \DashCArc(51.83,25.44)(18.84,111.27,258.26){1}
    \DashCArc(89.7,25)(18.3,-79.61,79.61){1}
    \DashLine(48,7)(93,7){1}
    \DashLine(45,43)(93,43){1}
    \Text(20,13)[br]{$\bar{\mathcal{A}}, 1$}
    \Text(25,54)[br]{$\mathcal{B}, 2$}
    \Text(71,54)[br]{$\mathcal{B}, 3$}
    \Text(125,31)[br]{$\bar{\mathcal{B}}, 4$}
  \end{picture}
  }
  \caption{Simplification of the $\bar{\mathcal{A}}$ expansion}
  \label{loop flipping}
  \end{figure}

Another issue regarding $i\epsilon$ prescription arises if we wish to
apply (\ref{eq-flipping}) to simplify $\bar{\mathcal{A}}$ expansions.
In the example illustrated below (Fig.\ref{loop flipping}) the first
two graphs are combined according to the identity
(\ref{eq-flipping2}).

\bq
&& \frac{1}{\frac{p_{3}^{2}+i\epsilon_{B}}{\hat{p}_{3}}+\frac{p_{4}^{2}+i\epsilon_{A}}{\hat{p}_{4}}+\frac{(p_{1}+p_{2})^{2}+i\epsilon_{B}}{\hat{p}_{1}+\hat{p}_{2}}}-\frac{1}{\frac{p_{1}^{2}+i\epsilon_{B}}{\hat{p}_{1}}+\frac{p_{2}^{2}+i\epsilon_{B}}{\hat{p}_{2}}+\frac{p_{3}^{2}+i\epsilon_{B}}{\hat{p}_{3}}+\frac{p_{4}^{2}+i\epsilon_{A}}{\hat{p}_{4}}} \nonumber \\
&& \hspace{1cm}
=\frac{1}{\frac{p_{3}^{2}+i\epsilon_{B}}{\hat{p}_{3}}+\frac{p_{4}^{2}+i\epsilon_{A}}{\hat{p}_{4}}+\frac{(p_{1}+p_{2})^{2}+i\epsilon_{B}}{\hat{p}_{1}+\hat{p}_{2}}}\,\frac{\frac{p_{1}^{2}+i\epsilon_{B}}{\hat{p}_{1}}+\frac{p_{2}^{2}+i\epsilon_{B}}{\hat{p}_{2}}+\frac{(p_{3}+p_{4})^{2}+i\epsilon_{B}}{\hat{p}_{3}+\hat{p}_{4}}}{\frac{p_{1}^{2}+i\epsilon_{B}}{\hat{p}_{1}}+\frac{p_{2}^{2}+i\epsilon_{B}}{\hat{p}_{2}}+\frac{p_{3}^{2}+i\epsilon_{B}}{\hat{p}_{3}}+\frac{p_{4}^{2}+i\epsilon_{A}}{\hat{p}_{4}}} \label{eq-flipping2} \eq

\newpage

However we see in (\ref{eq-flipping2}) the numerator generated from
subtraction has a different $i\epsilon$ associated with line
$(p_{3}+p_{4})$ and does not exactly cancel the factor
$1/\left(\frac{p_{1}^{2}+i\epsilon_{B}}{\hat{p}_{1}}+\frac{p_{2}^{2}+i\epsilon_{B}}{\hat{p}_{2}}+\frac{(p_{3}+p_{4})^{2}+i\epsilon_{A}}{\hat{p}_{3}+\hat{p}_{4}}\right)$
represented by the small dash line circle on the left.  The difference
can be accounted for if we introduce even more correction graphs
carrying infinitesimal vertices.

\begin{figure}[!h]
\centering
\subfigure{
  \begin{picture}(156,79) (6,7)
    \SetWidth{0.5}
    \Vertex(48,25){4}
    \Vertex(54,25){2}
    \SetWidth{0.375}
    \DashCArc(51,25)(12.21,145,505){1}
    \SetWidth{0.5}
    \Vertex(93,25){4}
    \SetWidth{0.375}
    \Line(15,25)(123,25)
    \Line(75,61)(93,25)
    \Line(30,61)(48,25)
    \SetWidth{0.5}
    \Vertex(99,25){2}
    \SetWidth{0.375}
    \DashCArc(93,25)(12.21,145,505){1}
    \DashCArc(51.83,25.44)(18.84,111.27,258.26){1}
    \DashCArc(89.7,25)(18.3,-79.61,79.61){1}
    \DashLine(48,7)(93,7){1}
    \DashLine(45,43)(93,43){1}
    \BCirc(73,25){4}
    \BCirc(73,25){2}
    \Text(20,13)[br]{$\bar{\mathcal{A}}, 1$}
    \Text(25,54)[br]{$\mathcal{B}, 2$}
    \Text(71,54)[br]{$\mathcal{B}, 3$}
    \Text(125,31)[br]{$\bar{\mathcal{B}}, 4$}
  \end{picture}
  }
\caption{Correction term to the $\bar{\mathcal{A}}$ expansion }
\end{figure}
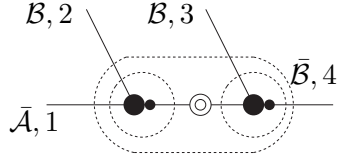

\bq
&& \frac{i(\epsilon_{A}-\epsilon_{B})}{\hat{p}_{3}+\hat{p}_{4}}\,\frac{1}{\frac{p_{3}^{2}+i\epsilon_{B}}{\hat{p}_{3}}+\frac{p_{4}^{2}+i\epsilon_{A}}{\hat{p}_{4}}+\frac{(p_{1}+p_{2})^{2}+i\epsilon_{B}}{\hat{p}_{1}+\hat{p}_{2}}}\vspace{2cm} \nonumber \\
&& \times \frac{1}{\frac{p_{1}^{2}+i\epsilon_{B}}{\hat{p}_{1}}+\frac{p_{2}^{2}+i\epsilon_{B}}{\hat{p}_{2}}+\frac{(p_{3}+p_{4})^{2}+i\epsilon_{A}}{\hat{p}_{3}+\hat{p}_{4}}} \, \frac{1}{\frac{p_{1}^{2}+i\epsilon_{B}}{\hat{p}_{1}}+\frac{p_{2}^{2}+i\epsilon_{B}}{\hat{p}_{2}}+\frac{p_{3}^{2}+i\epsilon_{B}}{\hat{p}_{3}}+\frac{p_{4}^{2}+i\epsilon_{A}}{\hat{p}_{4}}}
\eq

Again these corrections can generally be neglected except for
symmetrical tadpoles such as the graph constructed by contracting leg
$p_{3}$ and $p_{4}$.

Another way to deal with this problem without bothering with the $i
\epsilon$ is to change the orders of the LSZ procedure and the overall
delta function.  Let us look at diagrams:
\\
\centerline{\begin{picture}(150,80)
        \SetOffset(30,24)
        \DashCArc(0,0)(12,103,180){1}
        \DashLine(-3,12)(31,32){1}
\DashCArc(34,20)(12,283,103){1}
\DashCArc(0,0)(12,180,283){1}
        \Line(-25,0)(-2,0)\Text(-30,0)[br]{$p_1$}
        \Line(0,0)(34,20)
        \Vertex(0,0){4}\Vertex(5.6,3.3){2}
	\PhotonArc(17,10)(20,-150,30){1}{13.5}
	\Photon(34,20)(59,35){1.5}{8}
        \Vertex(34,20){4}\Vertex(37,13){2}
\DashCArc(34,20)(8,0,360){1}
        \Text(5,-10)[tr]{$p_4$}
        \Text(42,10)[tl]{$p_3$}
        \DashLine(3,-12)(37,8){1}
        \Text(72,32)[br]{$p_2$}
      \end{picture}
\begin{picture}(150,80)
        \SetOffset(30,40)
        \DashCArc(0,0)(12,-283,-103){1}
        \DashLine(-3,-12)(31,-32){1}
\DashCArc(34,-20)(12,-103,80){1}
        \Line(-25,0)(-2,0)\Text(-30,0)[br]{$p_1$}
        \Line(0,0)(34,-20)
        \Vertex(0,0){4}\Vertex(5,7){2}
	\PhotonArc(17,-10)(20,-30,150){1}{13.5}
	\Photon(34,-20)(59,-35){1.5}{8}
        \Vertex(34,-20){4}\Vertex(28.4,-16.7){2}
\DashCArc(34,-20)(8,0,360){1}
        \Text(5,10)[br]{$p_4$}
        \Text(42,-10)[bl]{$p_3$}
        \DashLine(3,12)(37,-8){1}
        \Text(72,-32)[br]{$p_2$}
      \end{picture}}
\\
We can impose $\delta(\hat p_1+\hat p_2)$ and the momentum
conservation on the right vertex, then apply LSZ procedure and impose
the $\delta(\bar p_1+\bar p_2)\delta(\tilde p_1+\tilde p_2)$ at last.
In the LSZ procedure we impose $p_2^2 \to 0$ first. The first diagram
is proportional to 
\begin{align}
&\lim_{p_1^2\to 0}\lim_{p_2^2\to 0}p_1^2 p_2^2\int_3 \frac {\hat 3}{\hat 1 (\Omega_1 +\Omega_2)}\bigg[\frac{(\bar
\zeta_3-\bar \zeta_1)(\bar \zeta_{23}-\bar
\zeta_2)}{-\Omega_{23}^2+\Omega_2+\Omega_3}\bigg]\frac 1{p_3^2}\frac 1{p_2^2}
\label{eq:barA-barA-1}\\
=&\lim_{p_1^2\to 0}\lim_{p_2^2\to 0}4p_1^2p_2^2\int_3 \frac{\hat 3}{\hat 1 (\frac {p_1^2}{\hat 1}
+\frac{p_2^2}{\hat 2})}
\bigg[\frac{(\bar
\zeta_3-\bar \zeta_1)(\bar \zeta_{23}-\bar
\zeta_2)}{-\frac{P_{23}^2}{\hat P_{23}}+\frac{p_2^2}{\hat 2}+\frac{p_3^2}{\hat
3}}\bigg]\frac 1{p_3^2p_2^2}
\label{eq:barA-barA-2}
\\=&4\int_3 \frac{\hat 3}{p_3^2}
\bigg[\frac{(\bar
\zeta_3-\bar \zeta_1)(\bar \zeta_{23}-\bar
\zeta_2)}{-\frac{P_{23}^2}{\hat P_{23}}+\frac{p_3^2}{\hat
3}}\bigg]
\label{eq:barA-barA-3}\\
=&2\left(\int_3 \frac{\hat 3}{p_3^2}
\bigg[\frac{(\bar
\zeta_3-\bar \zeta_1)(\bar \zeta_{23}-\bar
\zeta_2)}{-\frac{P_{23}^2}{\hat P_{23}}+\frac{p_3^2}{\hat
3}}\bigg]
-\int_3 \frac{\hat P_{23}}{P_{23}^2}
\bigg[\frac{(\bar \zeta_3-\bar \zeta_2)(\bar \zeta_{23}-\bar
\zeta_1)}{-\frac{P_{23}^2}{\hat P_{23}}+\frac{p_3^2}{\hat
3}}\bigg]\right)
\label{eq:barA-barA-4}\\
=&-2\int_3 \frac{\hat 3(\hat{2}+\hat 3)}{p_3^2P_{23}^2}
(\bar
\zeta_3-\bar \zeta_1)(\bar \zeta_{23}-\bar
\zeta_2)
\\&
+\frac{\hat P_{23}}{P_{23}^2\Big(-\frac{P_{23}^2}{\hat P_{23}}+\frac{p_3^2}{\hat
3}\Big)} 
\bigg[(\bar \zeta_3-\bar \zeta_1)(\bar \zeta_{23}-\bar
\zeta_2)-(\bar \zeta_3-\bar \zeta_2)(\bar \zeta_{23}-\bar
\zeta_1)\bigg]
\label{eq:barA-barA-5}
\\=&-2\int_3 \frac{\bar V^2(4,1,23)\bar V^2(41,2,3)}{p_3^2P_{23}^2}
\,,\label{eq:barA-barA-6}
\end{align}
where $\bar \zeta_{ij}=\bar P_{ij}/\hat P_{ij}$. This recovers the
light-cone integral. From (\ref{eq:barA-barA-1}) to
(\ref{eq:barA-barA-2}) we impose $\delta(\hat p_1+\hat p_2)$ and the
momentum conservation on the right vertex.   From
(\ref{eq:barA-barA-2}) to (\ref{eq:barA-barA-3}) we apply the LSZ
procedure. The would-be singularity of $1/(p_1^2/\hat 1+p_2^2/\hat 2)$
is cancelled by the $p_1^2$ factor from LSZ.  From
(\ref{eq:barA-barA-3}) to (\ref{eq:barA-barA-4}) we split the
integrand into two parts and change the integration variable to one
part.  (\ref{eq:barA-barA-4}) to (\ref{eq:barA-barA-5}) is simply
algebra and from (\ref{eq:barA-barA-5}) to (\ref{eq:barA-barA-6}) we
impose the last delta functions.  The second diagram can be worked out
similarly.  In fact, this integral is zero after integration as
required by helicity conservation.  So these kind of diagrams do not
contribute to the amplitude.

\section{Proof of recursion relation (\protect\ref{eq:NewXiR-})
}\label{sect:proof}
We start with the old recursion relation in momentum space:
\begin{eqnarray}
\begin{picture}(70,60)
        \SetOffset(30,30)
        \Line(-25,0)(-2,0)\Text(-26,0)[r]{$\bar 1$}
        \Line(0,0)(17,25)\Line(0,0)(25,0)\Text(26,0)[l]{$\bar\imath$}
        \Line(0,0)(17,-25)\Vertex(13,0){2}
        \GCirc(0,0){8}{0.75}\Text(0,0)[]{\tiny$n\!-\!1$}
        \Text(20,25)[l]{$\bar2$}
        \Text(20,-25)[l]{$\bar n$}
      \Text(30,16)[br]{.}\Text(30,11)[br]{.}\Text(30,-16)[br]{.}\Text(30,-11)[br]{.}
         \end{picture}
\begin{picture}(60,60)
        \SetOffset(30,30)\Text(0,-6)[]{$=-\sum\limits_{1\leq l\leq
n-2}$}
\end{picture}
\begin{picture}(40,60)
        \SetOffset(30,30)
        \Line(25,0)(2,0)
        \Line(0,0)(-17,25)\Line(0,0)(-25,0)\Text(-26,0)[r]{$\bar 1$}
        \Line(0,0)(-17,-25)
        \BCirc(0,0){8}\Vertex(12,0){2}
      \Text(-23,16)[br]{.}\Text(-23,11)[br]{.}\Text(-23,-16)[br]{.}\Text(-23,-11)[br]{.}
         \end{picture}
\begin{picture}(60,60)
        \SetOffset(30,30)
        \Line(-25,0)(-2,0)
        \Line(0,0)(17,25)\Line(0,0)(25,0)\Text(26,0)[l]{$\bar \imath$}
        \Line(0,0)(17,-25)\Vertex(13,0){2}
        \GCirc(0,0){8}{0.75}\Text(0,0)[]{$l$}
      \Text(30,16)[br]{.}\Text(30,11)[br]{.}\Text(30,-16)[br]{.}\Text(30,-11)[br]{.}
         \end{picture}
\begin{picture}(20,60)
        \SetOffset(10,25)
      \Text(0,0)[br]{.}
\end{picture}
 \end{eqnarray}

It is easy to see that for $n=3$
\\
\centerline{
\begin{picture}(60,70)
        \SetOffset(30,35)
        \Line(-25,0)(-2,0)\Text(-26,0)[r]{$\bar 1$}
        \Line(0,0)(25,15)\Text(26,15)[l]{$\bar 2$}
\Vertex(10,6){2}
        \Line(0,0)(25,-15)\Text(26,-15)[l]{$\bar 3$}
        \GCirc(0,0){8}{0.75}
  	\Text(0,0)[ ]{2}
      \end{picture}
\begin{picture}(20,70)
        \SetOffset(10,35)
 \Text(0,0)[]{$=-$}
\end{picture}
\begin{picture}(60,70)
        \SetOffset(30,35)
        \Line(25,0)(2,0)\Text(26,0)[l]{$\bar 2$}
        \Line(0,0)(-25,0)\Text(-26,0)[r]{$\bar 1$}
	\Vertex(12,0){2}
        \Line(0,0)(-25,-15)\Text(-26,-15)[r]{$\bar 3$}
        \BCirc(0,0){8}
  	\Text(0,0)[ ]{2}
\end{picture}
\begin{picture}(14,70)
        \SetOffset(7,35)
 \Text(0,0)[]{$=$}
      \end{picture}
\begin{picture}(60,70)
        \SetOffset(30,35)
        \Line(-25,0)(-2,0)\Vertex(8.3,5){2}
        \Line(0,0)(25,15)
        \Line(0,0)(25,-15)
       \DashCArc(0,0)(12,0,360){1}
        \Vertex(0,0){4}
      \end{picture}
\begin{picture}(10,70)
        \SetOffset(5,35)
	\Text(0,0)[ ]{$\,,$}
\end{picture}}
\\
\centerline{
\begin{picture}(60,70)
        \SetOffset(30,35)
        \Line(-25,0)(-2,0)\Text(-26,0)[r]{$\bar 1$}
        \Line(0,0)(25,15)\Text(26,15)[l]{$\bar 2$}
\Vertex(10,-6){2}
        \Line(0,0)(25,-15)\Text(26,-15)[l]{$\bar 3$}
        \GCirc(0,0){8}{0.75}
  	\Text(0,0)[ ]{2}
      \end{picture}
\begin{picture}(20,70)
        \SetOffset(10,35)
 \Text(0,0)[]{$=-$}
      \end{picture}
\begin{picture}(60,70)
        \SetOffset(30,35)
        \Line(25,0)(2,0)\Text(26,0)[l]{$\bar 3$}
        \Line(0,0)(-25,0)\Text(-26,0)[r]{$\bar 1$}
	\Vertex(12,0){2}
        \Line(0,0)(-25,15)\Text(-26,15)[r]{$\bar 2$}
        \BCirc(0,0){8}
  	\Text(0,0)[ ]{2}
      \end{picture}
\begin{picture}(14,70)
        \SetOffset(7,35)
 \Text(0,0)[]{$=$}
      \end{picture}
\begin{picture}(60,70)
        \SetOffset(30,35)
        \Line(-25,0)(-2,0)\Vertex(8.3,-5){2}
        \Line(0,0)(25,15)
        \Line(0,0)(25,-15)
       \DashCArc(0,0)(12,0,360){1}
        \Vertex(0,0){4}
      \end{picture}
\begin{picture}(10,70)
        \SetOffset(5,35)
	\Text(0,0)[ ]{$\,.$}
\end{picture}}
\\
For $n=4$, $\Xi^1(\bar1\bar2\bar3\bar4)$:
\begin{align}
\begin{picture}(60,60)
        \SetOffset(30,30)
        \Line(-25,0)(-2,0)\Text(-26,0)[r]{$\bar 1$}
        \Line(0,0)(17,25)\Vertex(8.5,12.5){2}
	\Line(0,0)(25,0)\Text(26,0)[l]{$\bar3$}
        \Line(0,0)(17,-25)
        \GCirc(0,0){8}{0.75}\Text(0,0)[]{$3$}
        \Text(20,25)[l]{$\bar2$}
        \Text(20,-25)[l]{$\bar 4$}
\end{picture}
\begin{picture}(14,60)
        \SetOffset(7,30)
 \Text(0,0)[]{$=$}
\end{picture}
&\begin{picture}(14,60)
        \SetOffset(7,30)
 \Text(0,0)[]{$-\bigg[$}
\end{picture}
\begin{picture}(90,60)
        \SetOffset(30,30)
        \Line(25,0)(2,0)
        \Line(0,0)(-25,0)
	\Vertex(12,0){2}
        \Line(0,0)(-25,-15)
        \BCirc(0,0){8}
  	\Text(0,0)[ ]{2}
        \SetOffset(60,30)
        \Line(-5,0)(-2,0)
        \Line(0,0)(25,0)
	\Vertex(10,6){2}
        \Line(0,0)(25,15)
        \GCirc(0,0){8}{0.75}
  	\Text(0,0)[ ]{$2$}
\end{picture}
\begin{picture}(14,60)
        \SetOffset(7,30)
 \Text(0,0)[]{$+$}
\end{picture}
\begin{picture}(55,60)
        \SetOffset(30,30)
        \Line(25,0)(2,0)
        \Line(0,0)(-25,0)
	\Vertex(12,0){2}
        \Line(0,0)(-21.7,-12.5)\Line(0,0)(-12.5,-21.7)
        \BCirc(0,0){8}
  	\Text(0,0)[ ]{3}
\end{picture}
\begin{picture}(14,60)
        \SetOffset(7,30)
 \Text(0,0)[]{$\bigg]$}
\end{picture}
\label{eq:proof-4p-1}\\
\begin{picture}(14,60)
        \SetOffset(7,30)
 \Text(0,0)[]{$=$}
\end{picture}
&\begin{picture}(14,60)
        \SetOffset(7,30)
 \Text(0,0)[]{$-\Bigg[$}
\end{picture}
\begin{picture}(90,60)
        \SetOffset(30,30)
        \Line(25,0)(2,0)
        \Line(0,0)(-25,0)
	\Vertex(12,0){2}
        \Line(0,0)(-25,-15)
        \BCirc(0,0){8}
  	\Text(0,0)[ ]{2}
        \SetOffset(60,30)
        \Line(-5,0)(-2,0)
        \Line(0,0)(25,0)
	\Vertex(10,6){2}
        \Line(0,0)(25,15)
        \GCirc(0,0){8}{0.75}
  	\Text(0,0)[ ]{$2$}
\end{picture}
\begin{picture}(20,60)
        \SetOffset(10,30)
 \Text(0,0)[]{$-\bigg[$}
\end{picture}
\begin{picture}(75,60)
        \SetOffset(30,35)
        \Line(45,0)(2,0)
        \Line(0,0)(-25,0)
	\Vertex(12,0){2}
        \Line(0,0)(-25,-15)
        \BCirc(0,0){8}
  	\Text(0,0)[ ]{2}
	\Vertex(30,0){4}\Vertex(37,0){2}
	\Line(30,0)(5,-15)
        \DashCArc(0,0)(12,90,270){1}
        \DashCArc(30,0)(12,-90,90){1}
	\DashLine(0,-12)(30,-12){1}
	\DashLine(0,12)(30,12){1}
\end{picture}
\begin{picture}(14,60)
        \SetOffset(7,30)
 \Text(0,0)[]{$+$}
\end{picture}
\begin{picture}(75,60)
        \SetOffset(40,35)
        \Line(35,0)(2,0)
        \Line(0,0)(-40,0)
        \Line(0,0)(-15,-15)
	\Vertex(0,0){4}\Vertex(7,0){2}
	\Line(-15,-15)(-35,-15)
        \Line(-15,-15)(-30,-30)
        \BCirc(-15,-15){8}
  	\Text(-15,-15)[ ]{2}
	\Vertex(-7,-7){2}
	\DashCArc(0,0)(12,-45,135){1}
        \DashCArc(-15,-15)(12,135,315){1}
	\DashLine(8.5,-8.5)(-6.5,-23.5){1}
	\DashLine(-8.5,8.5)(-23.5,-6.5){1}
\end{picture}
\begin{picture}(14,60)
        \SetOffset(7,30)
 \Text(0,0)[]{$\bigg]\Bigg]$}
\end{picture}
\label{eq:proof-4p-2}\\
\begin{picture}(14,60)
        \SetOffset(7,30)
 \Text(0,0)[]{$=$}
\end{picture}
&\begin{picture}(20,60)
        \SetOffset(10,30)
 \Text(0,0)[]{$-\Bigg[-$}
\end{picture}
\begin{picture}(90,60)
        \SetOffset(30,30)
        \Line(25,0)(2,0)
        \Line(0,0)(-25,0)
	\Vertex(12,0){2}
        \Line(0,0)(-25,-15)
        \GCirc(0,0){8}{0}
  	\Text(0,0)[ ]{2}
	\DashCArc(0,0)(12,90,270){1}
        \SetOffset(60,30)
        \Line(-5,0)(-2,0)
        \Line(0,0)(25,0)
	\Vertex(10,6){2}
        \Line(0,0)(25,15)
        \GCirc(0,0){8}{0.75}
	\DashCArc(0,0)(12,-90,90){1}
	\DashLine(0,-12)(-30,-12){1}
	\DashLine(0,+12)(-30,+12){1}
  	\Text(0,0)[ ]{$2$}
\end{picture}
\begin{picture}(14,60)
        \SetOffset(7,30)
 \Text(0,0)[]{$-$}
\end{picture}
\begin{picture}(75,60)
        \SetOffset(40,35)
        \Line(35,0)(2,0)
        \Line(0,0)(-40,0)
        \Line(0,0)(-15,-15)
	\Vertex(0,0){4}\Vertex(7,0){2}
	\Line(-15,-15)(-35,-15)
        \Line(-15,-15)(-30,-30)
        \BCirc(-15,-15){8}
  	\Text(-15,-15)[ ]{2}
	\Vertex(-7,-7){2}
	\DashCArc(0,0)(12,-45,135){1}
        \DashCArc(-15,-15)(12,135,315){1}
	\DashLine(8.5,-8.5)(-6.5,-23.5){1}
	\DashLine(-8.5,8.5)(-23.5,-6.5){1}
\end{picture}
\begin{picture}(14,60)
        \SetOffset(7,30)
 \Text(0,0)[]{$\bigg]$}
\end{picture}
\label{eq:proof-4p-3}\\
\begin{picture}(14,60)
        \SetOffset(7,30)
 \Text(0,0)[]{$=$}
\end{picture}
&
\begin{picture}(90,60)
        \SetOffset(30,30)
        \Line(25,0)(2,0)
        \Line(0,0)(-25,0)
	\Vertex(7,0){2}
        \Line(0,0)(25,-15)
        \GCirc(0,0){4}{0}
	\DashCArc(0,0)(12,90,270){1}
        \SetOffset(60,30)
        \Line(-5,0)(-2,0)
        \Line(0,0)(25,0)
	\Vertex(10,6){2}
        \Line(0,0)(25,15)
        \GCirc(0,0){8}{0.75}
	\DashCArc(0,0)(12,-90,90){1}
	\DashLine(0,-12)(-30,-12){1}
	\DashLine(0,+12)(-30,+12){1}
  	\Text(0,0)[ ]{$2$}
\end{picture}
\begin{picture}(14,60)
        \SetOffset(7,30)
 \Text(0,0)[]{$+$}
\end{picture}
\begin{picture}(75,60)
        \SetOffset(30,35)
        \Line(40,0)(2,0)
        \Line(0,0)(-30,0)
        \Line(0,0)(15,-15)
	\Vertex(0,0){4}\Vertex(7,0){2}
	\Line(15,-15)(35,-15)
        \Line(15,-15)(30,-30)
        \BCirc(15,-15){8}
  	\Text(15,-15)[ ]{2}
	\Vertex(7,-7){2}
	\DashCArc(0,0)(12,45,225){1}
        \DashCArc(15,-15)(12,-135,45){1}
	\DashLine(-8.5,-8.5)(6.5,-23.5){1}
	\DashLine(8.5,8.5)(23.5,-6.5){1}
\end{picture}\label{eq:proof-4p-4}
\begin{picture}(14,60)
        \SetOffset(7,25)
 \Text(0,0)[]{$\,.$}
\end{picture}
\end{align}
Equation (\ref{eq:proof-4p-1}) is just the old recursion relation.
From (\ref{eq:proof-4p-1}) to (\ref{eq:proof-4p-2}) we expand the
second term.  From (\ref{eq:proof-4p-2}) to (\ref{eq:proof-4p-3}) we
combine the first two terms using (\ref{eq:relation}). Then the
recursion relation for $\Xi^1(\bar1\bar2\bar3\bar4)$ is proven.
Similarly, one can also prove that $\Xi^2(\bar1\bar2\bar3\bar4)$,
$\Xi^3(\bar1\bar2\bar3\bar4)$ satisfy the recursion relation.

For $\Xi^s$ with general $n$ arguments, we suppose that for $\Xi^s$
with less than $n$ arguments the recursion relation is already proven.
Then at the first step we combine the following terms from the old
recursion relation 
\begin{equation}
\begin{picture}(14,60)
        \SetOffset(7,30)
 \Text(0,0)[]{$-\Bigg[$}
\end{picture}
\begin{picture}(95,60)
        \SetOffset(30,30)
        \Line(25,0)(2,0)
        \Line(0,0)(-20,0)
	\Vertex(12,0){2}
        \Line(0,0)(-20,-15)\Line(0,0)(-20,15)
	\Text(-20,0)[r]{\small $\bar 1$}
	\Text(-18,2)[b]{$\vdots$}\Text(-18,2.5)[t]{$\vdots$}
        \BCirc(0,0){8}
  	\Text(0,0)[ ]{\tiny $ n\!\!-\!\!2$}
        \SetOffset(60,30)
        \Line(-5,0)(-2,0)
        \Line(0,0)(20,0)\Text(26,0)[l]{\small$\bar \imath$}
	\Vertex(12,0){2}
        \Line(0,0)(20,15)\Text(21,15)[l]{\small$\overline{ i\!\!-\!\!1}$}
        \GCirc(0,0){8}{0.75}
  	\Text(0,0)[ ]{$2$}
\end{picture}
\begin{picture}(14,60)
        \SetOffset(7,30)
 \Text(0,0)[]{$+$}
\end{picture}
\begin{picture}(105,60)
        \SetOffset(30,30)
        \Line(25,0)(2,0)
        \Line(0,0)(-20,0)
	\Vertex(12,0){2}
        \Line(0,0)(-20,-15)\Line(0,0)(-20,15)
	\Text(-20,0)[r]{\small $\bar 1$}
	\Text(-18,2)[b]{$\vdots$}\Text(-18,2.5)[t]{$\vdots$}
        \BCirc(0,0){8}
  	\Text(0,0)[ ]{\tiny $ n\!\!-\!\!2$}
        \SetOffset(60,30)
        \Line(-5,0)(-2,0)
        \Line(0,0)(20,0)\Text(21,0)[l]{\small$\bar \imath$}
	\Vertex(12,0){2}
        \Line(0,0)(20,-15)\Text(25,-15)[l]{\small$\overline{ i\!\!+\!\!1}$}
        \GCirc(0,0){8}{0.75}
  	\Text(0,0)[ ]{$2$}
\end{picture}\begin{picture}(14,60)
        \SetOffset(7,30)
 \Text(0,0)[]{$+$}
\end{picture}
\begin{picture}(55,60)
        \SetOffset(30,30)
        \Line(20,0)(2,0)
        \Line(0,0)(-20,0)\Text(-21,0)[r]{\small$\bar 1$}
	\Vertex(12,0){2}
        \Text(-18,2)[b]{$\vdots$}\Text(-18,2.5)[t]{$\vdots$}
	\Line(0,0)(-20,-15)\Line(0,0)(-20,15)
        \BCirc(0,0){8}
  	\Text(0,0)[ ]{\tiny$n\!\!-\!\!1$}
\end{picture}
\begin{picture}(14,60)
        \SetOffset(7,30)
 \Text(0,0)[]{$\Bigg]\,.$}
\end{picture}\label{eq:step1}
\end{equation}
By expanding the third term using recursion for $\Upsilon$ and
combining terms, using the relation (\ref{eq:relation}), we obtain
\begin{align}
\begin{picture}(95,60)
        \SetOffset(30,30)
        \Line(25,0)(2,0)
        \Line(0,0)(-20,0)
	\Vertex(12,0){2}
        \Line(0,0)(-20,-15)\Line(0,0)(-20,15)
	\Text(-20,0)[r]{\small $\bar 1$}
	\Text(-18,2)[b]{$\vdots$}\Text(-18,2.5)[t]{$\vdots$}
        \GCirc(0,0){8}{0}
  	\Text(0,0)[ ]{\tiny $ n\!\!-\!\!2$}
        \DashCArc(0,0)(12,90,270){1}
        \DashCArc(30,0)(12,-90,90){1}
	\DashLine(0,-12)(30,-12){1}
	\DashLine(0,12)(30,12){1}
        \SetOffset(60,30)
        \Line(-5,0)(-2,0)
        \Line(0,0)(20,0)\Text(26,0)[l]{\small$\bar \imath$}
	\Vertex(12,0){2}
        \Line(0,0)(20,15)\Text(21,15)[l]{\small$\overline{ i\!\!-\!\!1}$}
        \GCirc(0,0){8}{0.75}
  	\Text(0,0)[ ]{$2$}
\end{picture}
\begin{picture}(12,60)
        \SetOffset(6,30)
 \Text(0,0)[]{$+$}
\end{picture}
\begin{picture}(98,60)
        \SetOffset(30,30)
        \Line(25,0)(2,0)
        \Line(0,0)(-20,0)
	\Vertex(12,0){2}
        \Line(0,0)(-20,-15)\Line(0,0)(-20,15)
	\Text(-20,0)[r]{\small $\bar 1$}
	\Text(-18,2)[b]{$\vdots$}\Text(-18,2.5)[t]{$\vdots$}
        \GCirc(0,0){8}{0}
  	\Text(0,0)[ ]{\tiny $ n\!\!-\!\!2$}
        \DashCArc(0,0)(12,90,270){1}
        \DashCArc(30,0)(12,-90,90){1}
	\DashLine(0,-12)(30,-12){1}
	\DashLine(0,12)(30,12){1}
        \SetOffset(60,30)
        \Line(-5,0)(-2,0)
        \Line(0,0)(20,0)\Text(21,0)[l]{\small$\bar \imath$}
	\Vertex(12,0){2}
        \Line(0,0)(20,-15)\Text(25,-15)[l]{\small$\overline{ i\!\!+\!\!1}$}
        \GCirc(0,0){8}{0.75}
  	\Text(0,0)[ ]{$2$}
\end{picture}
&\begin{picture}(30,60)
        \SetOffset(15,30)
 \Text(0,0)[]{$+\sum\limits_{m\ge 2}$}
\end{picture}
\begin{picture}(98,60)
        \SetOffset(30,30)
        \Line(25,0)(2,0)
        \Line(0,0)(-20,0)
	\Vertex(12,0){2}
        \Line(0,0)(-20,-15)\Line(0,0)(-20,15)
	\Text(-20,0)[r]{\small $\bar 1$}
	\Text(-18,2)[b]{$\vdots$}\Text(-18,2.5)[t]{$\vdots$}
        \BCirc(0,0){8}
        \DashCArc(0,0)(12,90,270){1}
        \DashCArc(50,0)(12,-90,90){1}
	\DashLine(0,-12)(50,-12){1}
	\DashLine(0,12)(50,12){1}
	\Vertex(25,0){4}
	\Vertex(31,-4.5){2}
	\Line(25,0)(45,-15)\Text(45,-15)[tl]{$\bar \imath$}
        \SetOffset(80,30)
        \Line(-25,0)(-2,0)
        \Line(0,0)(20,15)
        \Line(0,0)(20,-15)
	\Text(18,2)[ ]{$\vdots$}
        \BCirc(0,0){8}
  	\Text(0,0)[ ]{\small $m$}
	\Vertex(-12,0){2}
\end{picture}
\nonumber\\
&\begin{picture}(30,60)
        \SetOffset(15,30)
 \Text(0,0)[]{$+\sum\limits_{m\ge 2}$}
\end{picture}
\begin{picture}(100,60)
        \SetOffset(30,30)
        \Line(25,0)(2,0)
        \Line(0,0)(-20,0)
	\Vertex(12,0){2}
        \Line(0,0)(-20,-15)\Line(0,0)(-20,15)
	\Text(-20,0)[r]{\small $\bar 1$}
	\Text(-18,2)[b]{$\vdots$}\Text(-18,2.5)[t]{$\vdots$}
        \BCirc(0,0){8}
        \DashCArc(0,0)(12,90,270){1}
        \DashCArc(50,0)(12,-90,90){1}
	\DashLine(0,-12)(50,-12){1}
	\DashLine(0,12)(50,12){1}
	\Vertex(25,0){4}
	\Vertex(31,4.5){2}
	\Line(25,0)(45,15)\Text(45,15)[bl]{$\bar \imath$}
        \SetOffset(80,30)
        \Line(-25,0)(-2,0)
        \Line(0,0)(20,15)
        \Line(0,0)(20,-15)
	\Text(18,2)[ ]{$\vdots$}
        \BCirc(0,0){8}
  	\Text(0,0)[ ]{\small $m$}
	\Vertex(-12,0){2}
\end{picture}
\end{align}
The first two terms come from the first two terms in (\ref{eq:step1})
combined with two terms from the expansion of the last terms in
(\ref{eq:step1}).  The two sums are what is left from the expansion of
the last term in (\ref{eq:step1}).

For step $l-1$, $3\le l\le n-3$, we combine terms 
\begin{align}
\begin{picture}(24,60)
        \SetOffset(12,30)
 \Text(0,0)[]{$-\Bigg[\sum$}
\end{picture}
\begin{picture}(95,60)
        \SetOffset(30,30)
        \Line(25,0)(2,0)
        \Line(0,0)(-20,0)
	\Vertex(12,0){2}
        \Line(0,0)(-20,-15)\Line(0,0)(-20,15)
	\Text(-20,0)[r]{\small $\bar 1$}
	\Text(-18,2)[b]{$\vdots$}\Text(-18,2.5)[t]{$\vdots$}
        \BCirc(0,0){8}
  	\Text(0,0)[ ]{\tiny $ n\!\!-\!\!l$}
        \SetOffset(60,30)
        \Line(-5,0)(-2,0)
        \Line(0,0)(20,0)\Text(26,0)[l]{\small$\bar \imath$}
	\Vertex(12,0){2}
        \Line(0,0)(20,15)\Text(18,2)[b]{$\vdots$}\Text(18,2.5)[t]{$\vdots$}
	\Line(0,0)(20,-15)
        \GCirc(0,0){8}{0.75}
  	\Text(0,0)[ ]{$l$}
\end{picture}
\begin{picture}(24,60)
        \SetOffset(12,30)
 \Text(0,0)[]{$-\sum$}
\end{picture}
\begin{picture}(95,60)
        \SetOffset(30,30)
        \Line(25,0)(2,0)
        \Line(0,0)(-20,0)
	\Vertex(12,0){2}
        \Line(0,0)(-20,-15)\Line(0,0)(-20,15)
	\Text(-20,0)[r]{\small $\bar 1$}
	\Text(-18,2)[b]{$\vdots$}\Text(-18,2.5)[t]{$\vdots$}
        \GCirc(0,0){8}{0}
        \DashCArc(0,0)(12,90,270){1}
        \DashCArc(30,0)(12,-90,90){1}
	\DashLine(0,-12)(30,-12){1}
	\DashLine(0,12)(30,12){1}
        \SetOffset(60,30)
        \Line(-5,0)(-2,0)
        \Line(0,0)(20,0)\Text(26,0)[l]{\small$\bar \imath$}
	\Vertex(12,0){2}
        \Line(0,0)(20,15)\Text(18,2)[b]{$\vdots$}\Text(18,2.5)[t]{$\vdots$}
	\Line(0,0)(20,-15)
        \GCirc(0,0){8}{0.75}
  	\Text(0,0)[ ]{\tiny$l\!\!-\!\!1$}
\end{picture}
&\begin{picture}(30,60)
        \SetOffset(15,25)
 \Text(0,0)[]{$-\sum\limits_{m\ge 2}$}
\end{picture}
\begin{picture}(100,60)
        \SetOffset(30,15)
        \Line(25,0)(2,0)
        \Line(0,0)(-20,0)
	\Vertex(12,0){2}
        \Line(0,0)(-20,-15)\Line(0,0)(-20,15)
	\Text(-20,0)[r]{\small $\bar 1$}
	\Text(-18,2)[b]{$\vdots$}\Text(-18,2.5)[t]{$\vdots$}
        \BCirc(0,0){8}
        \DashCArc(0,0)(12,145,270){1}
        \DashCArc(50,0)(12,-90,45){1}
        \DashCArc(25,25)(12,45,135){1}
	\DashLine(0,-12)(50,-12){1}
	\DashLine(-8.5,8.5)(16.5,33.5){1}
	\DashLine(58.5,8.5)(33.5,33.5){1}
	\Vertex(25,0){4}
	\Vertex(32,0){2}
        \SetOffset(80,15)
        \Line(-25,0)(-2,0)
        \Line(0,0)(20,15)
        \Line(0,0)(20,-15)
	\Text(18,2)[ ]{$\vdots$}
        \GCirc(0,0){8}{0.75}
  	\Text(0,0)[ ]{\tiny $l\!\!-\!\!m$}
\SetOffset(55,40)
        \Line(0,-25)(0,0)
        \Line(0,0)(15,20)
        \Line(0,0)(-15,20)
	\Text(2,18)[ ]{$\cdots$}
        \BCirc(0,0){8}\Vertex(0,-12){2}
  	\Text(0,0)[ ]{\small $m$}
\end{picture}
\nonumber\\
&
\begin{picture}(30,60)
        \SetOffset(15,30)
 \Text(0,0)[]{$-\sum\limits_{m\ge 2}$}
\end{picture}
\begin{picture}(100,60)
        \SetOffset(30,45)
        \Line(25,0)(2,0)
        \Line(0,0)(-20,0)
	\Vertex(12,0){2}
        \Line(0,0)(-20,-15)\Line(0,0)(-20,15)
	\Text(-20,0)[r]{\small $\bar 1$}
	\Text(-18,2)[b]{$\vdots$}\Text(-18,2.5)[t]{$\vdots$}
        \BCirc(0,0){8}
        \DashCArc(0,0)(12,90,225){1}
        \DashCArc(50,0)(12,-45,90){1}
        \DashCArc(25,-25)(12,-135,-45){1}
	\DashLine(0,12)(50,12){1}
	\DashLine(-8.5,-8.5)(16.5,-33.5){1}
	\DashLine(58.5,-8.5)(33.5,-33.5){1}
	\Vertex(25,0){4}
	\Vertex(32,0){2}
        \SetOffset(80,45)
        \Line(-25,0)(-2,0)
        \Line(0,0)(20,15)
        \Line(0,0)(20,-15)
	\Text(18,2)[ ]{$\vdots$}
        \GCirc(0,0){8}{0.75}
  	\Text(0,0)[ ]{\tiny $l\!\!-\!\!m$}
\SetOffset(55,20)
        \Line(0,25)(0,0)
        \Line(0,0)(15,-20)
        \Line(0,0)(-15,-20)
	\Text(2,-18)[ ]{$\cdots$}
        \BCirc(0,0){8}\Vertex(0,12){2}
  	\Text(0,0)[ ]{\small $m$}
\end{picture}
\begin{picture}(14,60)
        \SetOffset(7,30)
 \Text(0,0)[]{$\Bigg]$}
\end{picture}\label{eq:stepl}
\end{align}
 where the first term is from the old recursion relation, the second
term and the $m=2$ terms in the last two sums come from step $l-2$ and
the other terms in the sums come from step $l-m$.  After expanding the
grey blob in the first term and the black blob in the second terms,
collecting terms using the relation (\ref{eq:relation}) and counting
in the other terms left from step $l-m$, we obtain
\begin{align}
\begin{picture}(14,60)
        \SetOffset(7,30)
 \Text(0,0)[]{$\sum$}
\end{picture}
\begin{picture}(95,60)
        \SetOffset(30,30)
        \Line(25,0)(2,0)
        \Line(0,0)(-20,0)
	\Vertex(12,0){2}
        \Line(0,0)(-20,-15)\Line(0,0)(-20,15)
	\Text(-20,0)[r]{\small $\bar 1$}
	\Text(-18,2)[b]{$\vdots$}\Text(-18,2.5)[t]{$\vdots$}
        \GCirc(0,0){8}{0}
        \DashCArc(0,0)(12,90,270){1}
        \DashCArc(30,0)(12,-90,90){1}
	\DashLine(0,-12)(30,-12){1}
	\DashLine(0,12)(30,12){1}
        \SetOffset(60,30)
        \Line(-5,0)(-2,0)
        \Line(0,0)(20,0)\Text(26,0)[l]{\small$\bar \imath$}
	\Vertex(12,0){2}
        \Line(0,0)(20,15)\Text(18,2)[b]{$\vdots$}\Text(18,2.5)[t]{$\vdots$}
	\Line(0,0)(20,-15)
        \GCirc(0,0){8}{0.75}
  	\Text(0,0)[ ]{$l$}
\end{picture}
\begin{picture}(60,60)
        \SetOffset(30,30)
 \Text(0,0)[]{$+\sum\limits_{m\ge 1}\sum\limits_{r\ge m+1}\Bigg[$}
\end{picture}
\begin{picture}(100,60)
        \SetOffset(30,15)
        \Line(25,0)(2,0)
        \Line(0,0)(-20,0)
	\Vertex(12,0){2}
        \Line(0,0)(-20,-15)\Line(0,0)(-20,15)
	\Text(-20,0)[r]{\small $\bar 1$}
	\Text(-18,2)[b]{$\vdots$}\Text(-18,2.5)[t]{$\vdots$}
        \BCirc(0,0){8}
        \DashCArc(0,0)(12,145,270){1}
        \DashCArc(50,0)(12,-90,45){1}
        \DashCArc(25,25)(12,45,135){1}
	\DashLine(0,-12)(50,-12){1}
	\DashLine(-8.5,8.5)(16.5,33.5){1}
	\DashLine(58.5,8.5)(33.5,33.5){1}
	\Vertex(25,0){4}
	\Vertex(32,0){2}
        \SetOffset(80,15)
        \Line(-25,0)(-2,0)
        \Line(0,0)(20,15)
        \Line(0,0)(20,-15)
	\Text(18,2)[ ]{$\vdots$}
        \GCirc(0,0){8}{0.75}
  	\Text(0,0)[ ]{\tiny $l\!\!-\!\!m$}
\SetOffset(55,40)
        \Line(0,-25)(0,0)
        \Line(0,0)(15,20)
        \Line(0,0)(-15,20)
	\Text(2,18)[ ]{$\cdots$}
        \BCirc(0,0){8}\Vertex(0,-12){2}
  	\Text(0,0)[ ]{\small $r$}
\end{picture}
\begin{picture}(14,60)
        \SetOffset(7,30)
 \Text(0,0)[]{$+$}
\end{picture}
\begin{picture}(100,60)
        \SetOffset(30,45)
        \Line(25,0)(2,0)
        \Line(0,0)(-20,0)
	\Vertex(12,0){2}
        \Line(0,0)(-20,-15)\Line(0,0)(-20,15)
	\Text(-20,0)[r]{\small $\bar 1$}
	\Text(-18,2)[b]{$\vdots$}\Text(-18,2.5)[t]{$\vdots$}
        \BCirc(0,0){8}
        \DashCArc(0,0)(12,90,225){1}
        \DashCArc(50,0)(12,-45,90){1}
        \DashCArc(25,-25)(12,-135,-45){1}
	\DashLine(0,12)(50,12){1}
	\DashLine(-8.5,-8.5)(16.5,-33.5){1}
	\DashLine(58.5,-8.5)(33.5,-33.5){1}
	\Vertex(25,0){4}
	\Vertex(32,0){2}
        \SetOffset(80,45)
        \Line(-25,0)(-2,0)
        \Line(0,0)(20,15)
        \Line(0,0)(20,-15)
	\Text(18,2)[ ]{$\vdots$}
        \GCirc(0,0){8}{0.75}
  	\Text(0,0)[ ]{\tiny $l\!\!-\!\!m$}
\SetOffset(55,20)
        \Line(0,25)(0,0)
        \Line(0,0)(15,-20)
        \Line(0,0)(-15,-20)
	\Text(2,-18)[ ]{$\cdots$}
        \BCirc(0,0){8}\Vertex(0,12){2}
  	\Text(0,0)[ ]{\small $r$}
\end{picture}
\begin{picture}(14,60)
        \SetOffset(7,30)
 \Text(0,0)[]{$\Bigg]\,.$}
\end{picture}\label{eq:stepl-result}
\end{align}
Iterate this procedure from (\ref{eq:stepl}), and at the last step
$l=n-2$, one can find the result (\ref{eq:stepl-result}) is just the
right hand side of the recursion relation to be proved.

\end{document}